\documentclass[prb, twocolumn, showpacs,10pt,superscriptaddress]{revtex4-1}

\usepackage{amssymb,amsfonts,amsmath} 
\usepackage{graphicx}
\usepackage{hyperref}
\usepackage{color}
\usepackage{graphics}
\usepackage[dvipsnames]{xcolor}

\bibpunct{[}{]}{,}{n}{}{}

\definecolor{lightblue}{RGB}{89, 171, 227}

\definecolor{change}{RGB}{255, 0, 0}

\begin{document} 

\title{The interacting Rice-Mele -- bulk and boundaries}

\author{Y.-T.~Lin}
\affiliation{Institut f{\"u}r Theorie der Statistischen Physik, RWTH Aachen University and
  JARA---Fundamentals of Future Information Technology, 52056 Aachen, Germany}
\author{D.~M.~Kennes}
\affiliation{Institut f{\"u}r Theorie der Statistischen Physik, RWTH Aachen University and
  JARA---Fundamentals of Future Information Technology, 52056 Aachen, Germany}
 \affiliation{Max Planck Institute for the Structure and Dynamics of Matter, Center for Free Electron Laser Science, 22761 Hamburg, Germany}
\author{M.~Pletyukhov}
\affiliation{Institut f{\"u}r Theorie der Statistischen Physik, RWTH Aachen University and
  JARA---Fundamentals of Future Information Technology, 52056 Aachen, Germany}
 \author{C. S. Weber}
\affiliation{Institut f{\"u}r Theorie der Statistischen Physik, RWTH Aachen University and
  JARA---Fundamentals of Future Information Technology, 52056 Aachen, Germany}
\author{H.~Schoeller}
\affiliation{Institut f{\"u}r Theorie der Statistischen Physik, RWTH Aachen University and
  JARA---Fundamentals of Future Information Technology, 52056 Aachen, Germany}
\author{V.~Meden}
\affiliation{Institut f{\"u}r Theorie der Statistischen Physik, RWTH Aachen University and
  JARA---Fundamentals of Future Information Technology, 52056 Aachen, Germany}

\begin{abstract}

We investigate the interacting, one-dimensional Rice-Mele model, a prototypical fermionic model of topological properties. 
To set the stage, we firstly compute the single-particle spectral function, the local density, and the boundary charge in the absence of interactions. We find that the fractional part of the boundary charge is fully determined by bulk properties of the lattice model. In a large parameter regime the boundary chargeagrees with the one obtained from an effective low-energy theory (arXiv:2004.00463). Secondly, we investigate the robustness of our results towards two-particle interactions. To resum the series of leading logarithms for small gaps, which dismantle plain perturbation theory in the interaction, we use an essentially analytical renormalization group approach. It is controlled for small interactions and can directly be applied to the microscopic lattice model. We benchmark the results against numerical density matrix renormalization group data. The main interaction effect in the bulk is a power-law renormalization of the gap with an interaction dependent exponent. The important characteristics of the fractional part of the boundary charge are unaltered and can be understood from the renormalized bulk properties. This requires a consistent treatment not only of the low-energy gap renormalization but also of the high-energy band width one. In contrast to low-energy field theories our renormalization group approach also provides the latter. We show that the interaction spoils the relation between the bulk properties and the number of edge states, consistent with the observation that the Rice-Mele model with finite potential modulation does not reveal any zero-energy edge states. 

\end{abstract}

\date{\today} 
\maketitle

\section{Introduction}
\label{sec:intro}

For noninteracting quantum many-body systems the relation between the topological properties and the behavior of typical observables is well understood \cite{volkov_pankratov_JETP_85,pankratov_etal_SSC_87,kane_mele_prl_95,bernevig_etal_science_06,fu_kane_mele_prl_07,koenig_etal_science_07,hsieh_etal_nature_08}; see Refs.~\cite{hasan_kane_RMP_10,qi_zhang_RMP_11,bernevig_book_13,tkachov_book_15,asboth_book_16} for reviews and textbooks. However, in many respects this understanding relies on ideas which make explicit use of the concept of independent particles. Insights on the relation between topology and the physics of interacting many-body systems are based either on case studies for specific models or on general considerations of how to extend the concept of topological invariants to the realm of correlated  systems \cite{gangadharaiah_etal_prl_11,Stoudenmire11,wang_zhang_prx_12,manmana_etal_prb_12,he_etal_prb_16,sbierski_karrasch_prb_18,Yahyavi18,Fidkowski_Kitaev_prb_10,Fidkowski_Kitaev_prb_11,Turner_Pollmann_Berg_prb_11,Morimoto_Furusaki_Mudry_prb_15,GM_prb_19}.

We here provide a case study for the interacting, one-dimensional (1d), and spinless Rice-Mele (RM) model \cite{Rice82}. For vanishing interaction and up to isolated points in the space of the single-particle parameters the model is an insulator with phases of distinct topological properties. It is one of the most elementary models with a band gap in the spectrum and was set up in the early eighties when investigating the electronic properties of linear polymers  \cite{Heeger01}. The model consists of two-site unit cells with an intra-cell hopping matrix element $t_1$ and alternating onsite energies $V_1$ and $V_2$. The unit cells are coupled by a nearest-neighbor inter-cell hopping $t_2$. For degenerate onsite energies it becomes the famous Su-Schriefer-Heeger (SSH) model \cite{Su79}. We add a nearest-neighbor two-particle interaction of amplitude $U$ to the Hamiltonian.  

 One of the hallmarks of topological systems is the bulk-boundary correspondence  \cite{fidkowski_etal_prl_11,mong_shivamoggi_prb_11,gurarie_prb_11,essin_gurarie_prb_11,ukui_etal_jpsjf_12,yu_wu_xie_npb_17,rhim_etal_prb_18,silveirinha_prx_19}. It is formulated in terms of a connection between topological bulk invariants and the appearance of edge states. For 1d systems, it is known that topological invariants are related to the number or parity of zero-energy edge states. Besides these topological edge states, there are other properties of a system close to a boundary which can solely be understood based on bulk characteristics. To investigate them we first solve the noninteracting infinite and semi-infinite RM model. We focus on three observables: The local single-particle spectral function, the local density, and the so-called boundary charge accumulated close to the boundary. Edge states show up as in-gap $\delta$-peaks in the local single-particle spectral function. The boundary charge, which is computed from the local density, is influenced by the number of edge states via an integer number. However, the {\it fractional} part of the boundary charge is an alternative and fundamentally different observable. It is influenced by the whole spectrum of extended states, which carry also important information from the boundary.

As shown for noninteracting and clean systems via the polarization in terms of the Zak-Berry phase \cite{kingsmith_vanderbilt_prb_93,vanderbilt_kingsmith_prb_93,resta,resta_revmodphys_94,marzari_etal_revmodphys_12,vanderbilt_book_2018,ortiz_martin_prb_94,rhim_etal_prb_17,miert_ortix_prb_17} and recently also for disordered and interacting systems \cite{pletyukhov_etal_preprint} the fractional part of the boundary charge shows characteristics which follow directly from bulk properties. Furthermore it is an interesting observable in its own right as it indicates various universal properties, such as the linear phase-dependence against continuous translations of the lattice \cite{park_etal_prb_16,thakurathi_etal_prb_18,pletyukhov_etal_short,pletyukhov_etal_long}, the possibility to realize rational quantization in the presence of symmetries \cite{pletyukhov_etal_preprint}, and a universal low-energy behavior for very small gaps \cite{pletyukhov_etal_preprint}.  Moreover, the fractional part of the boundary charge can be related to the bulk polarization which can be defined generically for any many-body system in terms of the phase of the ground state expectation value of an exponential containing the position operator \cite{resta_prl_98,resta_sorella_prl_99}. However, this quantity is quite hard to measure in an experiment, whereas the boundary charge is directly accessible and can be calculated easily from the density.

Our first important step is thus to compute the boundary charge for the noninteracting RM model and illustrate the above mentioned characteristics resulting from bulk properties. We show that results obtained from an effective low-energy theory for gaps much smaller than the band width \cite{pletyukhov_etal_preprint} hold in a surprisingly large parameter regime. In addition, we find an interesting $\frac{1}{4}$-quantization of the boundary charge in the limit of large gaps.

In 1d metallic systems two-particle interactions imply correlations which strongly alter the low-energy physics. They lead to Tomonaga-Luttinger liquid behavior \cite{Giamarchi03,Schoenhammer05} which can not be captured by perturbation theory in the interaction. In fact, perturbative approaches are plagued by logarithmic infrared divergences. One can expect that in systems with a band gap $2 \Delta$, and the chemical potential placed in the gap, these are cut off by $\Delta$ leading to dominant terms of the form $U^n \ln^n (\Delta/W)$, with the band width $2W$ and $n$ being the order of perturbation theory. In fact, such terms are found in plain perturbation theory for the interacting RM model (see below). They severely limit its applicability in the limit of small gaps as corrections to the leading term become exceedingly large. Thus, as for metallic systems \cite{Giamarchi03,Schoenhammer05}, in the past low-energy field theories and  field-theoretical methods, such as bosonization, were employed \cite{park_etal_prb_16,thakurathi_etal_prb_18,pletyukhov_etal_preprint,Kivelson85,Horovitz85,Wu86,gangadharaiah_etal_prl_12}. They circumvent logarithmic terms. However, if being interested in the properties of a microscopic lattice model, such as the interacting RM model, their application requires the additional approximate step of mapping the lattice model to a continuum field theory \cite{gangadharaiah_etal_prl_12,park_etal_prb_16,thakurathi_etal_prb_18,pletyukhov_etal_preprint}. They are furthermore bound to the low-energy limit.

To study the interacting RM model we follow an alternative route and use an essentially analytical but approximate truncated functional renormalization group (RG) approach \cite{Metzner12,Kopietz10}. This has the distinct advantage that it can directly be applied to the microscopic lattice model and consistently treats interaction effects on all energy scales, from the high-energy band width down to the low-energy gap. The approximations required to derive a finite set of RG flow equations for the components of the static self-energy are controlled for small interactions. Crucially, the solution of these leads to a proper resummation of the leading logarithms to a power law (for related examples, see Ref.~\cite{Metzner12}). We benchmark our functional RG results for the above observables to numerical density matrix renormalization group data (DMRG) \cite{Schollwoeck11}. DMRG was earlier applied to models of topological insulators \cite{Stoudenmire11,Gergs_Fritz_Schuricht_prb16,Milsted_Seabra_Fulga_Beenakker_Cobanera_prb_15,Yu_Li_Sacramento_Lin_prb_16,Pikulin_Chiu_Zhu_Franz_prb_15,GHF_prl_12,manmana_etal_prb_12,Rahmani_Zhu_Franz_Affleck_prb_15,Rahmani_Zhu_Franz_Affleck_prl_15, pletyukhov_etal_preprint}.
However, to reach the low-energy regime for systems with boundaries requires the study of exceedingly large systems which provides a computational challenge to this approach. 

We show that the interaction can induce in-gap $\delta$-peaks that is ``effective edge states'', in the local single-particle spectral function, which are absent for $U=0$. They originate from the local modulation of the self-energy close to the open boundary and cannot be explained based on renormalized bulk properties. Therefore, the appearance of edge states can not be related to bulk properties. These modulations also affect the local density close to the boundary. However, the characteristic features of the {\it fractional} part of the boundary charge remain unaffected and can be explained from the bulk properties of the interacting model. 

This paper is structured as follows. In Sect.~\ref{sec:modelphys} we present the lattice model and introduce the observables of interest
to us. For vanishing interactions, we compute all eigenenergies and wavefunctions for periodic as well as open boundary conditions---including possible (topological) edge states. From these we determine the local single-particle spectral function, the local density, as well as boundary charge accumulated close to the boundary. Details of these calculations are given in the Appendix. In Sect.~\ref{sec:methods} we next present the quantum many-body methods we employ to investigate the interacting RM model: Functional RG and DMRG. In addition, we introduce a field theoretical model to investigate the low-energy physics for small gaps. Our results for the bulk properties of the interacting model are presented in Sect.~\ref{sec:bulk}, while Sect.~\ref{sec:boundary} is devoted to the
study of the physics in the presence of an open boundary. In Sect.~\ref{sec:summary} we provide a---taken the length our paper---short summary of our results. As the individual sections end with summaries of the corresponding parts, we this way avoid a doubling. In addition, we present an outlook.  

\section{The model and its physics at vanishing interaction}
\label{sec:modelphys}

\subsection{The model}
\label{sec:model}

The noninteracting RM model \cite{Rice82} is one of the basic models discussed in connection with edge state physics and topological properties. In 1d, for spinless fermions, and in the Wannier state basis (with lattice site index $j$) it is given by the Hamiltonian
\begin{align}
  H_0 = \sum_j  \left( V_j n^{\phantom{}}_{j} 
  -  \left[t_j  c_{j+1}^\dag c_{j}^{\phantom{}} + \mbox{H.c.} \right] \right)
\label{eq:H0}  
\end{align}
with the site-density operator $n_j =  c_{j}^\dag c_j^{\phantom{}}$.    
Standard second quantized notation is used. The on-site potentials $V_j=V_{j+Z}$ and hoppings $t_j=t_{j+Z}$ are periodic with period $Z=2$, defining the number of lattice sites of the unit cell. With the 
average hopping $t=(t_1+t_2)/2$ and half the difference $\delta t = (t_1-t_2)/2$, we parametrize $V_j$ and $t_j$ by
\begin{align}
    V_1 = -V_2 = V,\quad  t_{1/2} = t \pm \delta t  > 0 . 
    \label{eq:Vt}
\end{align}
We take $t$ as our unit of energy and set $t=1$. In analytic calculations we still find it advantageous to introduce a symbol for an energy scale associated to this average hopping. We use $W=2t$, as it reminds us that $2t$ is half the band width of the gapless model with $\delta t=0=V$.  For compactness we refer to $2W$ as the band width.

As discussed in more detail in Sect.~\ref{sec:U_0results} the RM model displays two bands separated by a single particle gap of 
minimal size $2 \Delta$ (taken at wavevector $k=\pm \pi$, with the lattice constant $a=1$) with 
\begin{align}
    \Delta= \sqrt{V^2 + 4\delta t^2} .
    \label{eq:gap}
\end{align}
It is convenient to define a phase $\gamma \in [0,2\pi)$ via the complex gap parameter
\begin{align}
    \Delta \,e^{i\gamma} = V + i2\delta t . 
    \label{eq:gamma}
\end{align}
We vary $\gamma$ to modulate the staggered hopping and onsite energies such that the complex gap parameter stays on a circle in the complex plane defined by $(V,2\delta t)$. For $V=0$ the RM model becomes the SSH model \cite{Su79}. 

The Hamiltonian $H_0$ is complemented by a homogeneous two-particle
interaction of nearest neighbor type
\begin{align}
  H_{\rm int} = U \sum_j \left( n^{\phantom{}}_{j} -\frac{1}{2} \right)
  \left( n^{\phantom{}}_{j+1} -\frac{1}{2} \right)  ,
 \label{eq:Hint}  
\end{align}
with amplitude $U$. Subtracting $1/2$ from the local density operator $n_j$ the interaction is written in a particle-hole symmetric form.

We take the number of lattice sites $L$ to be even such that all unit cells remain intact. We are interested in the bulk properties as well as the boundary ones. In the former case we consider periodic boundary conditions (PBCs). The site
index $j$ in the sum of Eq.~(\ref{eq:H0}) and Eq.~(\ref{eq:Hint}) runs from $1$ to $L$ and
site indices are considered modulo $L$. For open boundary conditions (OBCs)  the sum in the first term of Eq.~(\ref{eq:H0}) runs from $1$ to $L$ while in the second one of Eq.~(\ref{eq:H0}) and in Eq.~(\ref{eq:Hint}) it only extends up to $L-1$.

For $\delta t= 0=V$ the elementary unit cell has a single site and $H=H_0 + H_{\rm int}$ is the Bethe ansatz solvable (single-band) lattice model of spinless fermions with nearest-neighbor hopping $t=1$ and nearest-neighbor
interaction $U$; see e.g.~Ref.~\cite{Giamarchi03}. For $|U|$ being smaller
than a filling dependent critical interaction it shows metallic behavior. E.g.~for half
filling the model remains gapless for $-2 < U < 2$. 
In this regime the system is a Tomonaga-Luttinger liquid with 
all low-energy excitations being of collective bosonic nature (instead of 
being fermionic quasi-particles) and correlation functions decay 
as power laws with interaction dependent exponents
\cite{Giamarchi03,Schoenhammer05}.  Outside the metallic Tomonaga-Luttinger
liquid phase correlations induce a gap. Here we are not interested in the interplay 
of the single-particle gap $2 \Delta$ of the noninteracting RM model 
and the interaction induced gap and always consider interactions so small that
the latter does not open. For results on this interplay, see Ref.~\cite{Yahyavi18}.  

Before discussing our results on the interaction effects of the spinless 1d RM
model we investigate its $U=0$ properties in the next section. For details, see the
Appendix. We also use this section to introduce our observables
of interest. A particular emphasize is put 
on the boundary charge accumulated close to an open boundary, as it is an 
interesting quantity with characteristics which can be understood solely based on bulk properties of the Hamiltonian.

\subsection{Spectra and the density}
\label{sec:U_0results}

For PBC the noninteracting Hamiltonian Eq.~(\ref{eq:H0}) can easily be diagonalized.
For this we rewrite the Wannier states as  
\begin{align}
  \left| j \right> = 
\left| n \right> \otimes \left| i \right>
\end{align}
with the unit cell index $n$ and the index $i=1,2$ for the two sites within the unit cell. They are related to the lattice site index via
\begin{align}
  j = 2(n-1) + i ,
  \label{eq:ijn}
\end{align}   
a relation which is used implicitly in the following. 

In the single-particle subspace $H_0$ can then be rewritten as \begin{align}
  H_0 = \sum_{n= -\infty}^{\infty}  \left[  \left|n\right> \left< n \right| \otimes
  h(0) + \left|n+1\right> \left< n \right| \otimes h(1) + \mbox{H.c.} \right],
  \label{eq:h0}  
\end{align}
with the $2 \times 2$-matrices 
\begin{align}
  h(0) = \left( \begin{array}{cc} V_1 & -t_1 \\ -t_1 & V_2 \end{array} \right), \quad
  h(1) = \left( \begin{array}{cc} 0 & -t_2 \\ 0 & 0  \end{array} \right)
\label{eq:h1_2}  
\end{align}
in the $i=1,2$ basis. Here we have already taken the thermodynamic limit $L \to \infty$ (infinite system, bulk properties). 
We next define $k$-states
\begin{align}
\left| k \right> = \frac{1}{\sqrt{2 \pi}} \sum_{n= -\infty}^\infty e^{{\rm i} k n} \left|n\right> ,
  \label{eq:kstates}  
\end{align}
with $k \in [-\pi , \pi)$. Taking these as our basis the Hamiltonian reads
\begin{align}
  H_0 = \int_{-\pi}^\pi dk \left|k \right> \left< k \right| \otimes h_k,
  \label{eq:h0_k}  
\end{align}
with
\begin{align}
    \nonumber
        h_k &= \sum_{\delta=0,\pm 1} h(\delta) e^{-ik\delta}\\
        &=\left( \begin{array}{cc} V_1 & -t_1 - t_2 e^{-{\rm i} k}
                 \\-t_1 - t_2 e^{{\rm i}k} & V_2
               \end{array} \right) .
  \label{eq:hk}  
\end{align}
The eigenenergies are given by the eigenvalues of $h_k$ as
\begin{align}
  \varepsilon_{k}^{(\alpha)} = \alpha \varepsilon_k
  = \alpha \sqrt{V^2 + t_1^2 + t_2^2 + 2 t_1 t_2 \cos{k}} ,
  \label{eq:disp}  
\end{align}
with the band index $\alpha = \pm $. We used $V_1 = -V_2=V$. We thus find two bands
separated by a single-particle energy gap which takes its minimal value $2\Delta$ at 
$k = \pm \pi$, with $\Delta$ defined in Eq.~(\ref{eq:gap}). 

The single-particle wave functions of the infinite (bulk) system are given by the Bloch states
\begin{align}
  \psi_{k,{\rm bulk}}^{(\alpha)}(j) = \frac{1}{\sqrt{2 \pi}} \chi_{k}^{(\alpha)}(i)  \, e^{{\rm i} k n} ,  
  \label{eq:wavefunc}  
\end{align}
where 
\begin{align}
  \chi_{k}^{(\alpha)}(1) = \frac{t_1 + t_2 e^{-{\rm i} k}}{\sqrt{N_{k}^{(\alpha)}}}, \quad 
  \chi_{k}^{(\alpha)}(2) = \frac{V - \alpha \varepsilon_k}{\sqrt{N_{k}^{(\alpha)}}} ,  
  \label{eq:wavefunc_defs1}  
\end{align}
with the normalization factor
\begin{align}
 N_{k}^{(\alpha)} = 2 \varepsilon_k (\varepsilon_k - \alpha V) .
 \label{eq:wavefunc_defs2}  
\end{align}
  
We here exclusively consider the case with the chemical potential $\mu$ lying in the gap between the valence and conduction band as well as  temperature $T=0$, such that the lower band is completely filled and the upper one is empty. 

Integrating over the absolute values squared of the wave functions in the lower band we obtain the bulk density. It is translationally invariant by two lattice sites and is given by
\begin{align}
  \rho_{\rm bulk}(j) &= \int_{-\pi}^{\pi} dk \left|\psi_{k,{\rm bulk}}^{(-)}(j)\right|^2  
   = \frac{1}{2\pi}\int_{-\pi}^{\pi} dk \left|\chi_k^{(-)}(i)\right|^2  \nonumber\\
   &=\frac{1}{2} + (-1)^j \frac{V}{4 \pi} \int_{-\pi}^{\pi} dk
  \frac{1}{\varepsilon_k} ,
  \label{eq:rho_bulk}  
\end{align}
where we made use of 
Eqs.~(\ref{eq:wavefunc})-(\ref{eq:wavefunc_defs2}) for the Bloch states and used  Eq.~(\ref{eq:ijn}) relating the indices $i$ and $j$. Closing the integration contour in the upper half we show in the Appendix that the bulk density can be calculated very efficiently from the integral
\begin{align}
    \rho_{\rm bulk}(j) = \frac{1}{2} + (-1)^j \frac{V}{2\pi}
    \int_0^\infty d\kappa \frac{1}{\sqrt{-R(\kappa)}}
    \label{eq:rho_bulk_2}
\end{align}
with 
\begin{align}
    R(\kappa) = V^2 + t_1^2 + t_2^2 - 2t_1 t_2 \cosh{(\kappa + \kappa_{\rm bc})} 
    \label{eq:R}
\end{align}
and 
\begin{align}
  \kappa_{\rm bc} =  \ln{ \frac{\Delta^2 + 2 t_1 t_2 + \Delta
  \sqrt{\Delta^2+ 4 t_1 t_2}}{2 t_1 t_2}} .
  \label{eq:kappa_bc}
\end{align}
As we will see below the length scale $\kappa_{\rm bc}^{-1}$ corresponds to the decay length of the exponential localization of the excess density for a semi-infinite system at the boundary. The fact that this length scale appears also in the calculation of the bulk density provides an interesting link between bulk and boundary quantities.   

The eigenstates of a semi-infinite chain with an open boundary (boundary properties), obtained by starting with OBC and taking $L \to \infty$, are given by
\begin{align}
  \psi_{k}^{(\alpha)} (j) = \frac{1}{\sqrt{2 \pi}} \left[ \chi_{k}^{(\alpha)}(i) \,e^{{\rm i} kn}
 -\chi_{-k}^{(\alpha)}(i)\, e^{-{\rm i} kn} \right] ,  
  \label{eq:wavefunc_semi}  
\end{align}
with $k \in [0,\pi]$. The dispersion remains the same as for the infinite chain; see Eq.~(\ref{eq:disp}).  
  
It is well established that for $t_1-t_2 <0$ the set of extended eigenstates of a semi-infinite chain Eq.~(\ref{eq:wavefunc_semi}) is complemented by an edge state with wavefunction
\begin{align}
   \psi_{\rm e} (j)  = \delta_{i,1} (-1)^{n+1} \left(\frac{t_2^2}{t_1^2}-1\right)^{1/2}\, 
  e^{- \kappa_{\rm e}n}  ,   
    \label{eq:wavefunc_edge}  
\end{align}
and 
\begin{align}
    \kappa_{\rm e} = \ln{\frac{t_2}{t_1}} .
    \label{eq:kappa_e}
\end{align}
It has weight exclusively on the sites with $i=1$ within the unit cell
and decays (purely) exponentially in the unit cell index $n$ away from the boundary.
The characteristic length scale is $\kappa_{\rm e}^{-1}$.  
The edge state is located at energy $V$ within the energy gap. Accordingly,
at $T=0$ the edge state is filled for $V <\mu$ and empty
for $V>\mu$. For $V=\mu$ it is half-filled. In the SSH model limit with $V=0$ the edge state is located
at vanishing energy and classified as topologically protected within the
standard nomenclature of topological insulators. 
The existence of the edge state follows from a property of the bulk parameters, namely $t_1-t_2<0$. 

\begin{figure}[t]
  \begin{center}
  \includegraphics[width=0.5\textwidth,clip]{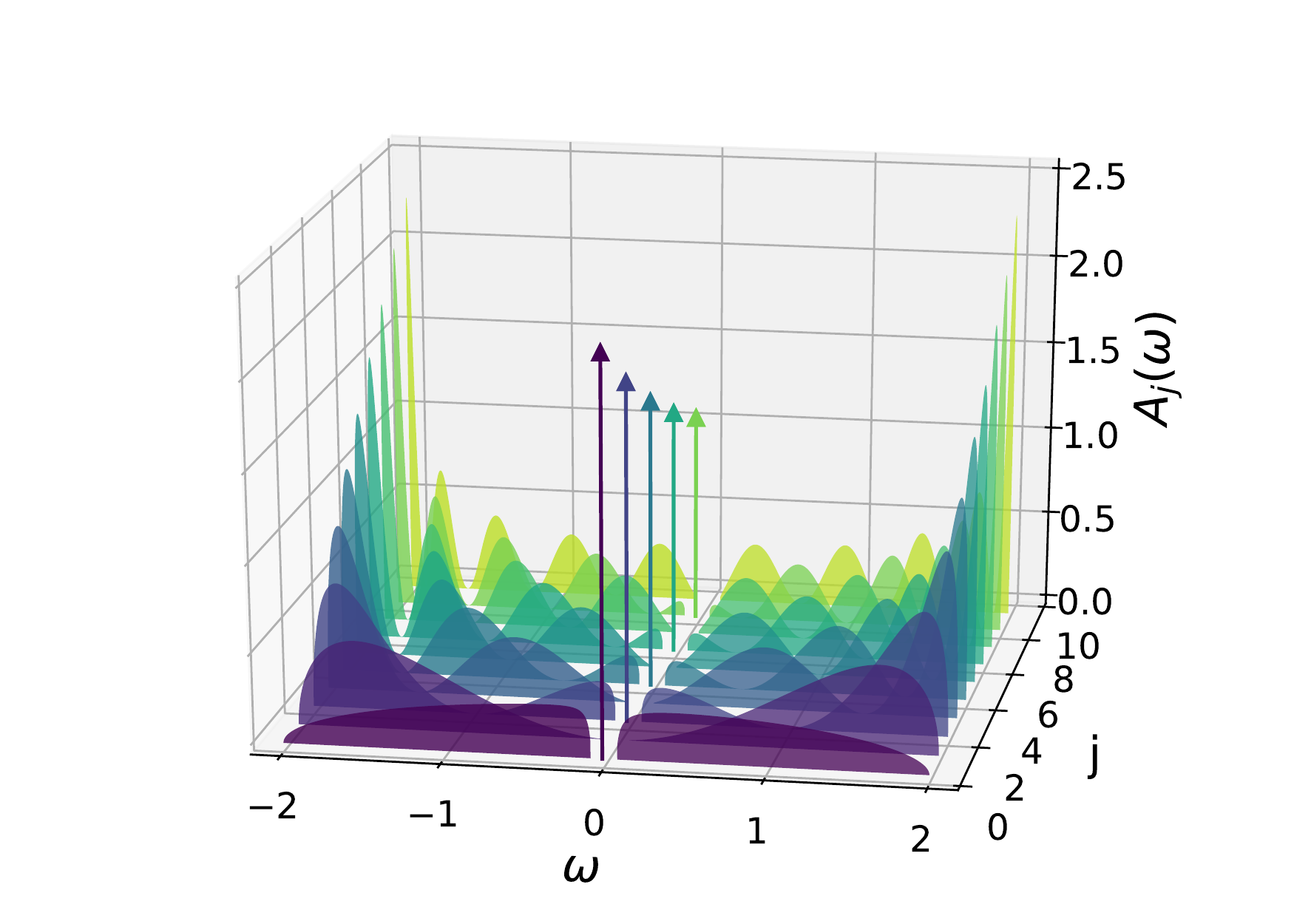}
  \caption{The local single-particle spectral function $A_j(\omega)$ of the noninteracting RM model as a function of energy $\omega$ for different lattice sites $j$ close to an open boundary. The in-gap edge state is indicated by a vertical arrow of height proportional to its weight. The parameters are  $\delta t=-0.04$, and $V=-0.01$.} 
  \label{fig:LDOS_U0}
  \end{center}
\end{figure}

From the eigenenergies and the eigenstates the 
local single-particle spectral function $A_j(\omega)$ of the
semi-infinite system can be computed as
\begin{align}
A_j(\omega) = & \sum_{\alpha = \pm} \int_{0}^\pi dk \left|   \psi_{k}^{(\alpha)} (j)  \right|^2
  \delta\left( \omega - \alpha \varepsilon_k \right) \nonumber        \\
  & +  \left| \psi_{\rm e} (j)
  \right|^2 \delta\left( \omega - V \right)  .
 \label{eq:specfu} 
\end{align}
Figure \ref{fig:LDOS_U0} shows results for different $j$. The parameters are $\delta t=-0.04$, 
and $V=-0.01$, thus from the regime featuring an edge state.
Therefore, the spectral function shows an in-gap $\delta$-peak at energy $\omega=\varepsilon_{\rm e}=V$ 
on odd sites with a weight which according to Eq.~(\ref{eq:wavefunc_edge}) decays exponentially 
for increasing $j=2(n-1)+i$. In the figure it is indicated as a vertical arrow. 
The height of the arrow  is proportional to the weight of the $\delta$-peak. 
The gap is clearly visible. 
Close to the boundary the spectral weight generically (for an 
exception, see Sect.~\ref{sec:subsecspec}) vanishes in a semi-circular way when the energy approaches the band edges. For larger $j$ inverse square-root-like van-Hove singularities typical for the density of states of 1d systems develop. For $\omega \to \pm \Delta$ this is visible only for larger $j$ than shown in Fig.~\ref{fig:LDOS_U0}.

We note that the results of Fig.~\ref{fig:LDOS_U0}   were computed for a finite system of $L=4096$ sites with PBC by numerical diagonalization. To obtain a smooth function out of the sum of $\delta$-peaks (finite system size) we averaged the spectral weight in the bands over several eigenenergies. Increasing the system size the curves do not change on the scale of the plot and the data can considered to be in the thermodynamic limit. 

The local density of the semi-infinite system can be written as
\begin{align}
   \rho(j) &=  \int_{0}^\pi dk \left|\psi_k^{(-)}(j)\right|^2 + \rho_{\rm e}(j) 
   \nonumber \\
   &=\rho_{\rm bulk}(j) + \rho_{\rm F}(j) + \rho_{\rm e}(j) ,
   \label{eq:rho_splitting}
\end{align}
where $\rho_{\rm e}(j)$ denotes the edge state density which is given by
\begin{align}
    \rho_{\rm e}(j)= \Theta(t_2-t_1)
  \left| \psi_{\rm e} (j) \right|^2 \frac{1}{2}\left[1+\text{sign}\left(\mu-V\right)\right] ,
\label{eq:rho_edge}
\end{align}
with $\text{sign}(0) =  0$. As outlined in the Appendix the Friedel density $\rho_{\rm F}$ 
can be split into a pole and branch cut contribution
\begin{align}
      \rho_{\rm F}(j) &= -\frac{1}{2\pi}\int_{-\pi}^\pi dk \left[\chi_k^{(-)}(i)\right]^2 e^{2ikn} 
      \label{eq:rho_F_chi}\\
      &=\rho_{\rm F}^{\rm (pole)}(j) + \rho_{\rm F}^{\rm (bc)}(j) ,  
  \label{eq:rho_F_splitting}
\end{align}
 given by
\begin{align}
  \rho_{\rm F}^{\rm (pole)}(j)  &= -\rho_{\rm e}(j)|_{\mu=0} ,
  \label{eq:rho_F_pole}\\
  \rho_{\rm F}^{\rm (bc)}(n,1)  &= -\frac{V}{2\pi}e^{-2\kappa_{\rm bc}n} \nonumber \\
  &\times\int_0^\infty d\kappa 
  \frac{(t_1-t_2 e^{\kappa + \kappa_{\rm bc}})^2}{\sqrt{-R(\kappa)}\left[V^2-R(\kappa)\right]} e^{-2\kappa n} ,
  \label{eq:rho_F_bc_1}\\
  \rho_{\rm F}^{\rm (bc)}(n,2)  &= -\frac{V}{2\pi}e^{-2\kappa_{\rm bc}n} \int_0^\infty d\kappa 
  \frac{1}{\sqrt{-R(\kappa)}} e^{-2\kappa n} ,
  \label{eq:rho_F_bc_2}
\end{align}
with $R(\kappa)$ and $\kappa_{\rm bc}$ defined in Eqs.~(\ref{eq:R}) and (\ref{eq:kappa_bc}). The pole contribution coincides with the negative edge state density at $\mu=0$. Therefore, for $\mu=0$ it exactly cancels the edge state density $\rho_{\rm e}(j)$ in Eq.~(\ref{eq:rho_splitting}). The second term of $\rho_{\rm F}(j)$ arises from a branch cut contribution and decays exponentially (to zero) for large $n$, i.e.~large $j=2(n-1)+i$, with the characteristic length scale $\kappa_{\rm bc}^{-1}$. Therefore, for $j \to \infty$ the total density Eq.~(\ref{eq:rho_splitting}) for the semi-infinite chain approaches the bulk values Eq.~(\ref{eq:rho_bulk}) as expected. This holds for any $\mu$ located in the gap. Inserting Eqs.~(\ref{eq:rho_F_splitting}) and (\ref{eq:rho_F_pole}) in Eq.~(\ref{eq:rho_splitting}) one finds that the only term depending on such a chemical potential is the difference 
\begin{align}
    \rho_{\rm e}(j)-\rho_{\rm e}(j)|_{\mu=0} = & \Theta(t_2-t_1) \left| \psi_{\rm e} (j) \right|^2   \label{eq:rho_edge_difference} \\
    & \times
    \frac{1}{2}\left[\text{sign}(\mu-V)-\text{sign}(-V)\right] .
\nonumber
\end{align}

In the following we mostly consider the case of vanishing chemical potential 
\begin{align}
    \mu = 0 ,
    \label{eq:mu_zero}
\end{align} 
in which the right hand side of Eq.~(\ref{eq:rho_edge_difference}) is zero. 
Therefore, the difference of the densities of the semi-infinite and infinite system is given by the branch cut contribution of the Friedel density
\begin{align}
  \delta\rho(j) &= \rho(j) - \rho_{\rm bulk}(j) \label{eq:delta_rho}\\
    &= \rho_{\rm F}(j) + \rho_{\rm e}(j) = \rho^{\rm (bc)}_{\rm F}(j) .
    \label{eq:delta_rho_mu=0}
\end{align}

In the SSH model limit with $V=0$ we find $\kappa_{\rm bc} = \kappa_{\rm e}$. In this case and for $\mu=0$ the total density of the semi-infinite chain is given by $1/2$ independent of the lattice site index $j$. This follows from particle-hole symmetry. 

In particular we are interested in the limit that the gap is smaller than the energy scale associated to the band width $2W$ of the gapless model:  $\Delta \ll W$. As shown in the Appendix in this case the branch cut contribution of the Friedel density decays asymptotically as
\begin{align}
  \rho_{\rm F}^{\rm (bc)}(j) \sim - \frac{1}{\sqrt{n}} e^{-2 \kappa_{\rm bc} n} , \quad n \gg
  \frac{W}{\Delta} \gg 1  \, 
  \label{eq:density_0_friedel_asym}
\end{align}
with the decay length $\kappa_{\rm bc}^{-1}$ and a prefactor which depends on $i$.

\begin{figure}[t]
  \begin{center}
  \includegraphics[width=0.5\textwidth,clip]{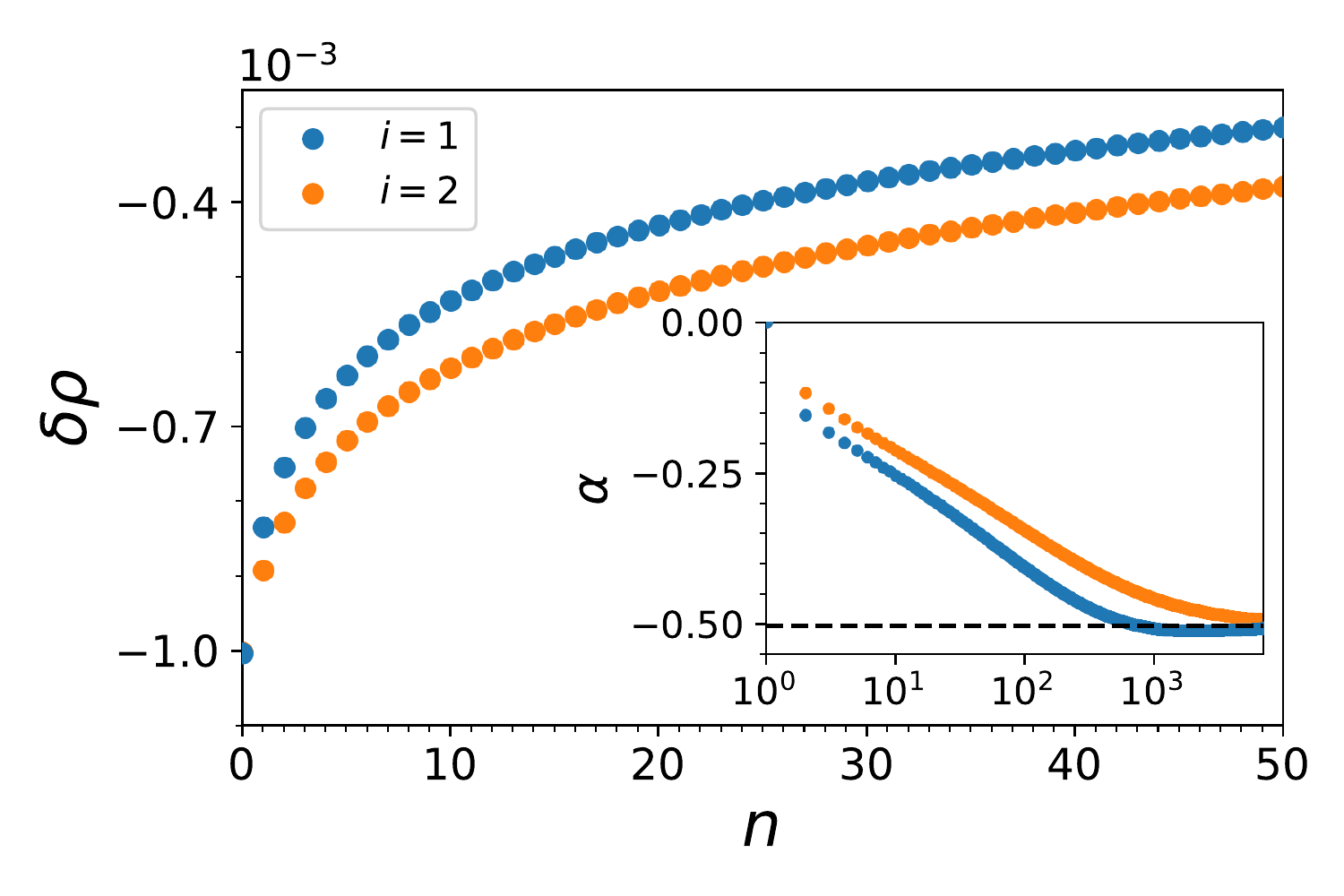}
  \caption{Main panel: The difference $\delta\rho(j)$ of the density of the semi-infinite and infinite system constructed from the extended eigenstates of the noninteracting RM model as a function of the unit cell index $n$ for chemical potential $\mu=0$. Data for the two different lattice sites $i=1,2$ within the unit cell are shown. The single-particle parameters are $\delta t=0.000125$ and $V=0.001$. Inset: The logarithmic derivative of the pre-exponential function computed according to Eq.~(\ref{eq:logderiv}). The asymptotic inverse square-root decay of the pre-exponential function is only reached for very large $n$.} 
  \label{fig:rho_F_non}
  \end{center}
\end{figure}

The main part of Fig.~\ref{fig:rho_F_non} shows $\delta\rho(j)$ for  $\delta t=0.000125$,
and $V=0.001$, that is, for a very small gap. A 
very large but finite system with OBC and $L=200000$ sites was considered. 
On the scale of the plot the data are free of finite size corrections and for
all practical purposes can considered to be in the thermodynamic limit.
In the inset the ``centered logarithmic differences''
\begin{align}
\alpha(n) = \frac{\ln{[f(n+1)]}- \ln{[f(n-1)]}}{\ln{(n+1)} - \ln{(n-1)}}
  \label{eq:logderiv}
\end{align}
with $f(n) = e^{2 \kappa_{\rm bc} n} \left|  \rho_{\rm F}(j) \right| $ are shown for $i=1,2$.
If $f(n)$ shows power-law scaling for large $n$, $\alpha(n)$ approaches a constant
in this limit with $\lim_{n \to \infty} \alpha(n)$ being the exponent. The inset
of Fig.~\ref{fig:rho_F_non} indicates that to identify the pre-exponential inverse square root
behavior of Eq.~(\ref{eq:density_0_friedel_asym}) fairly large $n$ must be accessed. 
Note that for $\mu=0$, $\rho_{\rm F}^{(\rm bc)} = \delta \rho$ holds.  
The Friedel density on the second sites of every unit cell ($i=2$)
takes longer to decay to zero as compared to the one on the first sites ($i=1$). We return to these observations in Sect.~\ref{sec:boundary} when studying
the interacting RM model.

\subsection{The boundary charge}
\label{sec:U_0_QB}

\begin{figure}
\centering
\includegraphics[width= \columnwidth]{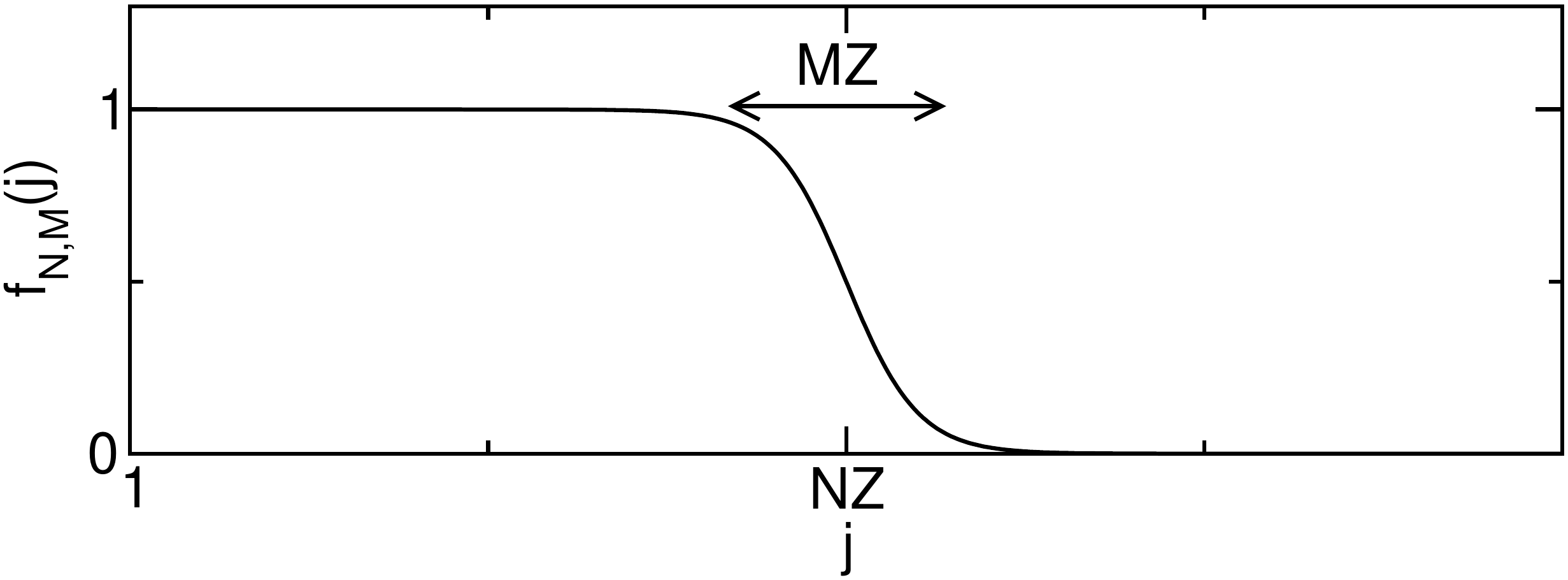} 
 \caption{Sketch of the envelope function $f(j)$ used to compute the boundary charge with $N\gg M \gg Z,\kappa_{\rm e}^{-1},\kappa_{\rm bc}^{-1}$, where $Z=2$ for the RM model.}
\label{fig:envelope}
\end{figure}

In this section we discuss the boundary charge $Q_{\rm B}$ of the noninteracting RM model. It is defined as the charge accumulated close to an open boundary. 
We closely follow the treatment of Ref.~\cite{pletyukhov_etal_long}. Here we summarize the most important results; see the Appendix for the technical details.
The boundary charge $Q_{\rm B}$ of the semi-infinite RM model for $\mu=0$ can be computed as 
\begin{align}
    Q_{\rm B} = \lim_{M\rightarrow\infty}\lim_{N\rightarrow\infty}\sum_{j=1}^\infty \left[ \rho(j)-{1\over 2}\right] f_{N,M}(j) ,
    \label{eq:QB_def}
\end{align}   
where $f_{N,M}(j)$ is an envelope function changing smoothly from $1$ to zero when going from the boundary towards the bulk. It  characterizes a macroscopic charge measurement probe; see Fig.~\ref{fig:envelope}, from which the definition of the parameters $M$ and $N$ is apparent. Using 
Eq.~(\ref{eq:delta_rho}), $Q_{\rm B}$ can be split as
\begin{align}
    Q_{\rm B} &= Q_{\rm P} + \delta Q_{\rm B} ,\label{eq:QB_splitting}\\
    Q_{\rm P} &= \lim_{M\rightarrow\infty}\lim_{N\rightarrow\infty}\sum_{j=1}^\infty \left[\rho_{\rm bulk}(j)-{1\over 2}\right] f_{N,M}(j)\nonumber\\
    &= -\frac{1}{2}\sum_{i=1,2} i\left[\rho_{\rm bulk}(i)-{1\over 2}\right] ,\label{eq:QP}\\
    \delta Q_{\rm B} &= \sum_{j=1}^\infty \delta\rho(j) .\label{eq:delta_QB}
\end{align}
Here, $Q_{\rm P}$ is the polarization charge determined by the bulk density. Using the translational invariance $\rho_{\rm bulk}(j=2[n-1]+i)=\rho_{\rm bulk}(i)$ and expanding the envelope function in $i$ one proceeds from the first to the second line of Eq.~(\ref{eq:QP}), see Ref.~\cite{pletyukhov_etal_long} for details. The term $\delta Q_{\rm B}$ involves the exponentially decaying part $\delta\rho(j)$ for which the $f_{N,M}(j)$ function can be set to $1$. Inserting Eqs.~(\ref{eq:rho_bulk}), (\ref{eq:rho_F_bc_1}), (\ref{eq:rho_F_bc_2}), and (\ref{eq:delta_rho_mu=0}) for the various parts of the density, together with the explicit solution Eq.~(\ref{eq:wavefunc_defs1}) for the Bloch states, we show in the Appendix that the total boundary charge can be calculated very efficiently as
\begin{align}
    Q_{\rm B} &= -\frac{1}{4} \text{sign}(V)  \nonumber\\
    & - \frac{V(t_2^2-t_1^2)}{4\pi}\int_0^\infty d\kappa
    \frac{1}{\sqrt{-R(\kappa)}\left[V^2-R(\kappa)\right]} ,
    \label{eq:QB_result_U=0}
\end{align}
where $R(\kappa)$ is defined in Eq.~(\ref{eq:R}). This holds for the special case $\mu=0$. For finite $\mu$ one has to add the difference of the edge state charge corresponding to Eq.~(\ref{eq:rho_edge_difference}) 
\begin{align}
    Q_{\rm e}(j)-Q_{\rm e}(j)|_{\mu=0} = &
    \Theta(t_2-t_1)  \label{eq:QB_edge_difference} \\ & \times \frac{1}{2}\left[\text{sign}(\mu-V)-\text{sign}(-V)\right] .
\nonumber
\end{align}

There are four characteristics of the boundary charge discussed in Refs.~\cite{pletyukhov_etal_long,park_etal_prb_16,thakurathi_etal_prb_18,pletyukhov_etal_short,pletyukhov_etal_preprint} which all can  be derived from properties of the bulk Hamiltonian. (i) Transformation property of $Q_{\rm B}$ when shifting the lattice by one site towards the boundary (also referred to as the universal linear slope of $Q_{\rm B}$ as a function of the phase variable $\gamma$). (ii) Transformation property of $Q_{\rm B}$ under local inversion. (iii) Low-energy behavior of $Q_{\rm B}$ for small gaps $\Delta\ll W$. (iv) Quantization of $Q_{\rm B}$ in the presence of local and nonlocal symmetries. These four features are specified in the following for the noninteracting RM model employing the above formulas and further alternatives to write Eq.~(\ref{eq:QB_result_U=0}) (see the Appendix).

{\it (i) Transformation of $Q_{\rm B}$ under translations.}
Using the parametrization of the single-particle parameters Eq.~(\ref{eq:gamma}) in terms of the phase variable $\gamma$, one can describe a translation of the lattice by one site towards the boundary as a phase change by $\pi$, which corresponds to $V_1\leftrightarrow V_2=-V_1$ (or $V\rightarrow -V$) and $t_1\leftrightarrow t_2$. Using Eq.~(\ref{eq:QB_result_U=0}) we find
\begin{align}
    \Delta Q_{\rm B}(\gamma) = Q_{\rm B}(\gamma+\pi) - Q_{\rm B}(\gamma) = \frac{1}{2}\text{sign}(V) .
    \label{eq:QB_translation_trafo}
\end{align}
This agrees with the general result derived in Refs.~\cite{pletyukhov_etal_short,pletyukhov_etal_long}
for all single-channel and nearest neighbor hopping models
that $Q_{\rm B}$ changes either by the average particle charge per site $\bar{\rho}$ or the average hole charge per site $\bar{\rho}-1$ \cite{footnote2}.  For the RM model at $\mu=0$ we have $\bar{\rho}=\frac{1}{2}$ leading to $\pm\frac{1}{2}$ for $\Delta Q_{\rm B}$ consistent with Eq.~(\ref{eq:QB_translation_trafo}). We note that, for finite $\mu$, we have to add the change of Eq.~(\ref{eq:QB_edge_difference}) under translation, which gives 
\begin{align}
    \Delta Q_{\rm B}(\gamma) = \frac{1}{2}&\left[\Theta(t_1-t_2)\text{sign}(\mu+V)  \right.
    \nonumber\\
    & - \left.\Theta(t_2-t_1)\text{sign}(\mu-V)\right] .
    \label{eq:QB_translation_trafo_finite_mu}
\end{align}
Again we see that the change of $Q_{\rm B}$ can only take the values $\pm\frac{1}{2}$. 

{\it (ii) Transformation of $Q_{\rm B}$ under local inversion.}
A local inversion for the RM model is defined within a unit cell by the transformation $V_1\leftrightarrow V_2=-V_1$ (or $V\leftrightarrow -V$) but leaving the hoppings invariant. In Ref.~\cite{pletyukhov_etal_preprint} it was shown that $Q_{\rm B}$ changes its sign under local inversion [$\text{mod}(1)$] for generic tight-binding models in 1d (for special cases see also Refs.~\cite{park_etal_prb_16,thakurathi_etal_prb_18,pletyukhov_etal_long}). Using Eq.~(\ref{eq:QB_result_U=0}) we find for the particular case of the RM model
\begin{align}
    Q_{\rm B}(-V) = - Q_{\rm B}(V) .
    \label{eq:QB_inversion_trafo}
\end{align}

{\it (iii) Low-energy theory for small gaps.}
In the low-energy limit of a very small gap $\Delta \ll W$ and using the definition Eq.~(\ref{eq:gamma}), we show in the Appendix that the boundary charge can be written approximately in the universal form
\begin{align}
    Q_{\rm B} \approx \frac{\gamma}{2\pi} - \frac{1}{4} - \Theta_{\frac{3}{2}\pi<\gamma<2\pi} ,
    \label{eq:QB_low_energy_limit}
\end{align}
for $0<\gamma<2\pi$ and periodic continuation to other intervals. Here, $\Theta_{a<x<b}=1$ for $a<x<b$ and zero otherwise. The universal linear behavior in $\gamma$ has been found in Ref.~\cite{pletyukhov_etal_preprint} for any single-channel and nearest-neighbor hopping model in the low-energy limit (note that in this reference $\gamma'=\gamma-\pi$ with $-\pi<\gamma'<\pi$ defines the phase of the gap parameter). 

{\it (iv) Quantization of $Q_{\rm B}$.}
In the presence of special symmetries the boundary charge is quantized to some rational number.  
For local inversion or local chiral symmetry $Q_{\rm B}$ is quantized in half-integer units. This was shown via the quantization of the Zak-Berry phase $\gamma_{\rm Z}$ in units of $\pi$ \cite{zak,schnyder_etal_njp_10}, which is related to the boundary charge by $Q_{\rm B}=-{\gamma_{\rm Z}\over2\pi}\,\text{mod}(1)$ \cite{kingsmith_vanderbilt_prb_93,vanderbilt_kingsmith_prb_93,resta,resta_revmodphys_94,marzari_etal_revmodphys_12,vanderbilt_book_2018,ortiz_martin_prb_94,rhim_etal_prb_17,miert_ortix_prb_17}. In the presence of nonlocal symmetries it was shown recently \cite{pletyukhov_etal_preprint} that any rational quantization of the boundary charge is possible in combinations of multiples of half of the average particle or hole charge per site $\frac{1}{2}\bar{\rho}$ or $\frac{1}{2}(\bar{\rho}-1)$. Since $\bar{\rho}=\frac{1}{2}$ for the RM model this means that both cases of $\frac{1}{2}$ and $\frac{1}{4}$ quantization can occur for $Q_{\rm B}$. 

For the RM model a local inversion or local chiral symmetry is present for $V_1=V_2=V=0$, which corresponds to the SSH model. Due to particle-hole symmetry at $\mu=0$ this gives $Q_{\rm B}=0$. For $|V|\ll |\delta t|$ we show in the Appendix [note that $\text{sign}(0)= 0$]
\begin{align}
    Q_{\rm B} = -\frac{1}{2}\Theta(t_2-t_1) \, \text{sign}(V) + {\mathcal O}\left(\frac{V}{\delta t}\right) .
    \label{eq:QB_quantization_half}
\end{align}
This gives half-integer quantization.

A nonlocal chiral symmetry occurs for the RM model for $t_1=t_2$. In this case one obtains for $Q_{\rm B}$ the novel quantization value $\frac{1}{4}$, see Ref.~\cite{pletyukhov_etal_preprint}. For $|\delta t|\ll |V|$ we show in the Appendix
\begin{align}
    Q_{\rm B} = -\frac{1}{4} \text{sign}(V) + {\mathcal O}\left(\frac{W\delta t}{V\text{max}\{|V|,W\}}\right) ,
    \label{eq:QB_quantization_quarter}
\end{align}
leading to the expected $\frac{1}{4}$-quantization. 

The main characteristics (i)-(iv) suggest the boundary charge to be an interesting quantity  with characteristics related to bulk properties. This has to be contrasted to the connection between topological bulk invariants and the appearance of edge states which, for 1d systems, have to be at zero energy \cite{fidkowski_etal_prl_11,mong_shivamoggi_prb_11,gurarie_prb_11,essin_gurarie_prb_11,ukui_etal_jpsjf_12,yu_wu_xie_npb_17,rhim_etal_prb_18,silveirinha_prx_19}. Our results of Sects.~\ref{sec:subsecspec} and \ref{sec:QB_interaction} indicate that the boundary charge might be a more robust signature related to bulk properties as compared to the number of edge states when the interaction is turned on.

\begin{figure}[t]
  \begin{center}
  \includegraphics[width=0.5\textwidth,clip]{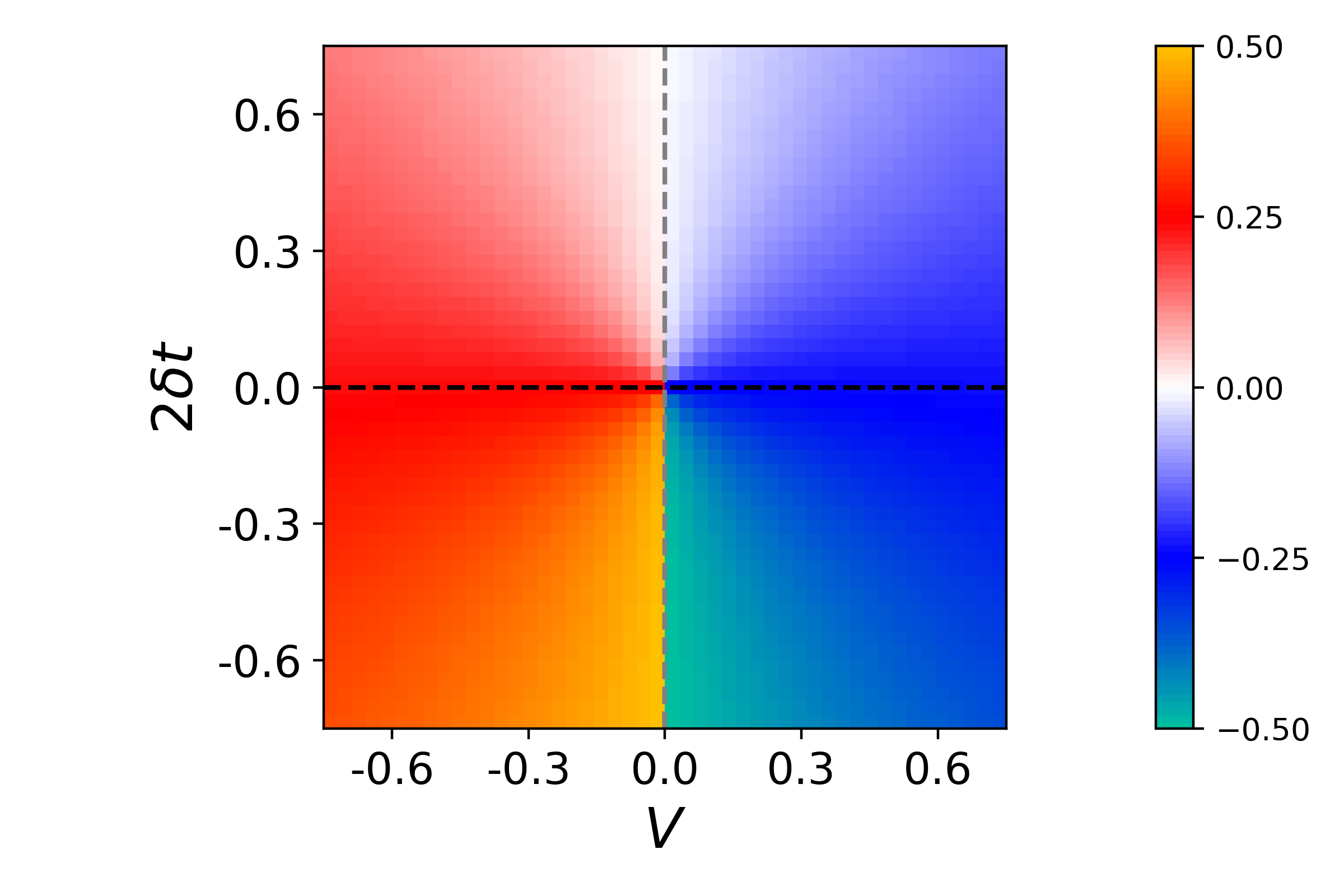}
  \caption{The boundary charge $Q_{\rm B}$ of the noninteracting RM model as a function of $V$ and $2 \delta t$ or the polar coordinates $\Delta$ and $\gamma$, see Eq.~(\ref{eq:gamma})} 
  \label{fig:phase_dia_non}
  \end{center}
\end{figure}

The features (i), (ii), and (iv) can be seen clearly in Fig.~\ref{fig:phase_dia_non}, where we show the boundary charge as function of the two parameters $V$ and $2\delta t$ defining the real and imaginary part of the quantity $\Delta e^{i\gamma}$ of  Eq.~(\ref{eq:gamma}). Therefore, $\gamma$ corresponds to the polar angle and $\Delta$ to the radial component in Fig.~\ref{fig:phase_dia_non}. 
The data were computed for $L=2000$ but are essentially free of finite size corrections.
A translation by one lattice site towards the boundary corresponds to a sign change of $V$ and $\delta t$, i.e., changes of the angle $\gamma$ by $\pi$. According to Eq.~(\ref{eq:QB_translation_trafo}) this leads to a change of $Q_{\rm B}$ by $\frac{1}{2}\text{sign}(V)$ which is consistent with Fig.~\ref{fig:phase_dia_non}. The transformation Eq.~(\ref{eq:QB_inversion_trafo}) under local inversion means that $Q_{\rm B}$ is antisymmetric when changing the sign of the variable $V$ in Fig.~\ref{fig:phase_dia_non}. The quantization rules  Eqs.~(\ref{eq:QB_quantization_half}) and (\ref{eq:QB_quantization_quarter}) can be seen on the axis $V=0$ and $\delta t=0$ in Fig.~\ref{fig:phase_dia_non}, respectively.

\begin{figure}[t]
  \begin{center}
  \includegraphics[width=0.5\textwidth,clip]{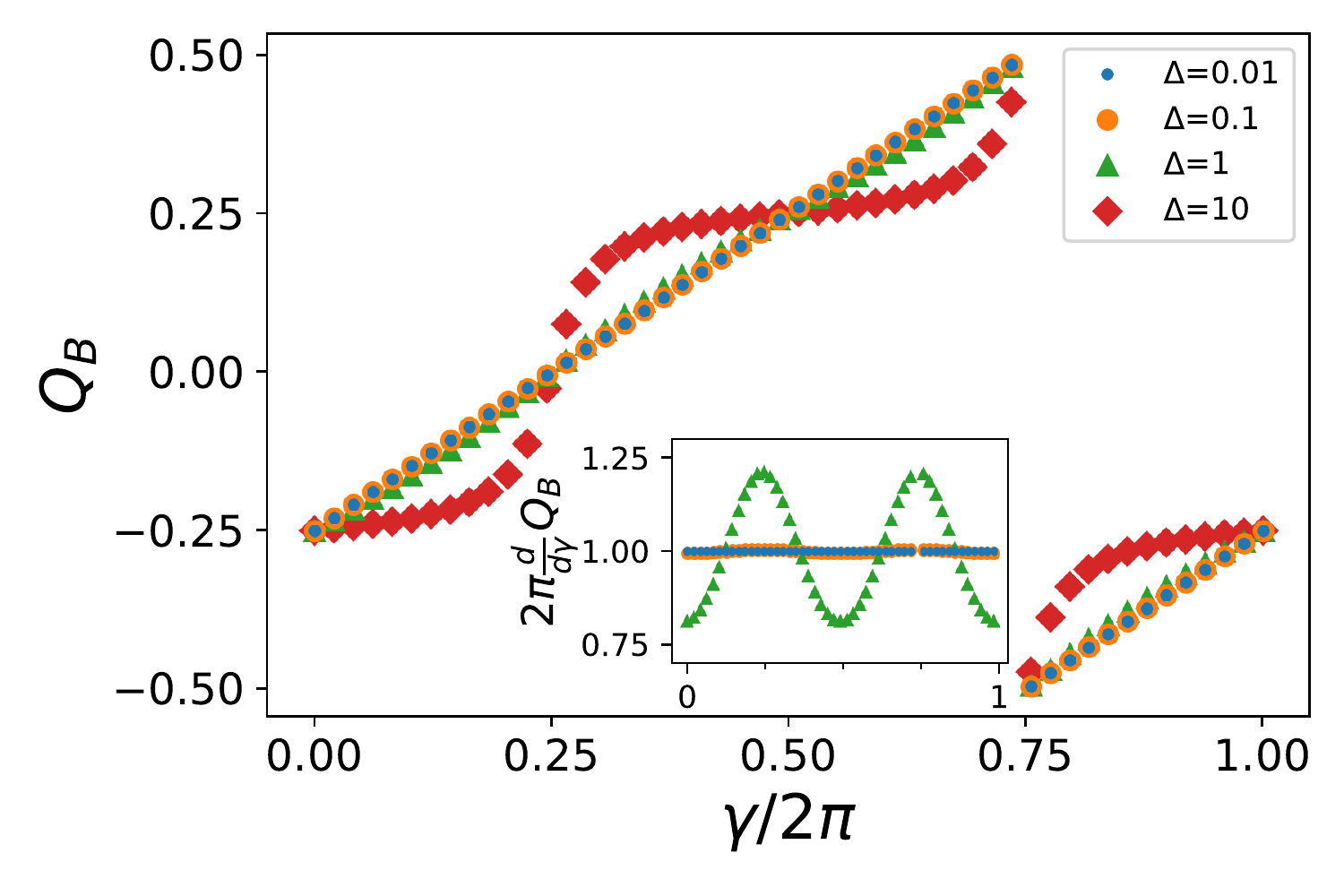}
  \caption{Main panel: The boundary charge as a function of the polar angle $\gamma$ [see Eq.~(\ref{eq:gamma})] for different $\Delta$. Inset: Derivative of the data of the main panel with respect to $\gamma$. This highlights the remarkable linearity even for sizable $\Delta$; see Eq.~(\ref{eq:QB_low_energy_limit}) and the discussion in the main text.} 
  \label{fig:BC_Delta_free}
  \end{center}
\end{figure}

Of particular interest is the validity range of the low-energy behavior (iii) of $Q_{\rm B}$ according to Eq.~(\ref{eq:QB_low_energy_limit}), i.e., the universal linear behavior as a function of the angle $\gamma$ if the gap $2 \Delta$ is very small compared to the band width. This is shown in Fig.~\ref{fig:BC_Delta_free} (again obtained for $L=2000$). Strikingly, the linear behavior is observed to a high accuracy in the whole parameter regime $\Delta<W$ extensively beyond the low-energy regime $\Delta\ll W$ where it is expected to hold. As shown in the Appendix the stability of the low-energy result up to values $\Delta\sim W$ can be explained by calculating the leading order correction to Eq.~(\ref{eq:QB_low_energy_limit}). According to Eq.~(\ref{eq:low_energy_correction}) it is given by $1/(8\pi)\sin(2\gamma)(\Delta/W)^2\ln(\Delta/W)$. This is in full agreement with the inset of Fig.~\ref{fig:BC_Delta_free}, showing the derivative of $Q_{\rm B}$ with respect to $\gamma$, where the corrections to the linear slope are zero for $\cos(2\gamma)=0$ and largest for $\cos(2\gamma)=\pm 1$.
Only for $\Delta\gtrsim W$ visible deviations from linear behavior occur and in the atomic limit $\Delta\gg W$ one obtains the universal result of $\frac{1}{4}$-quantization \begin{align}
     Q_{\rm B} &\approx- \frac14 \text{sign} (V)  \left[  1 - \frac{W \delta t}{V^2} \right],
     \label{eq:QB_atomic_limit}
\end{align}
see the Appendix for details. 

Therefore, we find two universal regimes of the boundary charge for the noninteracting RM model, given by the linear dependence in the phase $\gamma$ of the gap parameter for $\Delta<W$, and the $\frac{1}{4}$-quantization of $Q_{\rm B}$ for $\Delta\gg W$. Moreover, in Sect.~\ref{sec:QB_interaction} we demonstrate that this interesting behavior is stable against weak two-particle interactions.

\section{Many-body methods}
\label{sec:methods}

\subsection{Field theory}
\label{sec:field_theory}

Already in the early eighties it was suggested \cite{Jackiw83}
to use field theoretical models \cite{Jackiw76} to study the universal low-energy
physics of lattice models for linear polymers (such as the RM and the SSH models)
with small single-particle gaps.
Continuum models also provide a straightforward way to include two-particle
interactions \cite{Kivelson85,Horovitz85,Wu86}. It was shown that the
interaction leads to logarithmic terms of the form $g^n \ln^n (2 \Delta)$
in the first ($n=1$) and second ($n=2$) order 
perturbative expressions for the effective renormalized gap $2 \Delta^{\rm ren}$ as
a function of the bare one $2 \Delta$
\cite{Kivelson85,Horovitz85,Wu86}. Here $g$ denotes the coupling constant
of the field theory. 
As the gap is small this logarithmic dependence severely limits the applicability
of perturbation theory to tiny couplings $g$.  However, the leading-log series
can be resummed by either using field-theoretical RG \cite{Horovitz85}
or by adapting results from the Bethe ansatz solution of the massive Thirring
model \cite{Kivelson85}. In fact, in the field theoretical model the
effective gap depends on the bare one in a power-law fashion. Up to linear 
order in the coupling constant one finds $\Delta^{\rm ren}
\sim \Delta^{1-g/g_{\rm c}}$, with a characteristic interaction scale $g_{\rm c}$ \cite{footnote1}. 

We are not aware that this power-law renormalization has so far been verified directly for a microscopic lattice
model, i.e. without the intermediate approximate step of mapping it to a continuum field theory. However, expecting to find this and being interested in the entire space of
noninteracting parameters, including the small gap limit, we cannot use simple
perturbation theory to investigate the interaction effects in the RM model.
Instead we employ the functional RG \cite{Metzner12} in its lowest order
truncation. In addition, we benchmark our approximate results by comparing
to numerical ones obtained by DMRG. 

We note that recently the use of field theoretical models and tools (such as bosonization) to investigate the low-energy properties of (topological) insulators experienced a revival. 
In Ref.~\cite{gangadharaiah_etal_prl_12} they were used to not only study the gap renormalization but in addition the edge state and in Ref.~\cite{pletyukhov_etal_preprint} to investigate the boundary charge. In Refs.~\cite{gangadharaiah_etal_prl_12,park_etal_prb_16,thakurathi_etal_prb_18,pletyukhov_etal_preprint} it has furthermore been established how to express the parameters of a continuum Dirac model in $1+1$ dimension in terms of  microscopic lattice model parameters. However, by neglecting fast oscillating terms in these approaches one has to assume that the gap is much smaller than the band width and it is quite difficult to determine the quality of the low-energy results beyond this regime. The functional RG used here treats the microscopic details of the lattice model on all energy scales and thus can cover the entire parameter range from small to large gaps. Also, high-energy properties such as the renormalization of the band width are treated consistently in functional RG. This will turn out to be crucial to the discussion of the relation of the boundary charge to bulk properties in the presence of two-particle interactions. Field theories do not capture high-energy features and will thus fail in this respect.

 It was emphasized early on that the exponent of the renormalization of the gap by the two-particle interaction is independent of the details of the ultraviolet regularization of the field theory 
(``universal'') only to leading order in the coupling constant \cite{Kivelson85}. This implies that 
field theory can strictly speaking not provide a prediction for the exponent beyond leading order in the two-particle interaction. On this level many of  the details of the field theoretical model do not matter and one can e.g.~use the results from the massive Thirring model to predict the exponent of the RM model (see below). It is, however, still tempting to consider a field theoretical model which is closer to our lattice model and compute the exponent beyond leading order.

Using standard bosonization methods \cite{gogolin_etal_98,delft_schoeller_98,Giamarchi03} one can construct a continuum field theory capturing the low-energy physics of the RM model \cite{pletyukhov_etal_preprint,gangadharaiah_etal_prl_12}. It is of the sine-Gordon form 
\begin{align}
\nonumber
H&=\frac{v}{2}\int{dx}\Big\{ K \hat{\Pi}^2(x)+\frac{1}{K}[\partial_{x}\hat{\varphi}(x)]^{2}\Big\}\\
\label{eq:ham_FT}
&\hspace{0.5cm}
+\frac{\Delta}{\pi a_{\rm c}}\int{dx}\sin(\sqrt{4\pi}\hat{\varphi}(x)-\gamma),
\end{align}
where $K$ is the Tomonaga-Luttinger liquid parameter and $v$ denotes the renormalized Fermi velocity which, up to leading order in $U$, are given by $K=1-U/\pi$ and $v=v_F(1+U/\pi)$, with $v_F=2$. The canonically conjugate fields $\hat{\Pi}(x)=-\partial_x[\hat{\varphi}_{+}(x)-\hat{\varphi}_{-}(x)]$ and $\hat{\varphi}(x)=\hat{\varphi}_{+}(x)+\hat{\varphi}_{-}(x)$ are defined in terms of the chiral boson fields $\hat{\varphi}_{\pm}(x)$. The latter are related to the fermionic right and left movers via $\hat{\psi}_{\pm}(x)=\frac{1}{\sqrt{2\pi a_{\rm c}}}\,e^{\pm i\sqrt{4\pi}\hat{\varphi}_{\pm}(x)}$. Here, $1/a_{\rm c}$ denotes a phenomenological momentum cutoff which implies the high-energy cutoff $\lambda_0 = v/a_{\rm c}$.  

Changing the ultraviolet cutoff from $\lambda_0$, to a smaller value $\lambda$ a flow equation for the ratio of the gap and the cutoff, denoted by $\bar \Delta_l$, can be derived in a standard way from the scaling dimension of the nonlinear term of the sine-Gordon model
\begin{align}
\frac{d \bar \Delta_l}{dl} = (2-K) \bar \Delta_l .     
\label{eq:gap_flow_FT}    
\end{align}
The initial value is $\bar \Delta_0 = \Delta/\lambda_0$, with the bare gap $\Delta$, and the flow parameter $l$ is given by $l = \ln \frac{\lambda_0}{\lambda}$; it starts at zero and goes to infinity when $\lambda$ approaches zero. The right hand side of the flow equation for $K$ is of order $\bar \Delta^2$. For small gaps the flow of $K$ can thus be neglected. For repulsive interactions with $K < 1$ the gap grows under reduction of the cutoff. Stopping the flow if $\bar \Delta_l$ is of order one we find for the renormalized gap 
\begin{align}
\frac{\Delta^{\rm ren}}{\Delta} \sim \Delta^{(K-1)/(2-K)}.     
\label{eq:gap_ren_FT}    
\end{align}
The precise value at which the flow is stopped only enters the prefactor on the right hand side of this equation. 

For half-filling the relation between the Tomonaga-Luttinger liquid parameter $K$ and the interaction $U$ is known analytically 
beyond leading order from the Bethe ansatz solution of the gapless, interacting lattice model \cite{Giamarchi03}. Taking this value we obtain for the exponent
\begin{align}
\beta  = & \frac{K-1}{2-K} = \frac{1- \frac{2}{\pi} \arccos{(-U/2)}}{\frac{4}{\pi} \arccos{(-U/2)}-1}
\label{eq:exp_ex_FT} \\
 = &  -\frac{U}{\pi}+ 2 \left(\frac{U}{\pi}\right)^2 - 
\frac{96+\pi^2}{24} \left(\frac{U}{\pi}\right)^3 + 
\frac{48+\pi^2}{6} \left(\frac{U}{\pi}\right)^4 \nonumber \\
& +{\mathcal O}\left( \left[\frac{U}{\pi}\right]^5   \right).
\label{eq:exp_FT}    
\end{align}
To universal (see above), leading order the exponent is given by $-U/\pi$.
In comparison to the result from the massive Thirring model \cite{Kivelson85} we thus have to identify $g/g_{\rm c} \leftrightarrow U/\pi$.
We note that the coefficients of the power series Eq.~(\ref{eq:exp_FT}) in $U/\pi$ do not decay with the order and are alternating. This indicates that for increasing interactions corrections of order $U^2$ and higher will quickly become sizable and lead to a deviation from the universal linear interaction dependence. 
In Sect.~\ref{sec:num} we will return to this observation and investigate
how this result for the exponent $\beta$, obtained combining field theory, bosonization, and the Bethe ansatz result for $K$ of the gapless lattice model, compares to the one obtained if the renormalized gap is directly computed for the microscopic model by functional RG and DMRG.

\subsection{The functional RG}
\label{sec:FRG}

\subsubsection{The basic idea}

It was earlier shown that functional RG in its lowest-order
truncation \cite{Metzner12} can be used to properly resum leading logs in extended 1d models of
correlated fermions \cite{Meden08} as well as for quantum dot models with local
two-particle interactions \cite{Karrasch10}. 
In Sect.~\ref{sec:ana} it will be shown analytically that this also holds
in the (single-particle) gaped RM model with nearest-neighbor interaction.

The functional RG has the distinct advantage over other RG methods
that it is directly applicable to
microscopic lattice models. It does not require the intermediate (approximate)
step of the mapping to a field theory. It thus does not only capture the
low-energy physics but the one on all energy scales.
Functional RG still inherits the  RG idea of a successive
treatment of energy scales.
A comprehensive
account is given in Ref.~\cite{Metzner12} (see also Ref.~\cite{Kopietz10}). For
completeness we here present the basic idea and the important equations. 

The fundamental steps of the application of FRG to interacting fermionic systems are the
following:
\begin{enumerate}
\item Write the partition function as a coherent state functional integral
  (within the Matsubara formalism).
\item Replace the noninteracting propagator ${\mathcal G}_{0}(i\omega) $
  which inherits all the single-particle physics by one decorated
  by a cutoff $\Lambda$. For
  the initial value $\Lambda_{\rm i}$, the free propagation must vanish; for
  the final one $\Lambda_{\rm f}$, the original propagation must be restored.
  One often uses ${\mathcal G}_{0}^\Lambda({\rm i}\omega) = \Theta(|\omega| - \Lambda)
  {\mathcal G}_{0}({\rm i} \omega)$, $\Lambda_{\rm i}= \infty$,
  and $\Lambda_{\rm f}=0$.  When $\Lambda$ is sent from $\infty$ to $0$ (see below)
  this incorporates the RG idea of a successive treatment of energy scales. Here we
  will also use this cutoff. 
\item Differentiate the generating functional of one-particle irreducible vertex
  functions with respect to $\Lambda$.
\item Expand both sides of the functional differential equation with respect to
  the vertex functions. This leads to an infinite hierarchy of coupled differential
  equations for the vertex functions. The lowest order vertex function is the
  self-energy $\Sigma$.
\end{enumerate} 
The hierarchy of coupled flow equations presents an exact reformulation of the
quantum many-body problem. Integrating it from $\Lambda_{\rm i}$ to $\Lambda_{\rm f}$
leads to exact expressions for the vertex functions. From those observables, such
as the single-particle spectral function can be computed.

In practice, truncations of the hierarchy are required, resulting in a closed
finite set of equations. The integration of this leads to approximate expressions
for the vertices and, thus, for observables. We here employ the lowest-order
truncation in which the flowing two-particle vertex is replaced by the bare
interaction. What remains within this scheme is a set of coupled differential
equations for the matrix elements of a frequency independent self-energy.
This approximation contains all leading order in $U$ terms \cite{Metzner12} but
in addition selected higher order ones. As will be seen a posteriori in Sect.~\ref{sec:ana}
this includes all leading log terms of the form $U^n \ln^n (2 \Delta)$. In this
context we also show how to reproduce the perturbative results from the
functional RG.

The frequency independence of the self-energy has the distinct advantage that
it leads to an effective single-particle picture at the end of the RG flow.
All single-particle parameters of the model, that is all hoppings and onsite
energies get renormalized by the interaction
and in the final step of computing the renormalized propagator a single-particle Hamiltonian of the form Eq.~(\ref{eq:H0})
needs to be solved. We emphasize that for open boundaries the single-particle
parameters acquire a spatial dependence beyond the underlying unit cell
structure (see below). We will employ this effective single-particle
picture in the interpretation of our results, however, we already now
emphasize that it should not be overstressed. E.g.~the wave functions obtained
from diagonalizing the effective single-particle Hamiltonian do not have 
a direct physical meaning (similar to their role in Hartree-Fock or
density-functional theory). This includes energetically isolated
``effective single-particle edge states''.  

Within our approximation the 
local spectral function can be obtained by simply taking
\begin{align}
  A_j(\omega) = - \frac{1}{\pi} \mbox{Im} \, {\mathcal G}({\rm i}\omega \to
  \omega +  {\rm i} 0) ,
  \label{eq:specfu_int}
\end{align}
with $ {\mathcal G}({\rm i} \omega)$ computed using Eq.~(\ref{eq:dyson}) and $\Sigma$
taken at the end of the RG flow. 

As discussed in the introduction and Sect.~\ref{sec:U_0results} we are also
interested in the spatial dependence of the local density of the interacting RM
model with open boundaries. As it is well known the density on site $j$ can be
computed by integrating the $(j,j)$-matrix element of the full propagator (Dyson equation)
\begin{align}
  {\mathcal G}({\rm i}\omega) = \left\{ \left[ {\mathcal G}_{0}({\rm i}\omega) \right]^{-1} -
  \Sigma \right\}^{-1}
  \label{eq:dyson}
\end{align}
over the Matsubara frequency. However, truncated functional RG is not a so-called
conserving approximation. It is thus not guaranteed that computing the density
along this line will lead to the same result as computing it in a more consistent
way via its own RG flow equation. In fact, it was earlier shown that the above
frequency integration over the full approximate propagator does not capture
the typical Tomonaga-Luttinger liquid power-law decay of the Friedel density
oscillations away from an open boundary or an impurity for vanishing gap. 
In contrast, the (leading) interaction dependence of the exponent is properly captured
if a flow equation for the density is considered \cite{Andergassen04}. This
shows another limitation of the effective single-particle picture obtained
at the end of the RG flow. When computing the local density of the interacting RM model we thus set up its own flow equation. From the local
density the boundary charge can be computed as explained in Sect.~\ref{sec:U_0results}.

\subsubsection{The RG flow equations}
\label{sec:RGflow}

As described in the last subsection we focus on the lowest-order truncated
functional RG scheme featuring a static flowing self-energy  $\Sigma^{\Lambda}$
and consider a sharp frequency cutoff in Matsubara space \cite{Metzner12}.
As the two-particle interaction  Eq.~(\ref{eq:Hint}) is of nearest-neighbor type
the self-energy  matrix in real space has a tridiagonal form \cite{Metzner12}.
The flow equations
are given by
\begin{align}
  &\frac{\partial}{\partial \Lambda} \Sigma^{\Lambda}_{j,j} = - \frac{U}{2\pi} \sum_{\omega=\pm \Lambda} \sum_{r=\pm1} {\mathcal G}^{\Lambda}_{j+r,j+r}({{\rm i}\omega}),\nonumber \\
  &\frac{\partial}{\partial \Lambda} \Sigma^{\Lambda}_{j,j \pm1} = \frac{U}{2\pi} \sum_{\omega=\pm \Lambda}{\mathcal G}^{\Lambda}_{j,j\pm1}({{\rm i}\omega}),
  \label{eq:RG_SE_eq}
\end{align}
with a cutoff dependent propagator
\begin{align}
  {\mathcal G}^{\Lambda}({\rm i}\omega) =
  \left\{ \left[ {\mathcal G}_{0}({\rm i}\omega) \right]^{-1} -
  \Sigma^{\Lambda} \right\}^{-1}
  \label{eq:SSP}
\end{align}
For PBCs the self-energy only depends on the site index $i=1,2$ within the unit cell only.
The translation symmetry by two sites is preserved.
In contrast, for open boundaries $\Sigma^{\Lambda}$ acquires a nontrivial dependence
on $n$ in addition to the one on $i$.  
 
To consistently compute the local density $\rho(j)$ we set up according
flow equations for this observable 
\begin{align}
  &\frac{\partial}{\partial \Lambda} \rho^{\Lambda}(j) = -\frac{1}{2\pi}
    \sum_{\omega+\pm \Lambda} \mbox{tr} \left[ e^{\mbox{i}\omega 0^{+} }
    {{\mathcal G}}^{\Lambda}({{\rm i}\omega})R^{\Lambda}_{j}({{\rm i}\omega})
    \right] .
  \label{eq:RG_density_eq}
\end{align}
They involve a density response vertex $R^{\Lambda}_{j}$ which obeys the flow equation 
\begin{align}
   &\frac{\partial}{\partial \Lambda} R^{\Lambda}_{j;l,l} = -\frac{U}{2\pi} \nonumber\
     \sum_{\omega=\pm \Lambda} 
     \sum_{l'}\sum_{r=\pm1}\sum_{r^{'}=0,\pm1}  {{\mathcal G}}^{\Lambda}_{l+r,l'}({{\rm i}\omega}) \\ &\hspace{7.5em}
     \times 
 R^{\Lambda}_{j;l',l'+r'}{{\mathcal G}}_{l'+r',l+r}^{\Lambda}({{\rm i}\omega}),
                                                      \nonumber \\  
  &\frac{\partial}{\partial \Lambda} R^{\Lambda}_{j;l,l\pm1} = -\frac{U}{2\pi}
    \nonumber\ \sum_{\omega=\pm \Lambda} 
    \sum_{l'}\sum_{r'=0,\pm1}   {{\mathcal G}}^{\Lambda}_{l,l'}({{\rm i}\omega}) \\ &\hspace{7.5em} \times
                                        R^{\Lambda}_{j;l',l'+r'}
                                          {{\mathcal G}}_{l'+r',l\pm1}^{\Lambda}({{\rm i}\omega}).
  \label{eq:RG_vertex_eq}
\end{align}
Details on this can be found in Ref.~\cite{Andergassen04}.

The flow is uniquely determined by this set of coupled first order differential equations
and the initial conditions at $\Lambda=\infty$. 
However, the numerical solution of the equation have to start at a large but finite
cutoff $\Lambda_{0}$. One can integrate over the flow equation Eq.~(\ref{eq:RG_SE_eq})
from $\Lambda=\infty$ to $\Lambda=\Lambda_{0}$ analytically to obtain the initial
condition for the self-energy at this value of the cutoff \cite{Metzner12}. For PBC it is given by $\Sigma^{\Lambda_{0}}_{j,j}=U$
and $\Sigma^{\Lambda_{0}}_{j,j\pm1}=0$. For OBC the initial condition on the diagonal
of the self-energy matrix and the sites $j=1$ and $j=N$ has to be changed to
$\Sigma^{\Lambda_{0}}_{1,1}=\Sigma^{\Lambda_{0}}_{N,N}=U/2$.
Moreover, the initial condition for the local density and density response
vertex at $\Lambda_0$ are $\rho^{\Lambda_{0}}_{j}=\frac{1}{2}$ and $R^{\Lambda_{0}}_{j;l,l^{'}}
=\delta_{jl}\delta_{ll^{'}}$, respectively. The corrections
are of order $1/\Lambda_0$. 
To obtain the data shown below we set $\Lambda_0=10^{8}$. 

Note that $\Sigma^{\Lambda}$ and ${\mathcal G}_0^{-1}$ are both tridiagonal matrices in
real space. Using a particular algorithm \cite{Andergassen04}, the tridiagonal matrix elements of the the cutoff dependent propagator Eq.~(\ref{eq:SSP})
needed on the right hand side of the flow equations 
(\ref{eq:RG_SE_eq}) can be computed with an $\mathcal{O}(L)$
computational effort.
Similarly, the right hand side of the flow equation of the density response vertex which
involves the product of inverted tridiagonal matrices and the vertex itself can be
computed in $\mathcal{O}(L)$. We can therefore easily deal with
very large systems with $L \sim 10^6$ sites.

At  the end of the flow at $\Lambda=0$, one can decompose the self-energy into unit
cell index $n$ independent and dependent parts, labeled by ``bulk'' and ``F''
respectively. For $j=2(n-1)+i$,
\begin{align}
&\Sigma^{\Lambda=0}_{j,j}=\Sigma^{\text{bulk}}_i+\Sigma^{\text{F}}_i(n), \nonumber \\
   &\Sigma^{\Lambda=0}_{j,j+1}=  \left\{
   \begin{array}{ll}
        \Sigma^{\text{bulk}}_{\text{intra}}+\Sigma^{\text{F}}_{\text{intra}}(n) & \mbox{for $i = 1$} \, \\  
     \Sigma^{\text{bulk}}_{\text{inter}}+\Sigma^{\text{F}}_{\text{inter}}(n)
                                                                               & \mbox{for $i = 2$}. \,
\end{array} \right.
\end{align}
Finally, the renormalized onsite potentials and hoppings are determined by 
\begin{align}
&V^{\text{ren}}_{j=2(n-1)+i}= V^{\text{ren}}_{i}+V^{\text{F}}_{i}(n) , \nonumber  \\
&t^{\text{ren}}_{j=2(n-1)+i} = t^{\text{ren}}_{i}+t^{\text{F}}_{i}(n) ,
\label{eq:needed1}
\end{align} 
with
\begin{align}
&V^{\text{ren}}_{i}=V_{i}+\Sigma^{\text{bulk}}_{i},\\
&V^{\text{F}}_{i}(n)=\Sigma^{\text{F}}_{i}(n),\\
   &t^{\text{ren}}_{i}=  \left\{
   \begin{array}{ll}
        t_{1}-\Sigma^{\text{bulk}}_{\text{intra}} & \mbox{for $i = 1$} \,  \\
        t_{2}-\Sigma^{\text{bulk}}_{\text{inter}} & \mbox{for $i = 2$}, \, 
\end{array} \right. \\
   &t^{\text{F}}_{i}(n)=  \left\{
   \begin{array}{ll}
        -\Sigma^{\text{F}}_{\text{intra}}(n) & \mbox{for $i = 1$} \,  \\
        -\Sigma^{\text{F}}_{\text{inter}}(n) & \mbox{for $i = 2$}. \, 
\end{array} \right.
\label{eq:needed2}
\end{align}
For PBCs the unit cell index dependent interaction induced Friedel parts vanish. For OBC, however, they lead to a modulation of the potential and hopping landscape close to the boundary (and beyond the unit cell structure). The approach of the renormalized bulk values is dominated by an exponential decay in the unit cell index (see Fig.~\ref{fig:ren_t_V} below). We can thus expect that local properties, such as the weight of in-gap $\delta$-peaks of the single-particle spectral function, which for $U=0$ are associated to single-particle edge states, are altered by the two-particle interaction. We will even show that peaks can be generated which do not have any analog at $U=0$ and are thus purely interaction induced. The $\delta$-peaks are signatures of the edge states of the effective single-particle Hamiltonian to be diagonalized at the end of the RG procedure.

\subsection{The density matrix renormalization group}
\label{sec:DMRG}

We use a ``numerically exact'' DMRG approach set up in the
language of matrix product states \cite{Schollwoeck11} to compare to and to benchmark
the results obtained within the approximate functional RG method described above. The model
defined in Eqs.~\eqref{eq:H0} and \eqref{eq:Hint} can be mapped directly to a spin model
by a Jordan-Wigner transform \cite{Jordan28} rendering it amendable to standard
DMRG implementations, such as the one outlined in Ref.~\cite{Schollwoeck11}. 


An iterative two-site update sweeping procedure to obtain the ground state of
a system with OBC is employed. We follow precisely the procedure outlined in Sect.~6 of
Ref.~\cite{Schollwoeck11}. We use constant bond dimension and perform sweeps forth and back on the chain until the relative energy change per sweep falls below $10^{-8}$. Increasing the bond dimension we achieve a ``numerically exact''  approximation
to the ground state wavefunction. From this the site-dependent density
$\rho(j)= \left \langle n_j \right\rangle$ can be computed.  After the ground state wave function has been obtained we orthogonalize against this state and rerun the above procedure, which provides us with the first excited state of the system \cite{Schollwoeck11} in the same total particle number sector. The gap is then defined as the difference between the first excited and the ground state energy. 




Careful benchmarks in the non-interacting case show that we can converge the above described ground state and excited state DMRG calculations and provide confidence also for the interacting case.


\section{Bulk properties for $U>0$}
\label{sec:bulk}

We first discuss our results for the bulk properties of the interacting
RM model at $\mu=0$ obtained by truncated functional RG as well as by numerical DMRG.
In Sect.~\ref{sec:ana} we present the analytical solution of the
functional RG flow equations (\ref{eq:RG_SE_eq}) in the limit of small bare 
gaps $2 \Delta$. We
show that the renormalized gap, displayed in the (interacting) single-particle
spectral function, scales as a power law as a function of the bare gap with $U$
entering in the exponent. In
addition, we discuss how the first order in $U$ perturbative result can be
obtained from functional RG. These considerations prove that functional
RG in lowest order truncation captures the entire leading log series. 

In Sect.~\ref{sec:num} results for the effective gap obtained from a numerical
solution of the RG flow equations (\ref{eq:RG_SE_eq}) are presented and compared to
the DMRG results. In the limit of small bare gaps both confirm the analytical insight
of Sect.~\ref{sec:ana}. In addition, considering bulk properties we provide a first hint that Fig.~\ref{fig:phase_dia_non} is only altered 
quantitatively by small interactions.

\subsection{Analytical insights}
\label{sec:ana}

We consider periodic boundary conditions. In this case the chain is translational
invariant by two lattice sites. For analytic calculations 
it is advantageous to transform the right hand side 
of the flow  equations (\ref{eq:RG_SE_eq}) to $k$-space. The number of (independent) coupled  equations is reduced
to three: One for the onsite energy 
\begin{align}
	&\partial_{\Lambda} \Sigma^{\text{bulk},\Lambda}_{1/2}= \mp\frac{2U}{\pi} \int^{\pi}_{-\pi}\frac{dk}{2\pi}   \left\{ \frac{V^{\Lambda}_{1}}{a_{}^{\Lambda}+b_{}^{\Lambda}\cos k }
  \right\}, \label{eq:RG_sigma12_eq}
\end{align}
and one each for the intra and the inter unit cell hopping 
\begin{align}
	&\partial_{\Lambda} \Sigma^{\text{bulk},\Lambda}_{\text{intra}}= \frac{U}{\pi} \int^{\pi}_{-\pi}\frac{dk}{2\pi}   \left\{ \frac{t_{1}^{\Lambda}+t_{2}^{\Lambda}\cos k}{a_{}^{\Lambda}+b_{}^{\Lambda}\cos k } \right\},  \nonumber  \\
	&\partial_{\Lambda} \Sigma^{\text{bulk},\Lambda}_{\text{inter}}= \frac{U}{\pi} \int^{\pi}_{-\pi}\frac{dk}{2\pi}   \left\{ \frac{t_{1}^{\Lambda}\cos k+t_{2}^{\Lambda}}{a_{}^{\Lambda}+b_{}^{\Lambda}\cos(k) } \right\} . \label{eq:RG_t1/2_eq}
\end{align}
Here $a_{}^{\Lambda}=\Lambda^2+(V^{\Lambda}_{1})^2+(t^{\Lambda}_{1})^2+(t^{\Lambda}_{2})^2$, $b_{}^{\Lambda}=2t^{\Lambda}_{1}t^{\Lambda}_{2}$, and $V^{\Lambda}_{1}$, $t^{\Lambda}_{1/2}$ are defined as the effective onsite potential and hopping parameters during the flow, respectively. 

Remind that for the bare parameters $V_1 = - V_2$ holds. As the consequence,
Eq.~(\ref{eq:RG_sigma12_eq}) implies  $-V^{\Lambda}_2=V^{\Lambda}_1 = V^{\Lambda}$ during
the entire flow including the end $\Lambda=0$.   

 The $k$-integrals in Eqs.~(\ref{eq:RG_sigma12_eq}) and (\ref{eq:RG_t1/2_eq})
 can be performed analytically.
 Going over from flow equations for the self-energy to the ones for the renormalized 
 single-particle parameters one obtains 
 \begin{align}
	& \frac{ \partial_{\Lambda}\delta t^{\Lambda}}{ \delta t^{\Lambda}}= \frac{U}{\pi} \frac{1}{b_{}^{\Lambda}} \left\{ 1- \frac{a^{\Lambda}+b^{\Lambda}}{\sqrt{(a^{\Lambda})^2-(b^{\Lambda})^2}}\right\}, \label{eq:RG_v1_eq}  \\
	&\frac{ \partial_{\Lambda}V^{\Lambda}} {V^{\Lambda}}= -\frac{2U}{\pi}  \frac{1}{\sqrt{(a^{\Lambda})^2-(b^{\Lambda})^2}}, \label{eq:RG_t1-t2_eq}\\
	& \frac{ \partial_{\Lambda}t^{\Lambda}}{t^{\Lambda}}= -\frac{U}{\pi} \frac{1}{b_{}^{\Lambda}} \left\{ 1- \frac{a^{\Lambda}-b^{\Lambda}}{\sqrt{(a^{\Lambda})^2-(b^{\Lambda})^2}}\right\}, \label{eq:RG_t1+t2_eq}
\end{align}
with $\delta t^\Lambda = (t_1^\Lambda - t_2^\Lambda)/2$ and 
$t^\Lambda =  (t_1^\Lambda + t_2^\Lambda)/2$.
For weak interactions, one can expand the right hand sides of the RG equations in $U$. Due to the explicit prefactor $U$, the first order correction can be obtained by replacing the renormalized parameters in $a^{\Lambda}$ and $b^{\Lambda}$ by the bare ones. We will 
use this below. Note that this is an additional approximation which comes
on top of the truncation of the infinite hierarchy of functional RG 
flow equations to lowest order. In Sect.~\ref{sec:num} we avoid this 
and numerically integrate the full set of truncated flow equations. 

The self-energy or the (effective) single-particle parameters within standard first order perturbation theory (for the self-energy, not the Green function) can, as usual, be obtained from 
the lowest-order truncated RG flow equations by switching off the feedback of the self-energy \cite{Metzner12}. In Eqs.~(\ref{eq:RG_v1_eq}) to (\ref{eq:RG_t1+t2_eq}) we thus do not only have to replace the renormalized parameters by the bare ones in the expressions for   $a^{\Lambda}$ and $b^{\Lambda}$ on the right hand sides but in addition in the corresponding denominators on the left hand sides. We will return to this.

\subsubsection{Gap renormalization}
\label{sec:subsubgap}

 For bare gaps $2 \Delta$ much smaller than the band width $2W$ and keeping the leading order in $U$ only, the right hand sides of the Eqs.~(\ref{eq:RG_v1_eq}) and~(\ref{eq:RG_t1-t2_eq}) can systematically be expanded leading to
 \begin{align}
	\frac{ \partial_{\Lambda}V^{\Lambda}}{ V^{\Lambda}}= -\frac{U}{\pi \Lambda}, \hspace{1.2em}  
	\frac{ \partial_{\Lambda} \delta t^{\Lambda}}{\delta t^{\Lambda} }=-\frac{U}{\pi \Lambda}.
	 \label{eq:RG_v1_t1-t2_small_U_eq}
\end{align}
Integrating Eq.~(\ref{eq:RG_v1_t1-t2_small_U_eq}) from the high-energy cutoff $W$ down to the low-energy scale $\Delta$ leads to 
\begin{align}
	\frac{V^{\text{ren}}}{V} \thicksim\left(\frac{\Delta}{W}\right)^{-U/{\pi}} \thicksim \frac{ \delta t^{\text{ren}}}{ \delta t} \; 
	 \text{for \hspace{0.em} $\Delta \ll W$}.
	 \label{eq:solution_RG_v_t1-t2_eq}
\end{align}
Employing that the renormalized gap, as it shows up in the single-particle spectral function (see Sect.~\ref{sec:subsecspec}), can be obtained introducing the renormalized parameters
at the end of the flow into Eq.~(\ref{eq:gap}) we obtain  
 \begin{align}
	 \frac{\Delta^{\text{ren}}}{\hspace{-1em}\Delta^{}} = \left(\frac{ \Delta}{ W}\right)^{-U/{\pi}} \text{for  \hspace{0.em} $ \Delta\ll W$.} 
	 \label{eq:RG_gap_eq_smallgap} 
\end{align}
This result for the ratio of the renormalized and the bare gap as a function
of the bare one is fully consistent with the one obtained from field 
theory as mentioned in 
Sect.~\ref{sec:field_theory} \cite{Kivelson85,Horovitz85,Wu86}.
We are not aware that this power-law 
increase (for repulsive interactions) of the gap was earlier shown directly for a microscopic lattice model, that is without the intermediate approximate step of mapping the lattice model to a continuum field theory.

Evaluating  Eqs.~(\ref{eq:RG_v1_eq}) and~(\ref{eq:RG_t1-t2_eq}) in first order perturbation theory as described above we obtain the perturbative result
\begin{align}
\Delta^{\text{ren}}=\Delta \left(1-\frac{U}{\pi}\mbox{ln}\frac{\Delta}{W}\right) .
\label{eq:gap_perturb}
\end{align}
It coincides with the leading order in $U$ expansion of Eq.~(\ref{eq:RG_gap_eq_smallgap}).
This logarithmic divergence in the limit of small $\Delta$ is known from field theory 
\cite{Kivelson85,Horovitz85,Wu86}. 
For the present lattice model it can also be obtained directly by employing standard first order perturbation theory. 

To summarize this part, we have shown analytically that functional RG in its lowest-order truncation contains all infrared divergent leading log terms and is able to resum this series to a power law.
 
\subsubsection{Band width renormalization}
\label{sec:subsubband}

After the analysis of the gap renormalization, we next discuss the interaction effect on the band width. Similar to the renormalized gap it will be visible in the  local spectral function discussed in Sect.~\ref{sec:subsecspec}.
As for bare gaps $\Delta \ll W$, $V^{\text{ren}}\lesssim \Delta^{\text{ren}}\ll t^{\text{ren}}_1+t^{\text{ren}}_2$, it is 
meaningful to define half the renormalized band width as $W^{\text{ren}}=t^{\text{ren}}_1+t^{\text{ren}}_2 = 2 t^{\rm ren}$.
Keeping the term to leading order in $U$ on the right hand side of the RG equation and systematically expanding for $\Delta\ll  \, \mbox{min} \, \left\{W,\Lambda\right\}$, Eq.~(\ref{eq:RG_t1+t2_eq}) becomes
 \begin{align}
	\frac{ \partial_{\Lambda}t^{\Lambda}}{ t^{\Lambda}}= -\frac{U}{\pi}\frac{2}{W^2} \left( 1-\frac{\Lambda}{\sqrt{\Lambda^2+W^2}} \right), 	 \label{eq:RG_t1+t2_small_U_eq}
\end{align}
It can be integrated over $\Lambda$ from $\infty$ (which is possible as
the right hand side decays as $1/\Lambda^2$) to $\Delta$ with the solution 
 \begin{align}
	\frac{ W^{\text{ren}}}{\hspace{-1.0em}W} = \mbox{exp} \left\{\frac{U}{\pi}\left( \frac{\sqrt{\Delta^2+W^2}-\Delta}{W} \right) \right\}. 	 \label{eq:solution_t1+t2_small_U}
\end{align}
Expanding this result up to first order in $U$, (half) the renormalized bandwidth is given as
 \begin{align}
	 W^{\text{ren}}=W\left( 1+\frac{U}{\pi} \right).
	 \label{eq:ren_bandwidth}
\end{align}
This result is again consistent with the one known from first order perturbation theory (for the gapless model).

Note that the ``high-energy''  band width does not show any divergent behavior. A resummation of (logarithmically) divergent terms inherent to the functional RG procedure is not required. In accordance with the observation that our truncated RG does contain all regular (non-log-divergent) terms to leading order in $U$ only, the higher order terms of Eq.~(\ref{eq:solution_t1+t2_small_U}) 
are not systematic in the sense of perturbation theory. Accordingly, 
we cannot argue that this equation provides a better approximation 
to the unknown exact renormalization of the band width as compared
to the purely perturbative result Eq.~(\ref{eq:ren_bandwidth}).

\subsection{Numerical results}
\label{sec:num}

\begin{figure}[t]
  \begin{center}
  \includegraphics[width=0.5\textwidth,clip]{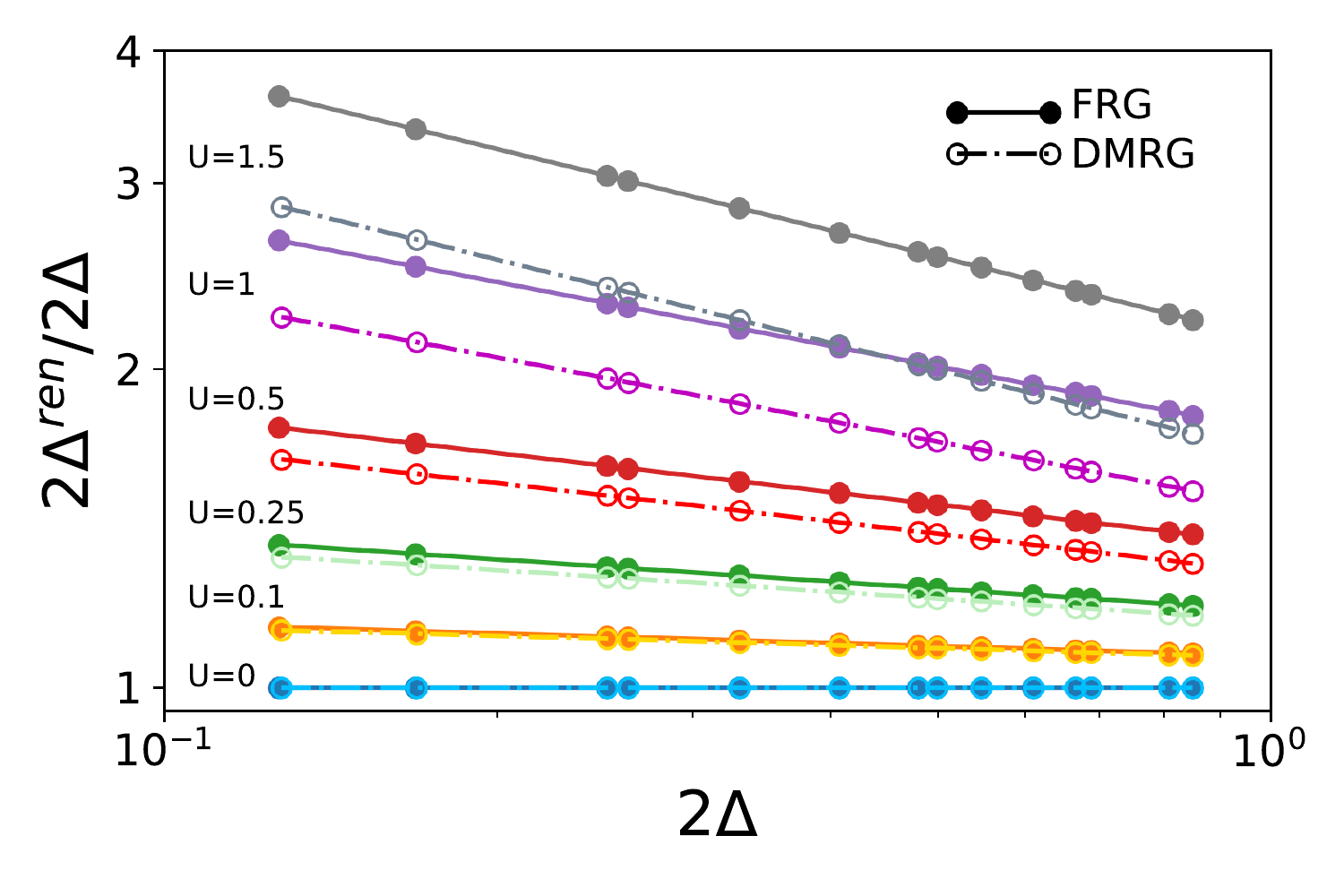}
  \caption{Main panel: The ratio of the renormalized gap and the bare one $2\Delta^{\rm ren}/(2 \Delta)$ as a function of the bare one. A comparison of functional RG (filled symbols) and DMRG data (open symbols) for different $U$ as indicated is shown. The single-particle parameters are $V=0.3\sin{\varphi} =2\delta t$ and $\varphi$ varies between $0$ and $\pi/2$. The system size is $L=1000$. Lines are guide to the eyes. A log-log scale is used.} 
  \label{fig:gap_ren}
  \end{center}
\end{figure}

\begin{figure}[t]
  \begin{center}
  \includegraphics[width=0.5\textwidth,clip]{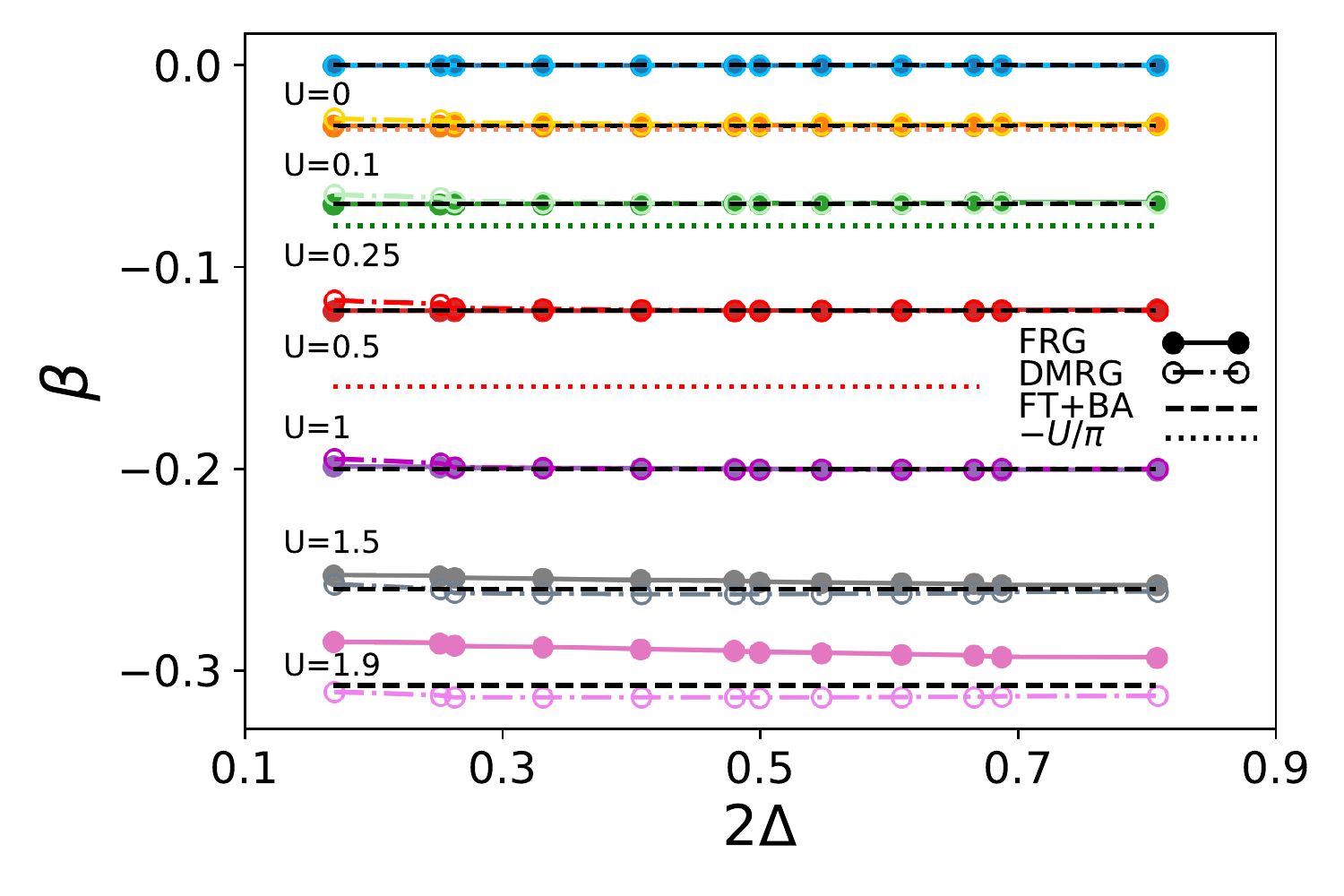}
  \caption{ Log-derivative [see Eq.~(\ref{eq:logderiv})] of the data of Fig.~\ref{fig:gap_ren}. The (color coded) dotted lines indicate the leading order exponent $-U/\pi$ (only shown for $U$ up to $0.5$). The (black) dashed lines indicate the exponent obtained by field theory employing the Bethe ansatz result for $K$ Eq.~(\ref{eq:exp_ex_FT}).} 
  \label{fig:gap_exp}
  \end{center}
\end{figure}

The effective
gap, as it shows up in the functional RG approximation of the spectral function (see Sect.~\ref{sec:subsecspec}), can be computed using
Eq.~(\ref{eq:gap}) with the bare $V$ and $\delta t$
replaced by the
renormalized values. The main panel of Fig.~\ref{fig:gap_ren} shows a comparison of
the renormalized gap devided by the bare one as a function of the bare gap
obtained by the numerical solution of the full truncated functional RG equations to DMRG data for different $U$. 
In contrast to our analytical considerations of Sect.~\ref{sec:ana} we do not employ 
any additional approximations besides the lowest order truncation when solving the RG flow equations.  
The DMRG data are obtained from the difference of the first excited and the ground state energy, as explained in Sect.~\ref{sec:DMRG}.
The single-particle parameters are $V=0.3\sin{\varphi} =2\delta t$ and $\varphi$
varies between $0$ and $\pi/2$. For interactions of up to $U=0.25$ the agreement between the functional RG and DMRG data is excellent. Both data sets show linear
behavior on a log-log scale indicating power-law scaling for small bare gaps as
discussed in Sects.~\ref{sec:ana} and \ref{sec:field_theory}. The slope and therefore the exponent depends on $U$. 

To further substantiate this we show the centered logarithmic differences
of $2\Delta^{\rm ren}/(2 \Delta)$ as a function of $2\Delta$ computed as in
Eq.~(\ref{eq:logderiv}) for different $U$ in Fig.~\ref{fig:gap_exp}.
The data obtained by both methods give a $U$ dependent constant which is the exponent $\beta$ of the power-law scaling of the 
renormalized gap. The deviations of the DMRG data from the plateau value at the smallest $\Delta$  indicate that convergence with respect to the bond dimension and/or the system size is not fully reached.  
For small $U$ the functional RG and DMRG data nicely approach
the expected leading order exponent $\beta=-U/\pi$ 
Eq.~(\ref{eq:RG_gap_eq_smallgap}) indicated as (color coded) dotted horizontal lines. 
However, the agreement between both methods persists even to interactions up to $U=1$ for which the exponent is apparently no longer dominated by the leading order expression. The plateau value obtained equally by functional RG and DMRG deviates significantly from the dotted line already for $U=0.5$. Only for very large interactions ($U \gtrapprox 1.5$) the exponents of both methods start to show visible differences on the scale of Fig.~\ref{fig:gap_exp}. This indicates that the  
higher-order corrections contained in the numerical solution of the full truncated functional RG equations (but not in the analytical solution of Sect.~\ref{sec:ana}
which required additional approximations) show the correct trend 
in comparison to the ones of the highly accurate DMRG exponent. 
The prefactors of a Taylor expansion of the functional RG 
exponent in powers of $U/\pi$ must be very close to the exact ones numerically determined by DMRG. However, within the lowest order truncated functional RG it is not possible to show analytically that the obtained exponent should agree with the exact one beyond leading order. 
Overall, this is a
rather stringent numerical confirmation that the analytical result Eq.~(\ref{eq:RG_gap_eq_smallgap}) gives
the exponent of the power-law renormalization of the gap to leading order.

We can compare the functional RG and DMRG results for $\beta$ to the result Eq.~(\ref{eq:exp_ex_FT}) for the exponent from the field theoretical model constructed for our particular lattice model by bosonization and using the Bethe ansatz result for the Tomonaga-Luttinger liquid parameter $K$. The corresponding values are shown as (black) dashed lines in  Fig.~\ref{fig:gap_exp}. They agree much better to the numerical functional RG and DMRG results than the leading order expression $-U/\pi$ (color-coded dotted lines only shown up to $U=0.5$). On the scale of the figure differences between the functional RG, the DMRG, and the field theoretical result are only visible for $U \gtrapprox 1.5$. 
This indicates that studying low-energy field theories which are designed as closely as possible to the microscopic lattice model of interest 
and using additional results available (Bethe ansatz for the gapless 
lattice model)
might be a very useful approach even beyond leading order considerations. In fact, one can raise the comparison to a higher level.

\begin{figure}[t]
  \begin{center}
  \includegraphics[width=0.5\textwidth,clip]{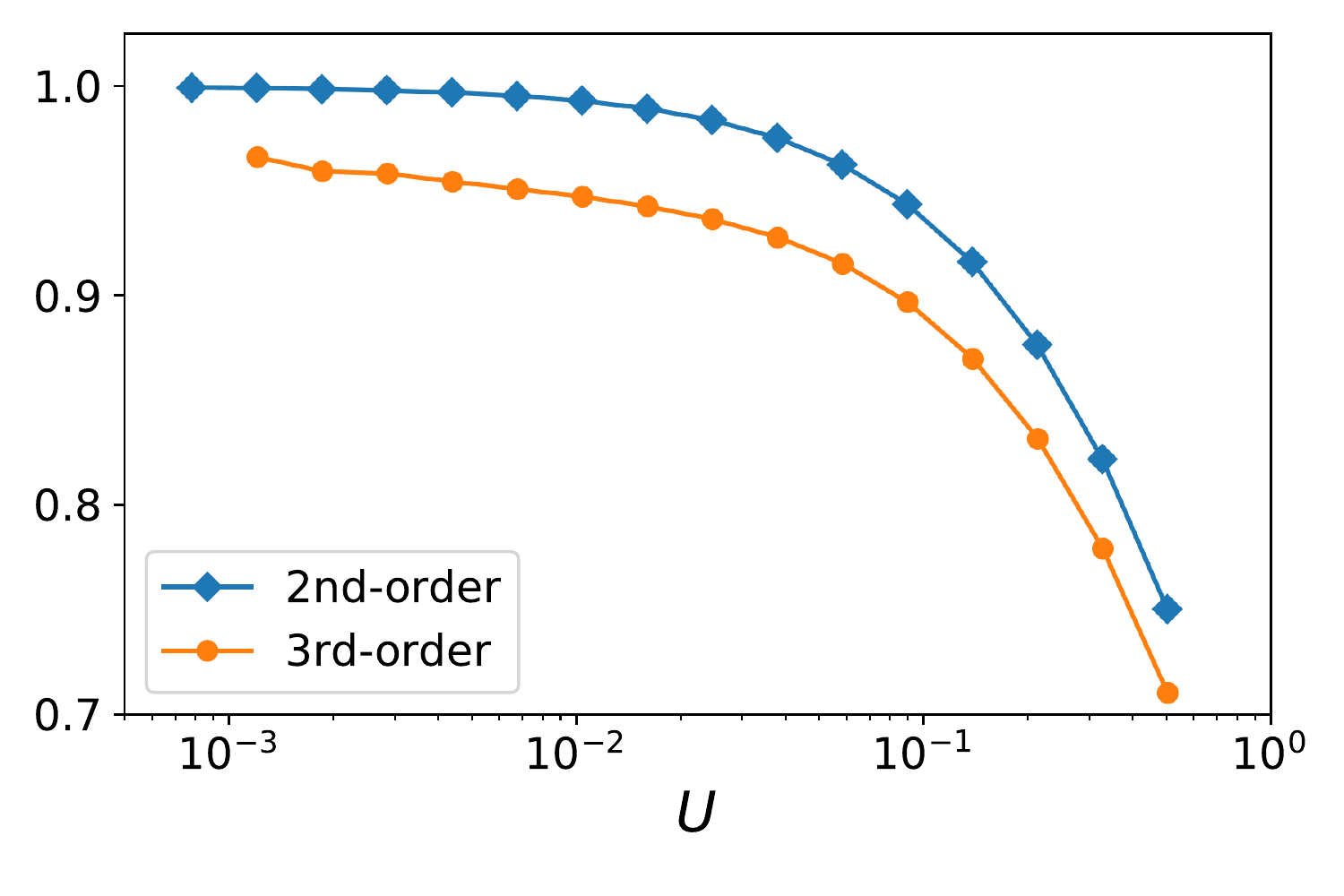}
  \caption{ The second and third order Taylor coefficient of $\beta$ (for $U \to 0$) from momentum space functional RG data, divided by the respective Taylor coefficients from the field theoretical result Eq.~(\ref{eq:exp_FT}). For dertails, see the text. }
  \label{fig:taylor_exp}
  \end{center}
\end{figure}

Using the momentum space 
functional RG flow equations (\ref{eq:RG_v1_eq})-(\ref{eq:RG_t1+t2_eq}) set up in the thermodynamic limit it is possible to extract highly accurate 
results for the exponent $\beta$ which are free of any finite size corrections by considering very small bare gaps $\Delta$ (of the order of $10^{-5}$ and smaller). This is not possible within DMRG due to finite size and bond dimension effects (see Fig.~\ref{fig:gap_exp}). From these data we subtract the leading order $-U/\pi$. After dividing the difference by $(U/\pi)^2$ and taking the limit $U \to 0$ we can read off the second order Taylor coefficient of the functional RG approximation to the exponent. 
For clarity we in addition divide by the field theoretical prediction 2 [see Eq.~(\ref{eq:exp_FT})].
As Fig.~\ref{fig:taylor_exp} shows this coefficient is indeed 2 and thus agrees with the field theoretical one.  
We proceed one step further, subtract $2 (U/\pi)^2$ and divide by $(U/\pi)^3$ as well as the prediction from field theory for the third order coefficient $-(96+\pi^2)/24$. The data are shown in  Fig.~\ref{fig:taylor_exp} as well. Now it is no longer obvious 
that the field theoretical prediction for the third order coefficient  Eq.~(\ref{eq:exp_FT}) is reached. However, the agreement is surprisingly good. In particular, functional RG and the field theoretical exponent both show an alternating structure of the power series. It is impossible to study smaller $U$ as with this highly sensitive analysis we reach machine precision. 
For the third order coefficient this is already visible at the smallest $U$ shown.
Having analyzed this in due detail we emphasize that 
it can neither be argued that the truncated functional RG nor the field theory (plus Bethe ansatz) provide the exact expression for the exponent beyond the leading order result $-U/\pi$.

In addition to the gap, the band width is renormalized by the
interaction. As discussed in Sect.~\ref{sec:subsubband} this can be computed
analytically using functional RG and simple perturbation theory. We note in passing that for sufficiently small $U$ the numerical functional RG data for the renormalized band width given by $2t^{\text{ren}}$  agree well with the perturbative result Eq.~(\ref{eq:ren_bandwidth}).   

\begin{figure}[t]
  \begin{center}
  \includegraphics[width=0.5\textwidth,clip]{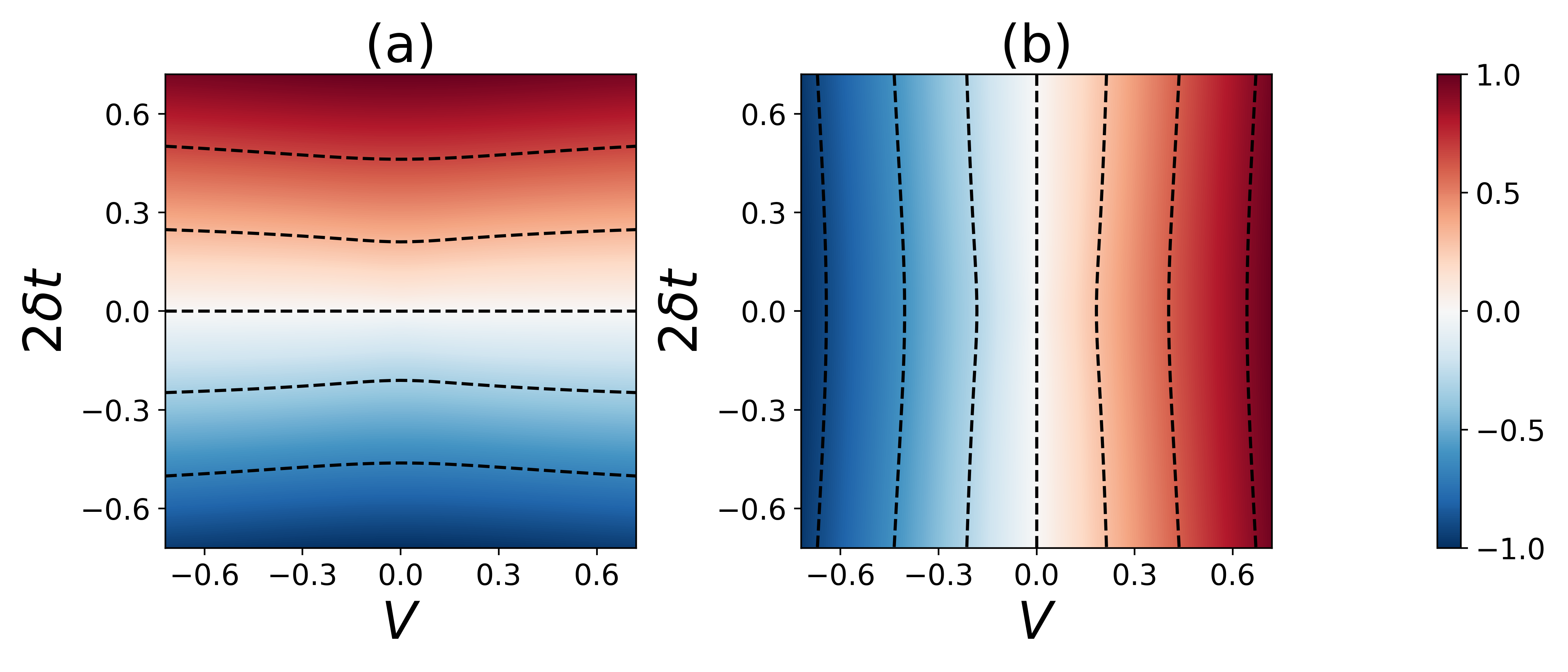}
  \caption{The renormalized parameter $\delta t^{\rm ren}$ (a) and $V^{\rm ren}$ (b) (color coded) as a function of the bare ones. The interaction is $U=0.5$. Dashed lines are equipotential lines.}
  \label{fig:ren_par}
  \end{center}
\end{figure}

In a first attempt to investigate if the topological properties are altered by the two-particle interaction we study the renormalized
$V^{\rm ren}$ and $\delta t^{\rm ren}$ at the end of the RG flow
as a function of the bare parameters $\delta t$ and $V$.  
In case $\delta t^{\rm ren}$ has
a sign opposite to $\delta t$ one would naively, that is within an effective 
single-particle picture, expect that the interaction
alters the topological properties as well as Fig.~\ref{fig:phase_dia_non}, highlighting the characteristic features of the boundary charge.
We did not observe this for any parameter set considered. Figure \ref{fig:ren_par} (a)
shows  $\delta t^{\rm ren}$ (color coded) as a function of
$V$ and $\delta t$. Similarly, $V^{\rm ren}$ stayed positive for all
positive $V$ and vice versa; see Fig.~\ref{fig:ren_par} (b). 
Still the renormalization leads to nontrivial structures indicated by the
bending of the (dashed) equipotential  lines. 

From the renormalized bulk properties
one would thus conclude that the number of  ``effective edge states'' showing up as in-gap $\delta$-peaks in the single-particle spectral function is not altered by the interaction.  We will return to this 
in Sect.~\ref{sec:subsecspec}.

\section{Systems with boundary for $U>0$}
\label{sec:boundary}

\begin{figure}[t]
  \begin{center}
  \includegraphics[width=0.5\textwidth,clip]{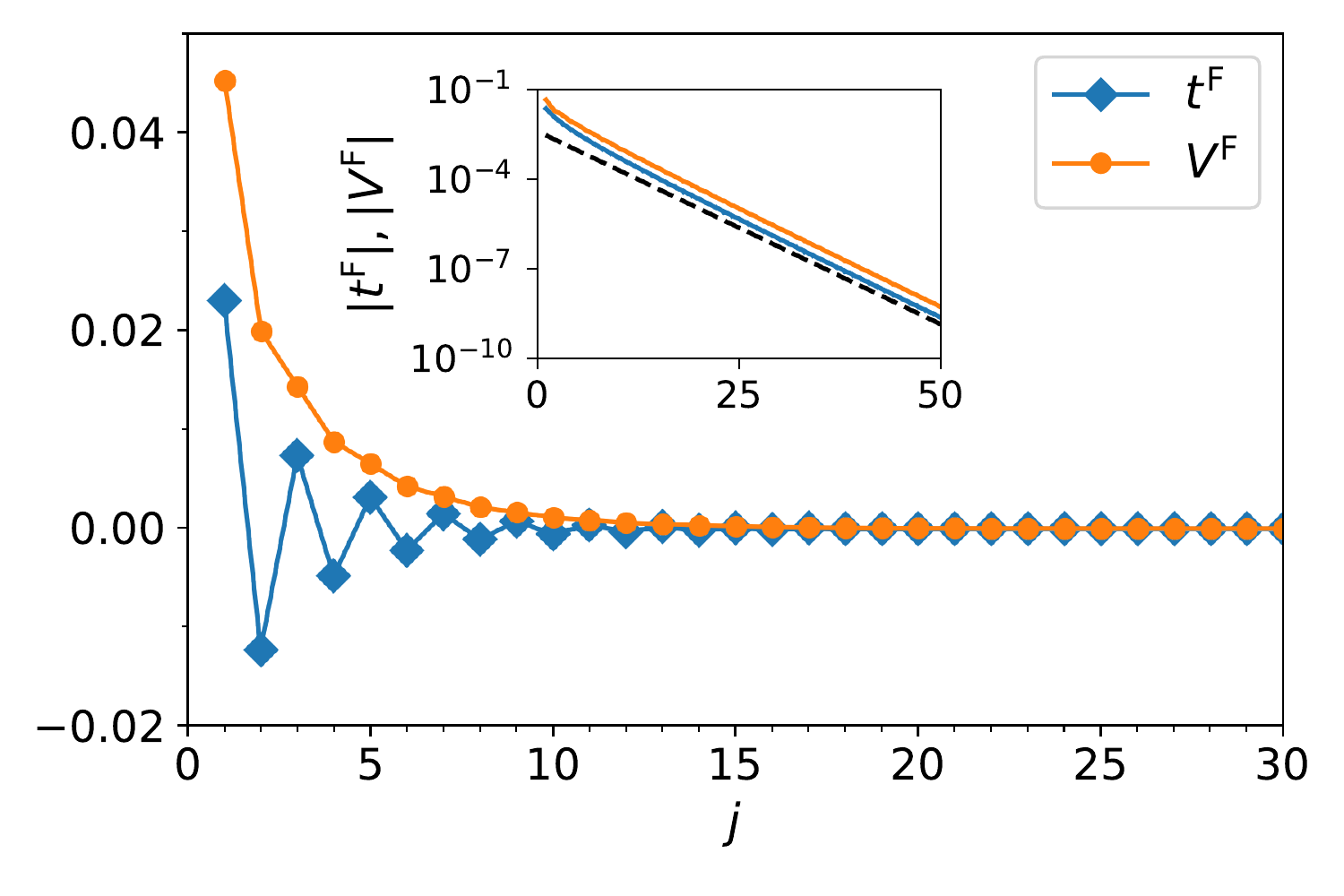}
  \caption{Main panel: Functional RG data for the Friedel-part of the renormalized single-particle parameters at the end of the RG flow; see Eqs.~(\ref{eq:needed1})-(\ref{eq:needed2}). The parameters are $V=-0.25$, $\delta t=0.001$, $U=0.25$ and the system size is $L=10000$. Inset: The absolute value of the data of the main panel on a linear-log sale highlighting the exponential decay. The dashed line shows an exponential function with decay length $1/\kappa_{\rm bc}^{\rm ren}$ which can 
be obtained by plugging the renormalized bulk values for
$\Delta$, $t_1$, and $t_2$ into Eq.~(\ref{eq:kappa_bc}).
} 
  \label{fig:ren_t_V}
  \end{center}
\end{figure}

 The nontrivial spatial structure of the frequency independent self-energy (or the renormalized  single-particle parameters) close to a boundary build up during the RG flow prohibits the analytical solution of the functional RG flow equations  (\ref{eq:RG_SE_eq}).
 This has to be contrasted to the case with PBC in which this was possible, at least in the limit of small bare gaps; see Sect.~\ref{sec:ana}. We thus have to rely on a numerical solution of the RG equations. 
 Figure~\ref{fig:ren_t_V} shows the Friedel-part of the renormalized single-particle parameters (or the static self-energy) at the end of the RG flow; see Eqs.~(\ref{eq:needed1})-(\ref{eq:needed2}). The parameters are $V=-0.015$, $\delta t=0.0025$, $U=0.25$ and the system size is $L=4000$.
 The inset illustrates that the decay towards the renormalized bulk values is exponential with a decay length $1/\kappa_{\rm bc}^{\rm ren}$ which can 
be obtained by plugging the renormalized bulk values for
$\Delta$, $t_1$, and $t_2$ into Eq.~(\ref{eq:kappa_bc}) (dashed line).
 We have verified that the renormalized values for the hoppings and onsite energies deep in the bulk of a chain with open
boundaries agree with the values computed for a chain with PBC.

In the above sections we explained how to obtain the observables of interest from
functional RG and DMRG. We focus on $\mu=0$ and start out with the local spectral function. 

\subsection{The local spectral function}
\label{sec:subsecspec}

\begin{figure}[t]
  \begin{center}
  \includegraphics[width=0.5\textwidth,clip]{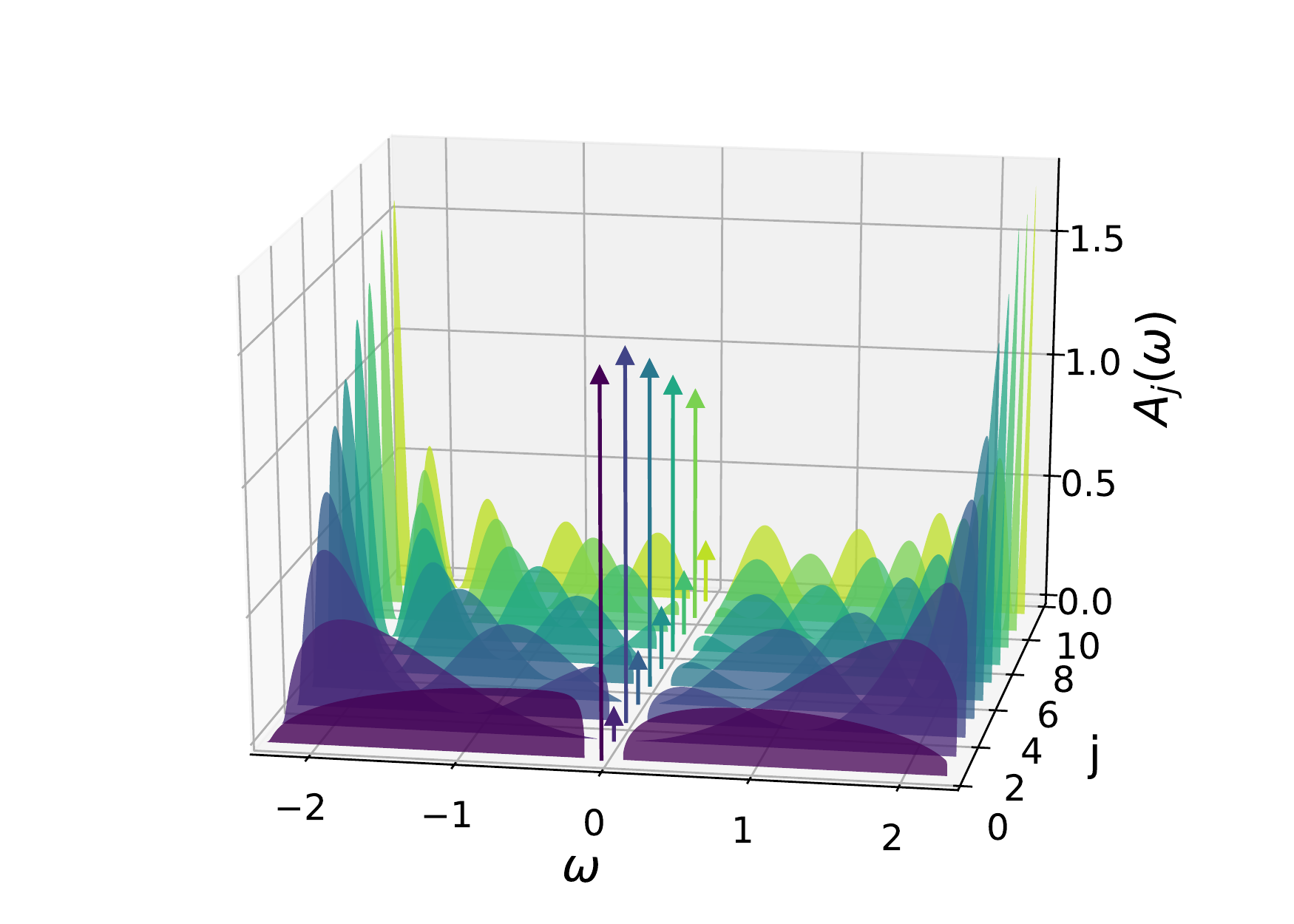}
  \caption{Functional RG data for the local single-particle spectral function of the interacting RM model for the same single-particle parameters as in Fig.~\ref{fig:LDOS_U0} and $U=0.5$. The height of the $\delta$-peaks (vertical arrows) on the $i=2$ sites is scaled up as compared to the one of the $i=1$ sites.} 
  \label{fig:LDOS_U0.5}
  \end{center}
\end{figure}

In Fig.~\ref{fig:LDOS_U0.5} we show functional RG results for the local single-particle spectral function of the interacting RM model with OBC computed using Eq.~(\ref{eq:specfu_int}). The single-particle parameters are as in Fig.~\ref{fig:LDOS_U0}, the interaction is $U=0.5$, and 
the system size $L=4096$. Similar to the procedure used in the  noninteracting case, to obtain a smooth function out of the sum of $\delta$-peaks (finite system size) we averaged the spectral weight in the bands over several eigenenergies of the effective single-particle Hamiltonian. Increasing the system size the curves do not change on the scale of the plot.

The interaction effects we expect based on our analysis of the bulk properties can clearly be observed in  Fig.~\ref{fig:LDOS_U0.5}. In comparison to  Fig.~\ref{fig:LDOS_U0} the gap size is increased and the bands extend to smaller (valence band) and larger (conduction band) energies (renormalization of the band width). The in-gap $\delta$-peak representing the edge state for $U=0$ still 
appears. We emphasize that strictly speaking the $\delta$-peak of the interacting spectral function does not have an interpretation as a single-particle (edge) state.

However, beyond these findings expected from the renormalized bulk properties we observe in Fig.~\ref{fig:LDOS_U0.5} that the $n$ dependence [$j=2(n-1)+i$]
of the weight of the $\delta$-peak on the first site of each unit cell ($i=1$) is modified as compared to the $U=0$ case [see Eq.~(\ref{eq:wavefunc_edge}) and Fig.~\ref{fig:LDOS_U0}].  For generic single-particle parameters and $U>0$ the spectral weight of the peak first increases before it starts to decrease when going from the boundary towards the bulk. This has to be contrasted to the purely exponential spatial decay of the noninteracting case. Furthermore, we observe the appearance of $\delta$-peak spectral weight
on the second sites of the unit cell ($i=2$). It is much smaller then the one on $i=1$ sites but also shows a nonmonotonic $n$ dependence. To render the weight on the $i=2$ sites visible they were all scaled up by a (arbitrary) factor as compared to the weights on $i=1$. Both these interaction effects are a consequence of the nontrivial interaction induced spatial dependence of the effective single-particle parameters close to the boundary (and beyond the unit cell structure) acquired during the RG flow.

\begin{figure}[t]
  \begin{center}
  \includegraphics[width=0.5\textwidth,clip]{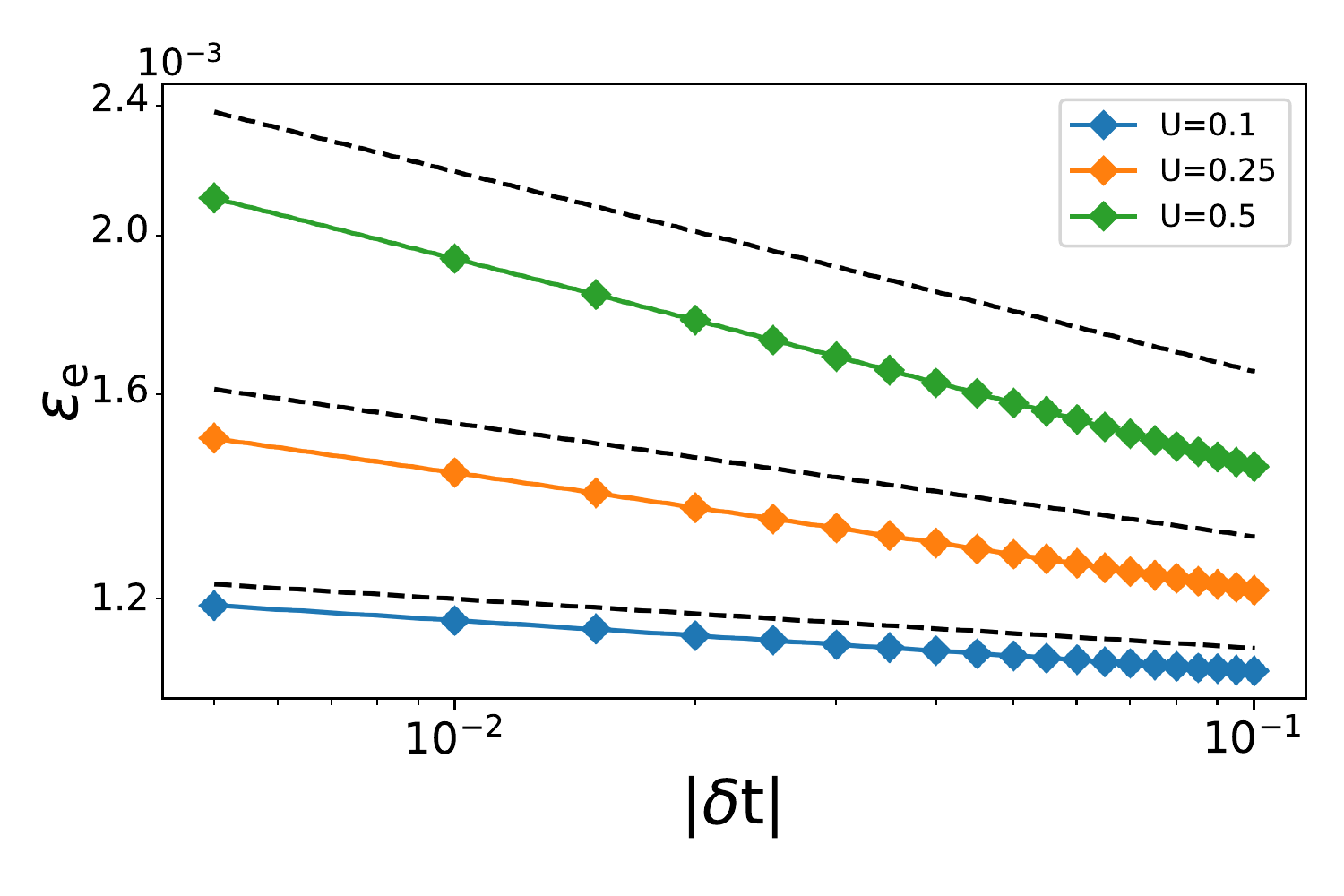}
  \caption{Functional RG data for the energy $\varepsilon_{\rm e}$ of the in-gap $\delta$-peak as a function of $|\delta t|$ in the $V\ll|\delta t|$ limit with fixed $V=0.001$ for different $U$ (symbols). It is compared to the renormalized bulk onsite potential $V^{\text{ren}}$ (dashed lines). Note the log-log scale.} 
  \label{fig:level_peak}
  \end{center}
\end{figure}

We observe that also for $U>0$, the energy $\varepsilon_{\rm e}$ of the $\delta$-peak, indicating the ``effective edge state'', turns out to be position independent.
Comparing Figs.~\ref{fig:LDOS_U0} and \ref{fig:LDOS_U0.5} one can barely see that $\varepsilon_{\rm e}$  is modified by the interaction.  To further illustrate this we show the dependence of the peak energy $\varepsilon_{\rm e}$ 
on $\delta t$ for different $U$ on a log-log scale in Fig.~\ref{fig:level_peak} (symbols). The 
single-particle parameters are $V=0.001$, and 
$\delta t$ varies from $-0.1$ to $-0.005$. The system size is
$L=2048$. As we are in the limit $|V| \ll |\delta t|$, according
to Eq.~(\ref{eq:gap}) $|\delta t|$ is a measure for the size of 
the bare gap. The energy of the $\delta$-peak thus scales as a power-law (straight line on the log-log scale) as a function of the bare gap with the leading order exponent
$-U/\pi$ known from the scaling of the renormalized gap
Eq.~(\ref{eq:RG_gap_eq_smallgap}). Consulting Eq.~(\ref{eq:solution_RG_v_t1-t2_eq}) for the renormalized bulk value of the onsite energy and taking into account that in the noninteracting case $\varepsilon_{\rm e}= V$ one might argue that this power-law dependence was to be expected. However, this ignores that the RG flow leads to a nontrivial spatial dependence of the renormalized onsite energies and bond hoppings close to the boundaries. This can be anticipated to affect all properties close to the boundaries. Indeed, for $U>0$ the energy $\varepsilon_{\rm e}$ of the in-gap $\delta$-peak does not coincide with the renormalized bulk value
$V^{\rm ren}$ of the onsite energy. The latter is shown as dashed lines in Fig.~\ref{fig:level_peak} [and shows power-law scaling as a function of $\delta t$ (respectively the bare gap) in accordance with Eq.~(\ref{eq:solution_RG_v_t1-t2_eq})].

Although first order perturbation theory for the self-energy misses the power-law renormalization of the gap as well as the power-law dependence of $\varepsilon_{\rm e}$ it leads to qualitatively the same interaction effects in the single-particle spectral function as discussed above. 

\begin{figure}[t]
  \begin{center}
  \includegraphics[width=0.5\textwidth,clip]{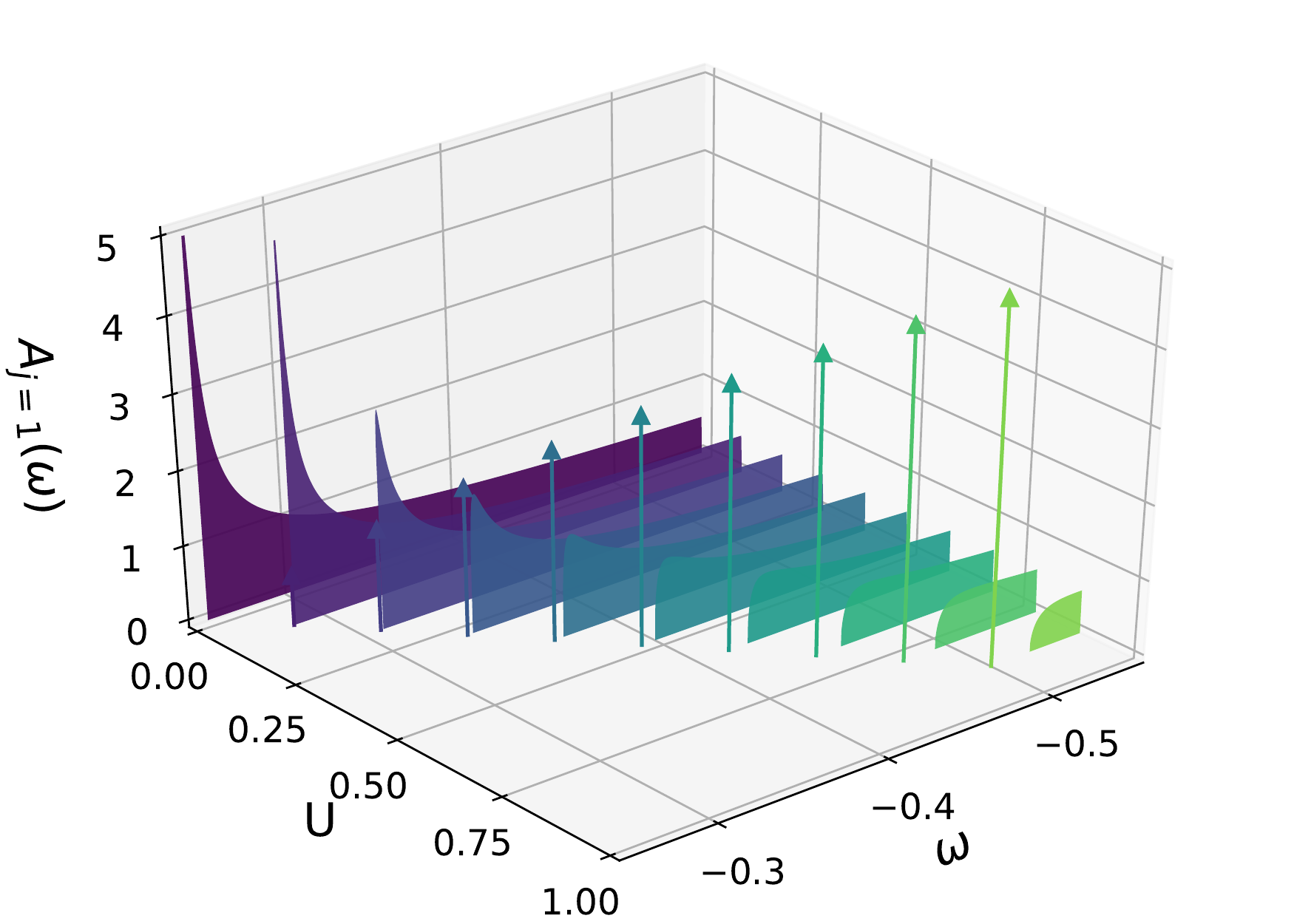}
  \caption{Functional RG data for the photoemission part of the single-particle spectral function $A_1(\omega)$ of  the interacting RM model on site $j=1$ with $\delta t=0.001$ and $V=-0.25$. Data for different $U$ are shown. In-gap $\delta$-peaks are indicated by vertical arrows.} 
  \label{fig:U_peak}
  \end{center}
\end{figure}

We find even more severe interaction effects in the single-particle spectral function associated to  ``effective edge states''. In the limit $|\delta t| \ll |V|$ the noninteracting gap Eq.~(\ref{eq:gap}) is dominated by $|V|$ and the spectral function on site $j=1$ shows a van-Hove singularity at $V$. This can be seen in the $U=0$ curve (deep purple) of Fig.~\ref{fig:U_peak}, which only displays the photoemission part $\omega<0$ of $A_1(\omega)$. For $\delta t >0$ no edge state appears. If in this regime of single-particle parameters an interaction is turned on an in-gap $\delta$-peak appears, which can be associated to an ``effective edge state''. In fact, it is an edge state of the effective single-particle Hamiltonian to be diagonalized at the end of the RG procedure. The appearance of the in-gap weight can be traced back to the interaction induced spatial modulation of the effective onsite energy and the hopping close to the boundary which can obviously alter local properties (such as ``effective edge states''). Increasing the interaction the weight of the $\delta$-peak increase as illustrated in Fig.~\ref{fig:U_peak}. It furthermore shows the characteristics  of an edge state as a function lattice site $j$; for large $j$ the weight decays exponentially. This is a property of the eigenstate of the effective single-particle Hamiltonian at the in-gap eigenvalue. However, similar to the peak of Fig.~\ref{fig:LDOS_U0.5} its weight first increases when going towards larger $j$ (not shown). 

We emphasize, that the appearance of the interaction induced ``effective edge states'' is not related to the ability of the lowest-order truncated functional RG to resum the series of leading logarithms. Accordingly, this effect can also be observed in first order perturbation theory for the self-energy.

We conclude that the interaction can alter the number of ``effective edge states'' (in-gap $\delta$-peaks of the single-particle spectral function). As discussed, this cannot be understood from the bulk properties of the system but follows from the interaction induced spatial modulation of the effective onsite energy and hopping close to the boundary. This insight shows that the number of ``effective edge states'' (in-gap $\delta$-peaks of the single-particle spectral function) in the interacting case cannot be predicted based on a bulk properties.
As we will discuss in Sect.~\ref{sec:QB_interaction} the main features of the boundary charge can be understood from the bulk properties even in the presence of interactions. 

\subsection{The local density}

\begin{figure}[t]
  \begin{center}
  \includegraphics[width=0.5\textwidth,clip]{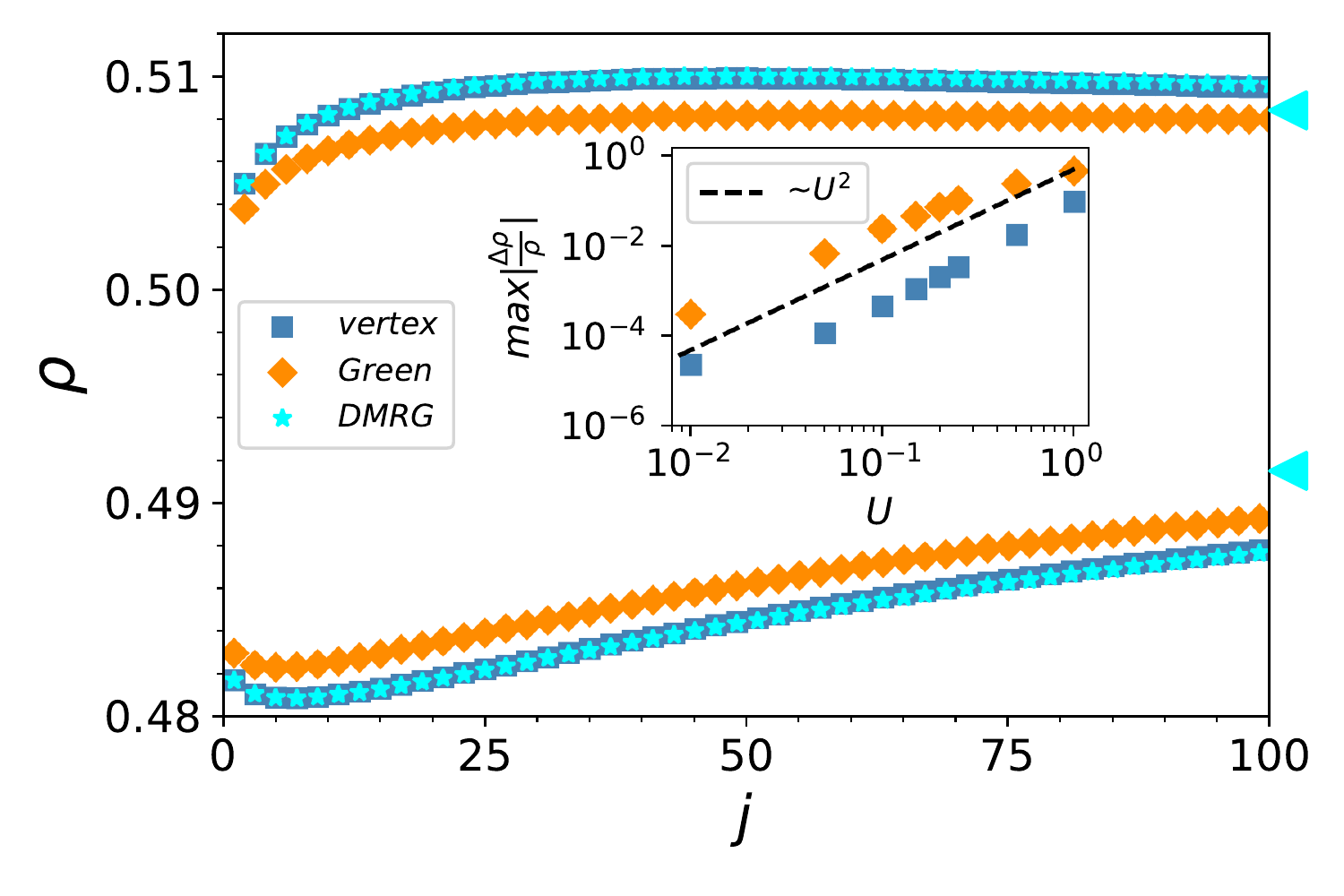}
  \caption{Main panel: Total density $\rho$ as a function of the site index $j$ for $V=0.0035$, $\delta t = -0.007$, $U=0.25$ and $L=1000$. The results from two different ways to compute the density within functional RG (labels ``vertex'' and ``Green'') are compared to the DMRG result. Filled triangles indicate the asymptotic bulk value of $\rho$ of the DMRG data. 
Inset: The largest absolute value of the relative difference between the functional RG and DMRG data taken over all lattice sites as a function of $U$. A log-log scale is taken. The dashed line indicates the power-law $U^2$ (line with slope 2 on the log-log scale).} 
  \label{fig:rho_FRG_DMRG}
  \end{center}
\end{figure}

In the discussion of our functional RG results for the local density modulations induced by an
open boundary, we start out with a comparison of functional RG and highly accurate DMRG data. The main panel
of Fig.~\ref{fig:rho_FRG_DMRG} shows results obtained for generic single-particle parameters in the small
gap limit $V=0.0035$, $\delta t = -0.007$, for a weak interaction $U=0.25$ and
system size $L=1000$.
Close to the boundary the density deviates from the bulk values, which, however, are 
approached for larger $j$. The renormalized (as compared to $U=0$) bulk value of the
density obtained from DMRG is indicated on the right by a triangle. The behavior
on the two sites of the unit cell ($i=1,2$) differs. Close to the boundary the density 
is nonmonotonic for the first site ($i=1$) in each unit cell and monotonic for the second
($i=2$). This nonmonotonicity is an
interaction effect (see below) which vanishes for $U \to 0$.
In addition, larger $j$ are required for the $i=1$ sites (odd $j$) to approach
their asymptotic bulk value as compared to the $i=2$ ones (even $j$).
This is opposite to the noninteracting case (see the discussion of Fig.~\ref{fig:rho_F_non})
and thus an interaction effect as well.
Within the approximate functional RG approach both these interaction effects can be traced back to the nontrivial spatial dependence of the effective single-particle parameters acquired during the RG flow.

Within the approximate functional RG approach the density was computed in two ways: By
integrating the $(j,j)$-matrix element of the Green function over Matsubara frequency
(label ``Green'', diamonds) and by its own
flow equation (label ``vertex'', squares).  In accordance with our discussion
in Sect.~\ref{sec:FRG} the density computed via the second way agrees better with the highly
accurate DMRG data. The inset shows the maximum (over all lattice sites) of the relative
difference between the functional RG and DMRG data as a function of $U$. Due to the truncation
this difference scales as $U^2$ (dashed line). Deviations from the $U^2$ scaling result from
the limited accuracy of the numerical solution of the functional RG flow equations as well
as the small errors inherent to the DMRG approach. We emphasize that using the flow equation
for the density one does not gain a power in $U$. Rather the difference to the exact prefactor
of the $U^2$ term is significantly smaller. From now on we refer
to functional RG density data obtained from their own flow equation.

\begin{figure}[t]
  \begin{center}
  \includegraphics[width=0.5\textwidth,clip]{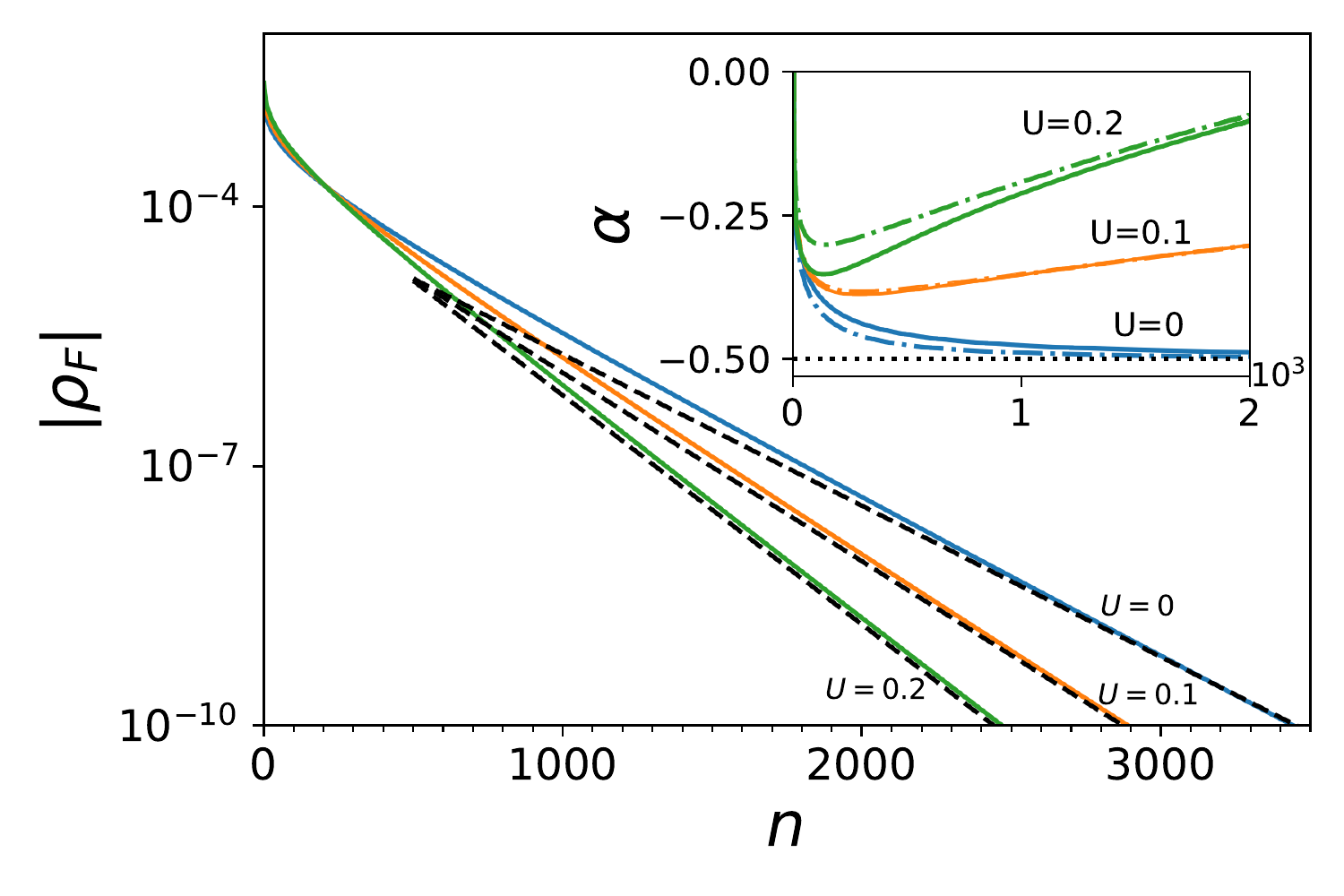}
  \caption{Main panel: Functional RG data for the  Friedel part $\rho_{\rm F}$ of the total density as a function of the unit cell index $n$ for $i=2$ and different $U$. The single-particle parameters are  $\delta t=0.0001$, and $V=0.002$. A fairy large chain with $L=20000$ sites is considered. A linear-log scale is taken to illustrate the dominating exponential decay. The slope of the dashed lines is computed plugging the renormalized bulk single-particle parameters into Eq.~(\ref{eq:kappa_bc}) for $-2 \kappa_{\rm bc}^{\rm ren}$. Inset: Logarithmic derivative [see Eq.~(\ref{eq:logderiv})]  of the pre-exponential function indicating that the interaction alters the $1/\sqrt{n}$ behavior (dotted line) of the noninteracting case. Solid lines are for $i=1$ and dashed-dotted ones for $i=2$.} 
  \label{fig:rho_exp_pre_exp}
  \end{center}
\end{figure}

As in the noninteracting case the approach of the bulk value of the density
on the two sites of the unit cell 
is dominated by an exponential factor. 
The bulk value itself agrees with the one obtained for PBC.
The functional RG decay rate $\kappa_{\rm bc}^{\rm ren}$ can 
for small $U$ be obtained by plugging the renormalized bulk values for
$\Delta$, $t_1$, and $t_2$ into Eq.~(\ref{eq:kappa_bc}). In other words, the effective single-particle picture can be used and the leading asymptotic decay is not altered by the spatial modulation of the effective single-particle parameters close to the boundary. This is shown in the main part of
Fig.~\ref{fig:rho_exp_pre_exp} for $\delta t=0.0001$, $V=0.002$, 
$L=20000$,  
and different $U$.  To avoid overloading the plot we focus on the 
unit cell index $i=2$ in the main part (solid lines). 
After subtracting the bulk value and on a linear-log
scale the data for sufficiently large $n$ are linear with the slope given by
$-2 \kappa_{\rm bc}^{\rm ren}$ as computed from the corresponding bulk
$\Delta^{\rm ren}$, $t_1^{\rm ren}$, and $t_2^{\rm ren}$ (see dashed lines).

The advantage of the functional RG approach as compared to DMRG is that it is easily possible
to study very large systems (see Sect.~\ref{sec:RGflow}). This is required if one is interested
in the spatial dependence of the density beyond the leading exponential behavior. By
subtracting the bulk values and factoring out the exponential term discussed in the last
paragraph we can extract the large $j=2(n-1)+i$ behavior of the pre-exponential function.
In the inset of Fig.~\ref{fig:rho_exp_pre_exp} we show centered
logarithmic differences of the pre-exponential function 
which were computed as in Eq.~(\ref{eq:logderiv}). The same parameters as in the main 
part are considered. We here show results for both unit cell indices $i=1$ (solid lines) and $i=2$ (dashed-dotted lines). In contrast
to the noninteracting case the $U>0$ data do not approach a plateau
at $-1/2$ (dotted line). The $1/\sqrt{n}$ decay of the pre-exponential function is thus altered by the
interaction. 
This is a qualitative change of the position dependence of the density due to the
interaction. 
However, this qualitative effect is hidden by an exponential decay and thus difficult to
observe. 
It results from a similar nontrivial pre-exponential function of the spatial dependence of the Friedel part of the renormalized self-energy (the dominant decay being exponential; see Sect.~\ref{sec:RGflow}).
The details of the behavior of the pre-exponential functions of the self-energy and the density for
$U>0$ are beyond the scope of the present paper.

We note in passing that we do not observe any remnants of the
Tomonaga-Luttinger liquid power-law decay of the Friedel oscillations of the density obtained for a 
vanishing single-particle gap. As discussed
in Sect.~\ref{sec:FRG}, if present, we should be able to observe this even within our
approximate functional RG approach. 

This completes our discussion of the spatial dependence of the density close to an open
boundary. We now turn to the boundary charge which can be computed from the density.  

\subsection{The boundary charge}
\label{sec:QB_interaction}

\begin{figure}[t]
  \begin{center}
  \includegraphics[width=0.5\textwidth,clip]{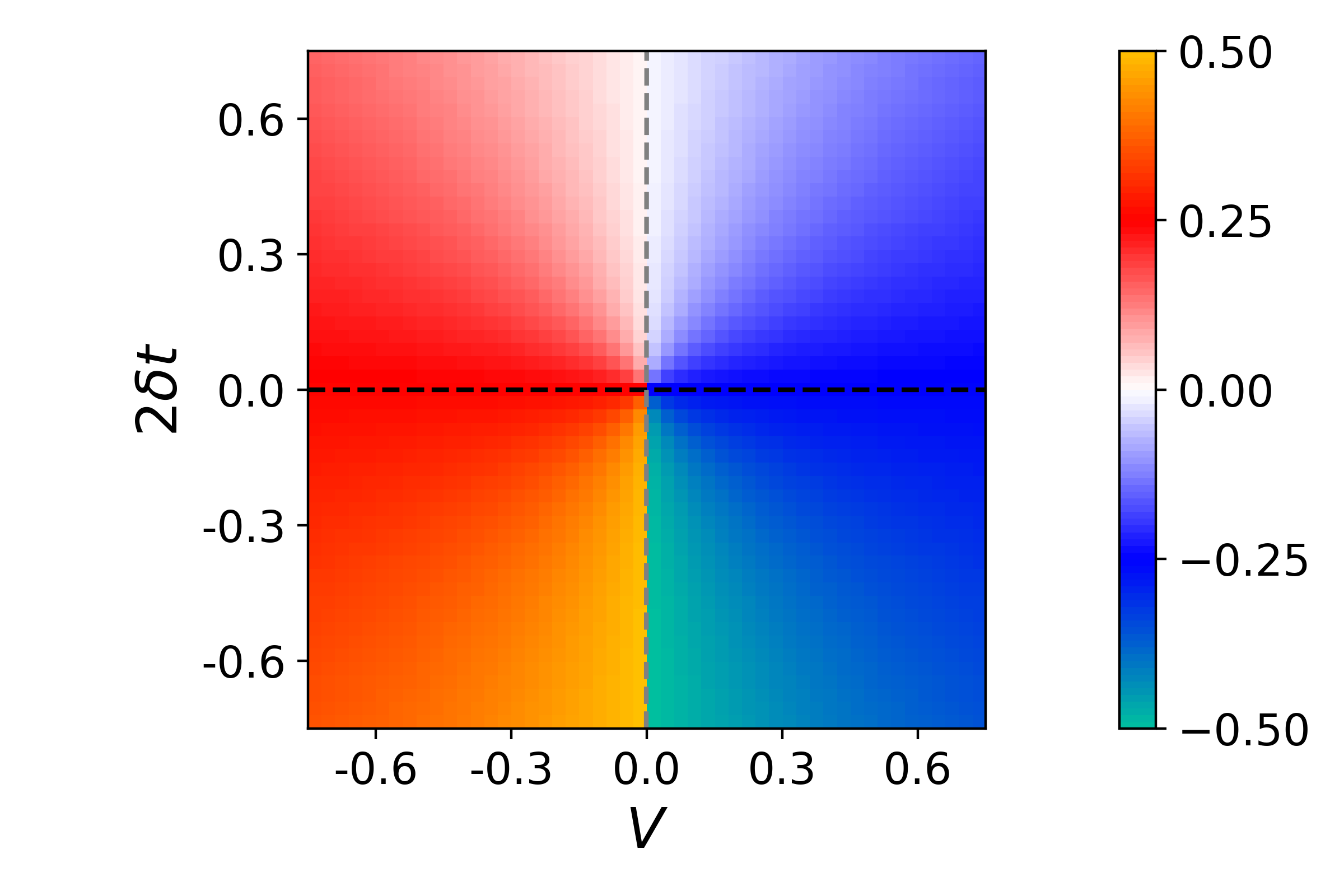}
  \caption{Functional RG data for the boundary charge $Q_{\rm B}$ of the interacting RM model as a function of $V$ and $2 \delta t$ or the polar coordinates $\Delta$ and $\gamma$, see Eq.~(\ref{eq:gamma}). The interaction is $U=0.25$.} 
  \label{fig:boundary_charge_vertex}
  \end{center}
\end{figure}

As our last observable of the interacting RM model with an open boundary we investigate the boundary charge. As in the noninteracting case it can be computed from the density by Eq.~(\ref{eq:QB_def}).

To get an overview of the interaction effects in Fig.~\ref{fig:boundary_charge_vertex} we show functional RG data for the boundary charge in the $(V,2\delta t)$ [or equivalently the $(\Delta, \gamma)$] plane for $U=0.25$. Barely any differences as compared to the noninteracting case Fig.~\ref{fig:phase_dia_non} are visible. As discussed in Sect.~\ref{sec:U_0_QB} this type of plot nicely 
illustrates the main characteristics  (i), (ii), and (iv) of the boundary charge for noninteracting models. Combined this already indicates that these characteristics are robust towards small two-particle interactions.

For the noninteracting model, the features (i)-(iv) of the boundary charge follow from bulk properties. Thus the apparent robustness of (i), (ii), and (iv) towards interactions  [for (iii), see below] in addition provides a first hint that this also holds for $U>0$. Crucially, the spatial modulations of the renormalized single-particle parameters close to the boundary do not seem to alter the general features of the boundary charge. This has to be contrasted to the number of  ``effective edge states'' (in-gap $\delta$-peaks of the single-particle spectral function) which in the interacting case cannot be predicted from bulk properties; see Sect.~\ref{sec:subsecspec}. 
Next we further substantiate the robustness of  (i)-(iv) towards two-particle interactions for $U>0$.

\begin{figure}[t]
  \begin{center}
  \includegraphics[width=0.5\textwidth,clip]{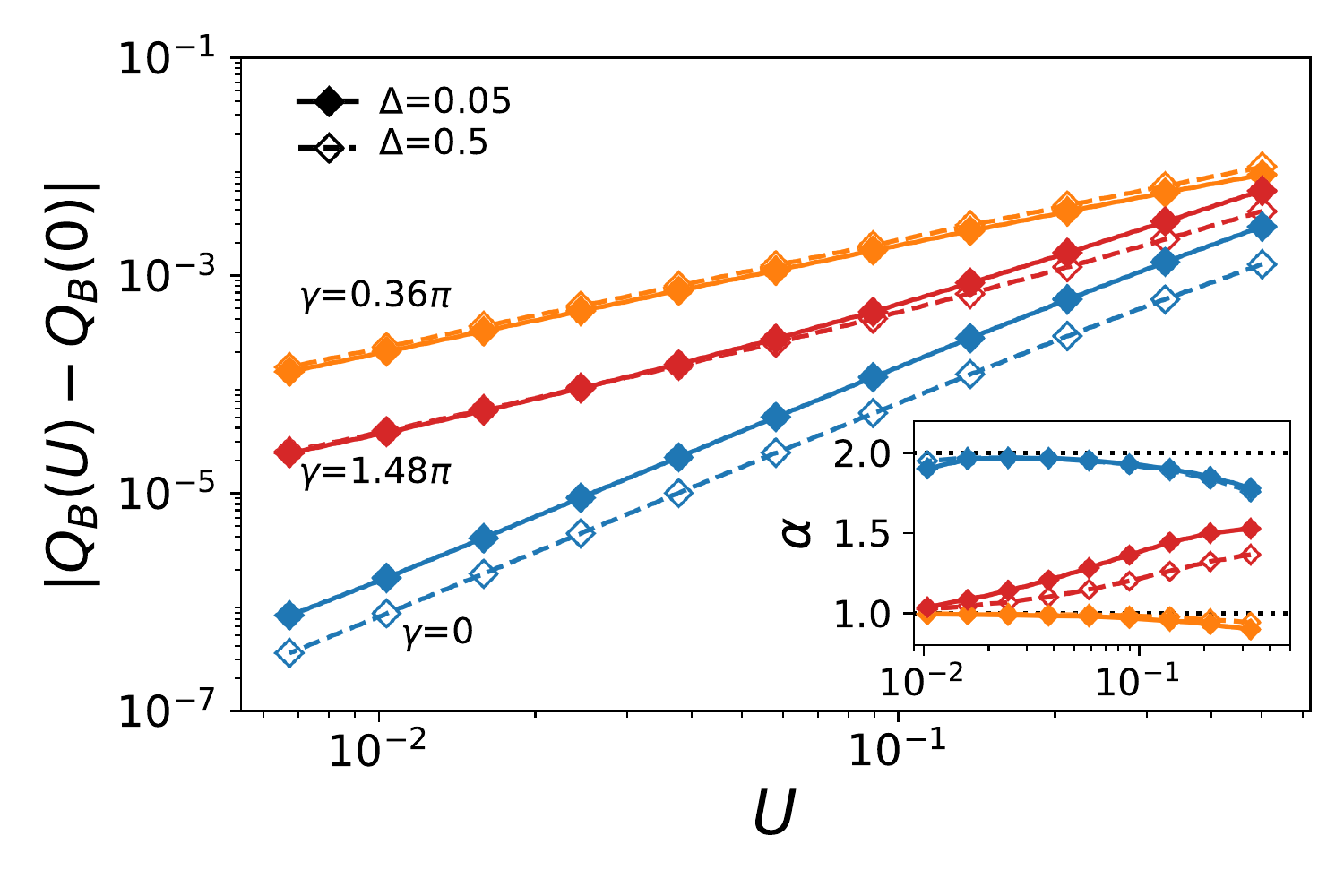}
  \caption{Main panel: Functional RG data for the interaction correction of the boundary charge with respect to the noninteracting one as a function of $U$. Different $\Delta$ and $\gamma$ are considered. Inset: Logarithmic derivative of the data, computed as in Eq.~(\ref{eq:logderiv}).} 
  \label{fig:BC_vertex_U}
  \end{center}
\end{figure}

In Fig.~\ref{fig:BC_vertex_U} we show the difference between the boundary charge with and without interaction as a function of $U$ for different $\Delta$ and $\gamma$ on a log-log scale. First of all, we realize that for small $U$ the corrections are very small. Still, for generic $\gamma$ they are of order $U$. This can be seen from the inset, which shows the logarithmic derivative 
of the data computed as in Eq.~(\ref{eq:logderiv}). 
Furthermore, the finite $U$ corrections depend on (the generic) $\gamma$ but are only weakly $\Delta$-dependent.
For $\gamma$ being a multiple of $\pi$ (blue symbols in  Fig.~\ref{fig:BC_vertex_U}), i.e.~$\delta t=0$ [see Eq.~(\ref{eq:gamma})], the corrections are of order $U^2$. We note that within our approximate functional RG procedure not all terms of order $U^2$ are kept. We thus do not control the value of these $U^2$ corrections. We associate the deviations from the exponent 2 for $U \lessapprox0.01$ (see the inset of Fig.~\ref{fig:BC_vertex_U}) to small errors of the data for $Q_{\rm B}$ obtained by the numerical integration of the RG flow equations. Note that the value of $\left| Q_{\rm B} (U) - Q_{\rm B}(0)\right| $ for $\gamma=0$ is already very small and that taking the logarithmic derivative significantly enhances small errors. 

\begin{figure}[t]
  \begin{center}
  \includegraphics[width=0.5\textwidth,clip]{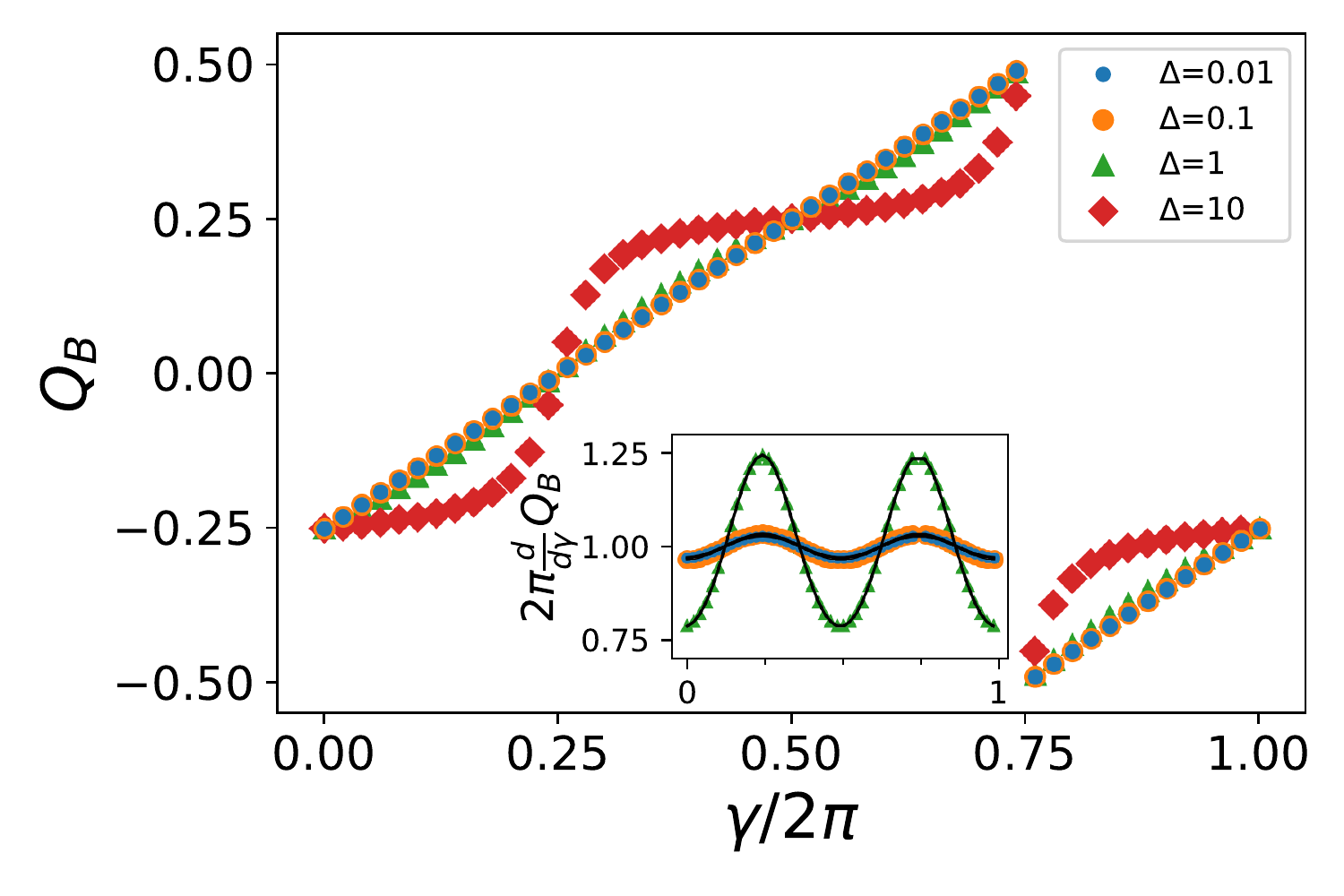}
  \caption{Main panel: Functional RG data for the boundary charge of the interacting RM model as a function of $\gamma$ for different $\Delta$. The interaction is $U=0.089$. Inset: Derivative of the data of the main panel with respect to $\gamma$.  Black lines (lying almost perfectly on top of the symbols) are data obtained for the noninteracting RM model but with the single-particle parameters $t_1$, $t_2$, and $V$ replaced by the bulk renormalized ones.} 
  \label{fig:BC_Delta_U}
  \end{center}
\end{figure}

In Fig.~\ref{fig:BC_Delta_U} we show $Q_{\rm B}$ as a function of $\gamma$ for $U=0.089$ and different $\Delta$ (symbols). The linearity in $\gamma$ for small $\Delta$ as derived analytically for $U=0$ [see Eq.~(\ref{eq:QB_low_energy_limit})] and illustrated in Fig.~\ref{fig:BC_Delta_free} is robust against small interactions. However, the interaction enhances the corrections to the linear behavior
and for small $\Delta$ they appear to be independent of the bare gap.
This can be seen most clearly by comparing the insets of  Fig.~\ref{fig:BC_Delta_U} and Fig.~\ref{fig:BC_Delta_free} which show the logarithmic derivatives of the data of the corresponding main panels with respect to $\gamma$. For $\Delta \gg W$ the same $\frac{1}{4}$-quantization of the boundary charge as derived for the noninteracting case in Eq.~(\ref{eq:QB_atomic_limit}) and discussed in connection with Fig.~\ref{fig:BC_Delta_free} can be found in Fig.~\ref{fig:BC_Delta_U}. Thus, also this feature is robust against small interactions.

The interaction effects found in Figs.~\ref{fig:BC_vertex_U} and \ref{fig:BC_Delta_U} at small $\Delta$ can all be understood from the behavior of the bulk renormalized parameters.  
Taking the analytical solution
Eq.~(\ref{eq:solution_RG_v_t1-t2_eq}) of the RG flow equation derived in the small gap limit, the renormalized $\gamma$, which is determined by the ratio of $\delta t^{\rm ren}$ and $V^{\rm ren}$ [(see Eq.~(\ref{eq:gamma})], is $U$-independent. Considering an effective single-particle picture the leading part of  Eq.~(\ref{eq:QB_low_energy_limit}) is thus unaffected by the interaction. However, using Eq.~(\ref{eq:low_energy_correction})
the correction  $1/(8\pi)\sin(2\gamma)(\Delta/W)^2\ln(\Delta/W)$ (to the noninteracting expression) acquires an interaction dependence via the renormalization of the gap  $\Delta \rightarrow \Delta^{\rm ren}$ as well as of the band width $W \to W^{\rm ren}$. 
For generic $\gamma$ this leads to a correction to the boundary charge which is linear in $U$. For small bare $\Delta$ the $U$-dependence of the renormalized band width $W^{\rm ren}$ dominates over the one of the renormalized gap and 
the interaction correction of the boundary charge becomes $\Delta$-independent.
Only for $\sin(2\gamma)=0$, i.e.~$\gamma$ being a multiple of $\pi$, the correction linear in $U$ vanishes. This explains the interaction effects seen in Figs.~\ref{fig:BC_vertex_U} and \ref{fig:BC_Delta_U}.  One can even go one step further and make this quantitative. For this we extracted the renormalized bulk values of the single-particle parameters $t_1$, $t_2$, and $V$ and inserted them in the expression for the boundary charge of the noninteracting RM model. The results shown as solid black lines in the inset of Fig.~\ref{fig:BC_Delta_U} perfectly match the functional RG data obtained for a chain with an open boundary. 

For $\Delta \gg W$ in Fig.~\ref{fig:BC_Delta_U} we find the same 
$\frac{1}{4}$-quantization of the boundary charge as in the noninteracting limit. However, as long as $\Delta \gg U$ it was to be expected that this feature of $Q_{\rm B}$ is robust against interactions. 

We can thus conclude that the interaction effects on the characteristic features (i)-(iv) of the boundary charge are weak and, most importantly, can fully be understood from the renormalized bulk properties. They are not altered by the interaction induced modulation of the onsite energies and hoppings close to the boundary. This must be contrasted to the number 
of ``effective edge states'' (in-gap $\delta$-peaks of the single-particle spectral function) discussed in Sect.~\ref{sec:subsecspec}. Therefore, the boundary charge might be the more appropriate indicator of the relation of boundary to bulk properties in the presence of two-particle interactions. We emphasize that it is only possible to show these properties of the boundary charge if in addition to the renormalization of the (low-energy) gap also the renormalization of the (high-energy) band width is properly captured. In contrast to low-energy field theories, which do not allow to compute the latter, the functional RG consistently provides the band width renormalization. This constitutes another advantage (besides the direct applicability to microscopic lattice models) of the functional RG over effective low-energy field theories.        

Our result of the stability of the boundary charge against short-ranged two-particle interactions is a microscopic manifestation of the important property of insulators that local fields (either external or interaction-induced ones) of arbitrary size lead only to local charge redistributions, i.e., charges can not be displaced beyond a characteristic length scale (given roughly by $W/\Delta$ for our model). This principle, also called the nearsightedness principle (NSP) \cite{kohn_prl_76,prodan_kohn_pnas_05}, is responsible for many universal properties of topological insulators as, e.g., charge pumping \cite{thouless_prb_83,niu_thouless_jphysA_84}, the  bulk-boundary correspondence
\cite{fidkowski_etal_prl_11,mong_shivamoggi_prb_11,gurarie_prb_11,essin_gurarie_prb_11,ukui_etal_jpsjf_12,yu_wu_xie_npb_17,rhim_etal_prb_18,silveirinha_prx_19}, and the exponential localization of the charge density at boundaries \cite{kallin_halperin_prb_84}. Recently, the NSP has also been used to derive the characteristic features (i), (ii) and (iv) of the boundary charge \cite{pletyukhov_etal_preprint} (see section \ref{sec:U_0_QB}). Therefore, the establishment of the NSP for an interacting microscopic lattice model is a very important step for a full understanding of the universal properties of insulators and their stability. In this regard the functional RG is a very useful tool as it can capture the microscopic details of the band structure and two-particle interactions on all energy scales. In contrast, other methods are either restricted to noninteracting systems (exact diagonalization) or to the regimes of small gaps (effective low-energy field theories). Computing the boundary charge for small gaps (low-energy limit) using the numerical DMRG is computationally very challenging. It requires the use of very large systems as the inverse system size sets a low-energy cutoff. 

\section{Summary and outlook}
\label{sec:summary}

We studied the local single-particle spectral function, the local density, as well as the boundary charge of the noninteracting and interacting RM model for periodic chains and such with open boundaries. For $U=0$ our main focus was on the boundary charge. We showed that results recently obtained in the low-energy limit $\Delta \ll W$ within an effective low-energy theory hold for surprisingly large gaps. In addition we found a universal $\frac{1}{4}$-quantization of the boundary charge for large gaps. We explicitly illustrated the four main characteristics of the boundary charge for the model under consideration which all follow from properties of the bulk Hamiltonian. We showed that this  relation to bulk properties is robust towards small two-particle interactions
employing a functional RG approach, which, for small interactions, provides reliable results on all energy scales. In contrast, interaction spoils the relation between the number of in-gap $\delta$-peaks, i.e.~the number of  ``effective edge states'', and renormalized bulk properties. Novel interaction induced peaks are generated by the spatial variation of the self-energy close to the boundary. These also affect the local density close to the boundary. Our results provide a hint that the fractional part of the boundary charge is an interesting quantity to study the relation of boundary physics to bulk properties.

For noninteracting and clean systems the relation to bulk properties for the fractional part of the boundary charge is established via its relation to the Zak-Berry phase (also called  ``surface charge theorem''). It is also applied within density functional theory (DFT) and mean-field theories (MFT) under the restrictive assumption that two-particle interactions can be treated within such methods \cite{vanderbilt_book_2018}. For 1d systems and in the limit of small gaps this assumption does not hold. The relation between the Zak-Berry phase and the fractional part of the boundary charge holds up to an unknown integer since there is a freedom of how to choose the gauge of the Bloch states. For systems with disorder or true many-body correlations, e.g.~interacting 1d systems in the limit of small $\Delta$, the Zak-Berry phase is not defined and one should directly study the physical observable, namely the boundary charge. Therefore, its determination in terms of renormalized bulk parameters and the stability analysis of its universal properties is a central task of many-body methods. The functional RG is a very useful tool in this respect since it can capture true many-body correlations on all energy scales not accessible by DFT and MFT. This is of particular importance for 1d systems where Tomonaga-Luttinger liquid physics is very important for vanishing $\Delta$. 

Furthermore, functional RG is very flexible and has the potential to be applied to a variety of interacting systems. It will be of interest to study the validity range of universal low-energy results for larger values of the wavelength $Z$ of the external modulation (as compared to $Z=2$ for the RM model) and for disordered systems. In addition, one can study systems with spin, e.g.~the 1d Hubbard model and multi-channel systems with several orbitals per site. Besides the boundary charge, the interface charge is expected to have comparable universal properties \cite{pletyukhov_etal_preprint} and can be directly calculated from the local density. Furthermore the functional RG can be used to study the density-density correlation function and the fluctuations of the boundary charge, and is in principle not restricted to one-dimensional systems. Therefore, we expect the functional RG to be a very useful tool to study topological properties in the presence of many-body correlations and disorder.

\section*{Acknowledgments}
We thank J. Klinovaja and D. Loss for fruitful discussions. This work was supported by the Deutsche Forschungsgemeinschaft  (DFG, German Research Foundation) via RTG 1995 and under Germany's Excellence Strategy - Cluster of Excellence Matter and Light for Quantum Computing (ML4Q) EXC 2004/1 - 390534769. DMK acknowledges support from the Max Planck-New York City Center for Non-Equilibrium Quantum Phenomena. Simulations were performed with computing resources granted by RWTH Aachen University.

\appendix*

\section{The noninteracting Rice-Mele model}
\label{app:RM_U=0}

In this Appendix we derive analytical expression for the density and the boundary charge of the noninteracting RM model. We start with the bulk density of the infinite system and prove Eq.~(\ref{eq:rho_bulk_2}). Using Eq.~(\ref{eq:rho_bulk}) we close the integration contour over $k$ in the upper half of the complex plane 
\begin{align}
  \rho_{\rm bulk}(j) = \frac{1}{2}+(-1)^j\frac{V}{4\pi}\oint_{\cal{C}} dk \frac{1}{\epsilon_k} .
    \label{eq:rho_bulk_complex_plane}
\end{align}
Here, $\cal{C}$ is a closed curve defined via straight lines on the segments $-\pi\rightarrow\pi\rightarrow\pi+i\infty\rightarrow -\pi+i\infty\rightarrow -\pi$. This can be done since the additional segments do not contribute. The two segments $\pi\rightarrow\pi+i\infty$ and $-\pi+i\infty\rightarrow -\pi$ cancel each other due to periodicity under the shift of $k$ by $2\pi$. The segment $\pi+i\infty\rightarrow -\pi+i\infty$ is zero due to the infinite imaginary part of $k$. Using Eq.~(\ref{eq:disp}) for $\epsilon_k$ one finds a branch cut starting at the branching point $k_{\rm bc}$ where $\epsilon_{k_{\rm bc}}=0$, leading to $k_{\rm bc}=\pi + i\kappa_{\rm bc}$ and $\kappa_{\rm bc}$ given by Eq.~(\ref{eq:kappa_bc}). Choosing the branch cut in the direction of the positive imaginary axis and closing the integration contour around the branch cut, we find for the bulk density
\begin{align}
      \rho_{\rm bulk}^{\rm (bc)}(j) = \frac{1}{2} - (-1)^j\frac{V}{2\pi}\text{Im} \int_0^\infty d\kappa \,
      \frac{1}{\epsilon_{k_{\rm bc}+i\kappa + 0^+}} . 
      \label{eq:rho_bulk_integral}
  \end{align}
Using
\begin{align}
    \epsilon_{k_{\rm bc}+i\kappa + 0^+} = i\sqrt{-R(\kappa)}, 
    \label{eq:epsilon_bc}
 \end{align}
with $R(\kappa)$ defined in Eq.~(\ref{eq:R}), we arrive at Eq.~(\ref{eq:rho_bulk_2}).

To calculate the Friedel density from Eq.~(\ref{eq:rho_F_chi}) we again close the integration contour over $k$ in the upper half of the complex plane 
\begin{align}
  \rho_{\rm F}(j) &= -\frac{1}{2\pi}\oint_{\cal{C}} dk \left[\chi_k^{(-)}(i)\right]^2 e^{2ikn} .
    \label{eq:rho_F_complex_plane}
\end{align}
Using the form Eq.~(\ref{eq:wavefunc_defs1}) of the Bloch states we find a pole of the integrand for $\epsilon_k=-V$ and a branch cut starting at $k_{\rm bc}$. The pole is only present for $t_2>t_1$ and $V<0$ and the residuum can be shown to be such that the contribution to the integral Eq.~(\ref{eq:rho_F_complex_plane}) cancels the edge state density Eq.~(\ref{eq:rho_edge}) for $\mu=0$, see Ref.~\cite{pletyukhov_etal_long} for details. This proves Eq.~(\ref{eq:rho_F_pole}). Closing the integration contour around the branch cut, we find for the branch cut contribution to the Friedel density
\begin{align}
      \rho_{\rm F}^{\rm (bc)}(n,i) = \frac{1}{\pi}e^{-2\kappa_{\rm bc}n}\, \text{Im} \int_0^\infty d\kappa \,
      \chi_{k_{\rm bc}+i\kappa + 0^+}^{(-)}(i)^2 e^{-2\kappa n} . 
      \label{eq:rho_F_bc_integral}
  \end{align}
Inserting Eq.~(\ref{eq:wavefunc_defs1}) and using
\begin{align}
    N_{k_{\rm bc}+i\kappa + 0^+}^{(-)} &= 2 R(\kappa) + 2iV \sqrt{-R(\kappa)} , \label{eq:N_bc}\\
    \text{Im}\frac{1}{N_{k_{\rm bc}+i\kappa + 0^+}^{(-)}} & = -\frac{V}{2\sqrt{-R(\kappa)}\left[V^2-R(\kappa)\right]} ,\\
    \text{Im}\frac{(V + \epsilon_{k_{\rm bc}+i\kappa + 0^+})^2}{N_{k_{\rm bc}+i\kappa + 0^+}^{(-)}} & = -\frac{V}{2\sqrt{-R(\kappa)}} ,
\end{align}
we find Eqs.~(\ref{eq:rho_F_bc_1}) and (\ref{eq:rho_F_bc_2}). 

To prove the asymptotic behavior Eq.~(\ref{eq:density_0_friedel_asym}) of the branch cut contribution
\begin{align}
  \rho_{\rm F}^{\rm (bc)}(n,i) \approx - \frac{c_i}{\sqrt{n}} e^{-2 \kappa_{\rm bc} n} , \quad n \gg
  \frac{W}{\Delta} \gg 1   ,
  \label{eq:rho_F_bc_asymp_i}
\end{align}
we consider the regime of small gap $\Delta\ll W=2t$ and note that $\kappa_{\rm bc}\approx \frac{2\Delta}{W}$ in this case. Therefore, for $n\gg \frac{W}{\Delta}=2\kappa_{\rm bc}^{-1}$, we get $\kappa\sim\frac{1}{n}\ll\kappa_{\rm bc}$ for the integration variable in Eqs.~(\ref{eq:rho_F_bc_1}) and (\ref{eq:rho_F_bc_2}). Expanding $R(\kappa)$ for $\kappa\ll\kappa_{\rm bc}$ by using Eq.~(\ref{eq:R}) we find 
\begin{align}
    R(\kappa) \approx -W\Delta \kappa .
    \label{eq:R_small_gap}
\end{align}
Inserting this result in Eq.~(\ref{eq:rho_F_bc_2}) for $\rho_F^{\rm (bc)}(n,2)$ and performing the integration we obtain
Eq.~(\ref{eq:rho_F_bc_asymp_i}) for $i=2$ with
\begin{align}
    c_2 = \frac{V}{\sqrt{\pi W\Delta}} .
    \label{eq:c_2}
\end{align}
To prove Eq.~(\ref{eq:rho_F_bc_asymp_i}) for $i=1$, we consider the case $V\gtrsim\delta t$ such that $\Delta\sim V$ and
\begin{align}
    R(\kappa)\sim W\Delta \kappa \sim \frac{W\Delta}{n}\ll \Delta^2 \sim V^2 .
\end{align}
Therefore, we can use $V^2-R(\kappa)\approx V^2$ in the integrand of Eq.~(\ref{eq:rho_F_bc_1}) and, together with Eq.~(\ref{eq:R_small_gap}), can calculate the integral with the result Eq.~(\ref{eq:rho_F_bc_asymp_i}) for $i=1$ and 
\begin{align}
    c_1 = \frac{(2\delta t - \Delta)^2}{V\sqrt{\pi W\Delta}} .
    \label{eq:c_1}
\end{align}

To prove Eq.~(\ref{eq:QB_result_U=0}) for the boundary charge we split $Q_{\rm B}=Q_{\rm P}+\delta Q_{\rm B}$ via Eq.~(\ref{eq:QB_splitting}). To calculate $Q_P$ we insert Eq.~(\ref{eq:rho_bulk_2}) in Eq.~(\ref{eq:QP}) and get 
\begin{align}
    Q_{\rm P} = -\frac{V}{4\pi} \int_0^\infty d\kappa \frac{1}{\sqrt{-R(\kappa)}} .
    \label{eq:QP_integration_kappa}
\end{align}
To obtain $\delta Q_{\rm B}$ we use Eq.~(\ref{eq:delta_rho_mu=0}) for $\delta\rho(j)=\rho_F^{\rm (bc)}(j)$ in Eq.~(\ref{eq:delta_QB}), and use Eqs.~(\ref{eq:rho_F_bc_1}) and (\ref{eq:rho_F_bc_2}) for the branch cut contribution of the Friedel density. Adding $Q_{\rm P}$ from Eq.~(\ref{eq:QP_integration_kappa}), we find after a lengthy but straightforward calculation
\begin{align}
    Q_{\rm B} = I_1 + I_2 ,
    \label{eq:QB_I12}
\end{align}
with
\begin{align}
    I_1 &= -\frac{V(t_2^2-t_1^2)}{4\pi}\int_0^\infty d\kappa 
      \frac{1}{\sqrt{-R(\kappa)}\left[V^2-R(\kappa)\right]} , \label{eq:I1} \\
    I_2 &= -\frac{Vt_1 t_2}{2\pi}\int_0^\infty d\kappa 
      \frac{\sinh(\kappa_{\rm bc} + \kappa)}{\sqrt{-R(\kappa)}\left[V^2-R(\kappa)\right]} . \label{eq:I2} 
\end{align}
Inserting Eq.~(\ref{eq:R}) for $R(\kappa)$, the integral $I_2$ can be analytically calculated with the result 
\begin{align}
    I_2 = -\frac{1}{4}\text{sign}(V) .
    \label{eq:I2_result}
\end{align}
Taking Eqs.~(\ref{eq:QB_I12}), (\ref{eq:I1}), and (\ref{eq:I2_result}) together we arrive at Eq.~(\ref{eq:QB_result_U=0}). 

Alternatively, one can write Eq.~(\ref{eq:QB_result_U=0}) for the boundary charge also via an integration over the real axis
\begin{align}
    Q_{\rm B} = - \frac{1}{2} \Theta(t_2-t_1) \text{sign}(V) + \tilde{I} ,
    \label{eq:QB_real_integration}
\end{align}
with
\begin{align}
    \tilde{I} &=- \frac{W V \delta t}{8\pi t_1 t_2} \int_{-\pi}^\pi  \frac{d k }{\varepsilon_k (\frac{2 \delta t^2}{t_1 t_2} + 1 +\cos{k})} 
    \label{eq:tilde_I} \\
    &= - \frac{ V \delta t}{\pi W \sqrt{4 t_1 t_2 + \Delta^2}} \Pi \left( \frac{4 t_1 t_2}{W^2} , \frac{2 \sqrt{t_1 t_2}}{\sqrt{4 t_1 t_2 + \Delta^2}} \right),
\end{align}
where $\Pi$ is the complete elliptic integral of the third kind. Closing the integration contour of Eq.~\eqref{eq:tilde_I} in the upper half of the complex plane, we split this integral into a pole and a branch cut contributions
\begin{align}
    \tilde{I} = \tilde{I}^{\rm (pole)} + \tilde{I}^{\rm (bc)} .
    \label{eq:tilde_I_splitting}
\end{align}
A straightforward calculation gives for the pole contribution 
\begin{align}
    \tilde{I}^{\rm (pole)} = \frac{1}{2} \Theta(t_2-t_1) \text{sign}(V) -\frac{1}{4} \text{sign}(V) ,
    \label{eq:tilde_I_pole}
\end{align}
while the branch cut contribution $\tilde{I}^{\rm (bc)}$ is identical to $I_1$, see above. Taking all together we find the equivalence of Eqs.~(\ref{eq:QB_real_integration}) and (\ref{eq:QB_result_U=0}). 

Using the representation Eq.~\eqref{eq:tilde_I} we study the limit $|V| \ll |\delta t|$. Approximating $\varepsilon_k \approx \sqrt{4 \delta t^2 +2 t_1 t_2 (1+\cos k)}$, we immediately get
\begin{align}
    Q_{\rm B} &\approx - \frac12 \Theta (t_2 - t_1) \text{sign} (V) - \frac{V}{4 \pi \delta t} E \left(\frac{\sqrt{4 t_1 t_2}}{W} \right),
\end{align}
where $E$ is the complete elliptic integral of the second kind. This proves Eq.~(\ref{eq:QB_quantization_half}). 
Assuming additionally $|\delta t| \ll W$, we can use the low-energy result Eq.~(\ref{eq:QB_low_energy_limit}) and get
\begin{align}
    Q_{\rm B} &\approx - \frac12 \Theta (t_2 - t_1) \text{sign} (V) - \frac{V}{4 \pi \delta t} .
\end{align}

For large $|V|\gg W,|\delta t|$ (atomic limit) we approximate
\begin{align}
    \frac{1}{\varepsilon_k} &= \frac{1}{|V| \sqrt{ 1+ \frac{2 t_1 t_2}{V^2} (\frac{2 \delta t^2}{t_1 t_2} +  1+\cos k)}} \nonumber \\
    & \approx \frac{1}{|V|} [ 1- \frac{ t_1 t_2}{V^2} (\frac{2 \delta t^2}{t_1 t_2} +  1+\cos k)].
\end{align}
It follows
\begin{align}
    \tilde{I} &=- \frac{W \text{sign} (V) \delta t}{8\pi t_1 t_2} \int_{-\pi}^\pi  \frac{d k }{\frac{2 \delta t^2}{t_1 t_2} + 1 +\cos{k}}  \label{eq:It_lead} \\
    &+  \frac{W \text{sign} (V) \delta t}{4 V^2} +  \frac{ \text{sign} (V) \delta t}{ W} {\mathcal O} \left(\frac{W^4}{V^4}\right).
\end{align}
Evaluating Eq.~\eqref{eq:It_lead} with the result
\begin{align}
    - \frac{ \text{sign} (V) \, \text{sign} (\delta t)}{4},
\end{align}
we obtain for the boundary charge in this parameter regime
\begin{align}
     Q_{\rm B} &\approx- \frac14 \text{sign} (V)  \left[  1 - \frac{W \delta t}{V^2} \right].
     \label{eq:QB_atomic_limit_app}
\end{align}
This proves Eq.~(\ref{eq:QB_atomic_limit}). 

The case $|\delta t|\ll |V| \ll W$ is treated by approximating
\begin{align}
    \frac{1}{\varepsilon_k} \approx \frac{1}{\sqrt{V^2 + \frac12 W^2(1+\cos k)}}
\end{align}
and
\begin{align}
    \tilde{I} &\approx - \frac{ V \delta t}{4\pi W^2} \int_{-\pi}^\pi  \frac{d k }{ \frac{ 4 \delta t^2}{ W^2} + \frac12 (1 +\cos{k})} \nonumber \\ 
    & \qquad \times \frac{1}{\sqrt{\frac{V^2}{W^2} + \frac12 (1+\cos k)}}.
\end{align}
It is necessary to estimate the latter integral for the two small parameters $\frac{|\delta t|}{W} \ll \frac{|V|}{W} \ll 1$. The main contribution is received from the vicinity of $k= \pi$. Expanding $\frac12 (1+\cos k) \approx \frac{x^2}{4}$, with $x= k-\pi$, and extending the integration limits to infinities, we obtain
\begin{align}
    \tilde{I} \approx - \frac{ 2 V \delta t}{\pi W^2} \int_{-\infty}^{\infty}  \frac{d x }{ \frac{16 \delta t^2}{W^2} + x^2} \frac{1}{\sqrt{\frac{4 V^2}{W^2} + x^2}}.
\end{align}
To perform this integral we deform the integration contour in the complex upper half-plane to embrace the pole $x= i \frac{4 |\delta t|}{W}$ and the branch cut starting at $x=i \frac{2 |V|}{W}$. Thus we obtain to the order ${\mathcal O }(\delta t /V)$
\begin{align}
    \tilde{I} \approx - \frac{ \text{sign} (V) \text{sign} (\delta t)}{4 } + \frac{\delta t}{\pi V}.
\end{align}
Adding the other contributions, we obtain the result
\begin{align}
     Q_{\rm B} &\approx- \frac14 \text{sign} (V)  + \frac{\delta t}{\pi V}
\end{align}
for this parameter regime. Together with Eq.~(\ref{eq:QB_atomic_limit}) this proves Eq.~(\ref{eq:QB_quantization_quarter}).

Finally, to derive the low-energy result Eq.~\eqref{eq:QB_low_energy_limit} for small gap $|\Delta|\ll W$, a convenient starting point is the representation Eq.~\eqref{eq:tilde_I_splitting} together with $I_1=\tilde{I}^{\rm (bc)}$ expressed as
\begin{align}
I_1 &= \frac{W \Delta^2 \sin 2 \gamma}{ 16 \pi t_1 t_2 \sqrt{2 t_1 t_2}} \int_{k_{bc}}^{\infty} \frac{d \kappa}{\cosh \kappa - \cosh \kappa_{\rm bc} + \frac{V^2}{2 t_1 t_2}} \nonumber \\
& \times \frac{1}{ \sqrt{\cosh \kappa - \cosh \kappa_{\rm bc}}} .
\label{eq:tilde_I_bc}
\end{align}
In particular, by introducing the new integration variable $x= \frac{\sqrt{2 t_1 t_2}}{\Delta}\sqrt{\cosh \kappa - \cosh \kappa_{\rm bc}}$ in Eq.~\eqref{eq:tilde_I_bc}, we cast it to
\begin{align}
    I_1 &=  \frac{W \sin 2 \gamma}{4 \pi \Delta} \int_{0}^{\infty} \frac{d x}{ x^2 + \cos^2 \gamma} \nonumber \\ & \times \frac{1}{\sqrt{(x^2+1) (x^2+1 + \frac{4 t_1 t_2}{\Delta^2})}}.
\label{eq:tilde_I_bc_x}
\end{align}
In the low-energy limit we have $W \approx \sqrt{4 t_1 t_2}$ as well as 
\begin{align}
    I_1 &\approx  \frac{\sin 2 \gamma}{4 \pi} \int_{0}^{\infty} \frac{d x}{ x^2 + \cos^2 \gamma} \frac{1}{\sqrt{x^2+1}} \nonumber \\
    &= \frac{\gamma}{2 \pi} - \frac12 \Theta_{\frac{1}{2} \pi< \gamma < \frac{3}{2} \pi} - \Theta_{\frac{3}{2} \pi <\gamma< 2 \pi},
\label{eq:tilde_I_bc_low}
\end{align}
where the last equality holds for $0 < \gamma < 2 \pi$. Combining this result with the other contributions, namely with
\begin{align}
    - \frac14 \text{sign}(V) = - \frac14 +\frac12 \Theta_{\frac{1}{2} \pi <\gamma < \frac{3}{2} \pi }, 
\end{align}
we arrive at Eq.~\eqref{eq:QB_low_energy_limit}. 

On the basis of Eq.~\eqref{eq:tilde_I_bc_x} we also estimate the leading correction to Eq.~\eqref{eq:QB_low_energy_limit}, which amounts to
\begin{align}
\frac{\sin 2 \gamma}{8 \pi} \left( \frac{\Delta}{W} \right)^2 \ln \frac{\Delta}{W} .
\label{eq:low_energy_correction}
\end{align}
Due to the large denominator, this correction is negligible even for $\Delta \sim W$, and therefore the low-energy result Eq.~\eqref{eq:QB_low_energy_limit} remains quantitatively accurate up to these gap values.


\begin{thebibliography}{}
\bibitem{footnote2} 
For generic tight-binding models in 1d see Ref.~\cite{pletyukhov_etal_preprint}, where it was shown that the boundary charge changes generically by $\bar{\rho}\,\text{mod}(1)$ under the translation by one lattice site towards the boundary.
\end{thebibliography}
\end{document}